\documentclass[aps,prb,footinbib,showpacs,twocolumn,showkeys,amssymb,amsmath,superscriptaddress,floatfix]{revtex4-1}
\pdfoutput=1

\usepackage{hyperref}

\usepackage{amsmath}
\usepackage{braket}
\usepackage{graphicx}
\usepackage{graphics}
\usepackage{color}
\usepackage{stackengine}

\usepackage{verbatim}

\begin{document}

\title{Entanglement entropy and out-of-time-order correlator in the long-range Aubry-Andr\'e-Harper model}

\author{Nilanjan Roy}\affiliation{Department of Physics, Indian Institute of Science Education and Research, Bhopal, Madhya Pradesh 462066, India}
\affiliation{Department of Physics, Indian Institute of Science, Bangalore 560012, India}
\author{Auditya Sharma}
\affiliation{Department of Physics, Indian Institute of Science Education and Research, Bhopal, Madhya Pradesh 462066, India}

\date{\today}

\begin{abstract}
We investigate the nonequilbrium dynamics of entanglement entropy and
out-of-time-order correlator (OTOC) of noninteracting fermions at
half-filling starting from a product state to distinguish the
delocalized, multifractal (in the limit of nearest neighbor hopping),
localized and mixed phases hosted by the quasiperiodic
Aubry-Andr\'e-Harper (AAH) model in the presence of long-range
hopping. For sufficiently long-range hopping strength a secondary
logarithmic behavior in the entanglement entropy is found in the mixed
phases whereas the primary behavior is a power-law the exponent of
which is different in different phases. The saturation value of
entanglement entropy in the delocalized, multifractal and mixed phases
depends linearly on system size whereas in the localized phase (in the
short-range regime) it is independent of system size. The early-time
growth of OTOC shows very different power-law behaviors in the
presence of nearest neighbor hopping and long-range hopping. The
late time decay of OTOC leads to noticeably different power-law
exponents in different phases. The spatial profile of OTOC and its
system-size dependence also provide distinct features to distinguish
phases. In the mixed phases the spatial profile of OTOC shows two
different dependences on space for small and large distances
respectively. Interestingly the spatial profile contains large
fluctuations at the special locations related to the quasiperiodicity
parameter in the presence of multifractal states.
\end{abstract}

\maketitle

\section{Introduction}
The nature of correlations between different parts of a system is of
fundamental interest in physics. Entanglement entropy has been a
popular measure of quantum correlations in many-body
systems~\cite{laflorencie2016quantum}. The study of entanglement in
stationary, equilibrium and nonequilibrium states has proven to be
insightful in a wide variety of contexts~\cite{eisert,vidal_entanglement,serbyn2013universal,alet,abanin2019colloquium}. In recent years,
\emph{out-of-time-order} correlators or OTOC, which have emerged as a
useful probe of quantum chaos~\cite{hashimoto2017out}, have gained
importance in a diverse set of fields ranging from high energy
physics~\cite{shenker2015stringy,maldacena2016bound,roberts2016lieb,roberts2014two}
to condensed matter
physics~\cite{swingle2018unscrambling,swingle2017slow,rozenbaum2017lyapunov,chen2017out,fan2017out,lewis2019unifying}
to quantum
information~\cite{garttner2018relating,halpern2019entropic}.  Devised
originally as a theoretical
measure~\cite{larkin1969quasiclassical,aleiner1996divergence},
considerable excitement has been generated from the recent
experimental measurement of OTOC using nuclear magnetic
resonance~\cite{li2017measuring,wei2018exploring,niknam2020sensitivity}
and trapped
ions~\cite{garttner2017measuring,landsman2019verified}. The OTOC is
generically defined as:
\begin{eqnarray}
C(x,t)=\langle [ \hat{W}(x,t), \hat{V}(0,0) ]^\dagger [\hat{W}(x,t), \hat{V}(0,0)] \rangle,
\label{otoc}
\end{eqnarray}
where $\hat{W}$ and $\hat{V}$ are arbitrary local operators separated
by a displacement $x$ and commute at $t=0$. Here $\langle.\rangle$
typically refers to a thermal average, although the expectation value
in specific states may also be of interest. Choosing both $W$ and $V$ to be both
Hermitian and unitary is particularly advantageous as Eq.~\ref{otoc} reduces to the compact
expression:
\begin{eqnarray}
C(x,t)= 2(1-Re[F(x,t)]),
\label{otoc2}
\end{eqnarray}
where $F(x,t)=\langle \hat{W}(x,t) \hat{V}(0) \hat{W}(x,t)
\hat{V}(0)\rangle$. At $t=0$ $C(x,t)$ is zero. Then it increases for
$t>0$ due to non-commutativity of $\hat{W}(x,t)$ and $\hat{V}(0)$.

Models that exhibit localization are a natural setup for investigation
of OTOC, in condensed matter systems.  A particularly important class
of such models is the family of models with quasi-periodic disorder,
that have sustained interest over several
decades~\cite{kohmoto_rev,kohmoto2,kohmoto3,rmp}. Unlike with Anderson
localization where even an infinitesimal random disorder results in
localization, a quasiperiodic disorder of finite strength is essential
for localization of a single particle even in one
dimension~\cite{aubry,harper}.  There has been a revival of interest
in quasiperiodic systems since their experimental realization using
ultra-cold
atoms~\cite{roati2008anderson,lahini,lye2005bose,lucioni2011observation}. Furthermore
the possibility of \emph{many-body} localization in such models has
triggered a lot of interest both from a
theoretical~\cite{huse2007,Pal2010,iyer} and an
experimental~\cite{Bordia} perspective. Apart from the delocalized and
localized phases, quasiperiodic systems can also host other nonergodic
phases~\cite{deng2019one,roy2020prescription} with their
characteristic properties. In this study, we numerically probe the
different phases using quantum dynamics of out-of-time-order
correlators. We also study the quantum dynamics of entanglement entropy to
complement and contrast against OTOC.

If $C(x,t)$ remains non-zero for an extended period of time one says
that the system has `scrambled'. For early time approach to scrambling
one expects $C(x,t)\sim e^{\lambda_{quant}(t-x/v_B)}$ where $\lambda_{quant}$ is
the `quantum Lyapunov exponent' which is bounded by $\lambda_{quant}\leq2\pi
k_BT/\hbar$ as conjectured in~\onlinecite{maldacena2016bound}. $v_B$
is called the `Butterfly velocity' which is also bounded by the
Lieb-Robinson bound~\cite{lieb}. Quantum systems in which $\lambda_{quant}$
approaches its bound are called fast
scramblers~\cite{sekino2008fast,sachdev2015bekenstein}. However, many
condensed matter systems exhibit a much slower growth and hence are
called slow scramblers. This includes the many-body localized systems
showing a power law
growth~\cite{swingle2017slow,chen2017out,huang2017out,fan2017out}
which itself may be contrasted with Anderson localized systems where
$C(x,t)$ is expected to be a constant~\cite{fan2017out}.  It should be
noted that $\lambda_{quant}$, although inspired by classical chaos is quite
different from its classical counterpart $\lambda_L$ that
characterizes chaotic motion in classical
systems~\cite{rozenbaum2017lyapunov,chavez2019quantum,jalabert2018semiclassical}.
The OTOC corresponding to classical chaos was found to grow as
$C(t)=\langle [q(t),p]^2\rangle\sim e^{2\lambda_Lt}$, where
$\lambda_L$ may become arbitrarily large.

 Also the late time dynamics of $C(x,t)$ has turned out to be quite
 interesting. An inverse power-law behavior has been seen in
 integrable quantum spin chains~\cite{lin2018out,bao2019out} and
 many-body localized systems~\cite{swingle2017slow}. Recently late
 time behavior of $C(x,t)$ has been proposed as a diagnostic to
 distinguish regular and chaotic quantum
 systems~\cite{fortes2020signatures,yan2019similar}. Although OTOC has
 been studied extensively in quantum systems, not many disordered
 integrable models have been
 addressed~\cite{riddell2019out,riddell2020out} in the context of the
 delocalization-localization transition. In addition to studies that
 look at the evolution of an initial thermal state, studies involving
 an initial product state in a nonequilibrium setting have also been
 carried
 out~\cite{chen2017out,lee2019typical,bordia2018out,riddell2019out}. Here
 we study OTOC starting from a CDW-type initial product state. We also
 study entanglement entropy which has been one of the most popular
 tools to characterize different many-body phases, especially in
 disordered quantum systems~\cite{alet}.

This paper is organized as follows. In Section~\ref{sec2} we introduce
the model and briefly discuss the various single particle phases shown
by it~\cite{deng2019one,roy2020prescription}. In Section~\ref{sec3} we
describe the results obtained from the nonequilbrium dynamics of the
entanglement entropy. In Section~\ref{sec4} we study the nonequlibrium
dynamics of OTOC. This section consists of two subsections:
Subsection~\ref{app_otoc} where we briefly describe the formalism for
noninteracting fermions and Subsection~\ref{results} where we discuss
the results for our model. Finally we conclude in
Section~\ref{conclusion}.

\section{The model}\label{sec2}
The one dimensional long-range Harper (LRH) model is given by the
Hamiltonian:
\begin{eqnarray}
H = -\sum\limits_{i<j}^{N} \bigg(\frac{J}{r_{ij}^\sigma} \hat{c}_i^\dagger \hat{c}_j + H.c.\bigg)\nonumber + \lambda\sum\limits_{i=1}^{N} \cos(2\pi\alpha i + \theta_p)\hat{n}_i,\nonumber\\
\label{ham}
\end{eqnarray}
where $\hat{c}_i^\dagger$ $(\hat{c}_i)$ represents the single particle
creation (destruction) operator at site $i$ and
$\hat{n}_i=\hat{c}_i^\dagger \hat{c}_i$, the number operator acting at
site $i$. We consider a lattice of total number of sites $N$, where
$r_{ij}$ is the geometric distance between the sites $i$ and $j$ in an
open chain. Here $\lambda$ is the strength of the quasi-periodic
potential with the quasiperiodicity parameter $\alpha$ which is a Diophantine
irrational number~\cite{modugno2009exponential}
e.g. $\alpha_g=(\sqrt{5}-1)/2$, $\alpha_s=(\sqrt{2}-1)$,
$\alpha_b=(\sqrt{13}-3)/2$ etc~\cite{diophantine,continuedfraction},
also known as the `golden mean', `silver mean', `bronze mean'
etc. $\theta_p$ is an arbitrary global phase. The strength of the long
range hopping is controlled by $J$ and the long range parameter in the
hopping $\sigma$. We set our units such that $J=1$ throughout this
article. In the $\sigma\to\infty$ limit, this model is the well-known
Aubry-Andr\'e-Harper(AAH) model~\cite{aubry,harper}. The AAH model has
a self-dual point $\lambda=2J$, where the model in position space maps
to itself in momentum space. As a consequence, all the eigenstates are
delocalized in position space for $\lambda<2J$ and localized for
$\lambda>2J$~\cite{modugno2009exponential}. Some filling-fraction
dependent properties of the AAH model have also been
reported~\cite{roy2019study,roy2020prescription}.

The single particle phase diagram of the LRH model has been chalked
out recently~\cite{deng2019one,roy2020prescription}. Along with the
delocalized and localized phases the phase digram contains mixed
phases where a certain fraction of delocalized eigenstates coexists
with multifractal or localized eigenstates. For the `golden mean'
$\alpha_g$ the mixed phases can be denoted as $P_q$ $(q=1,2,3...)$
where $\alpha_g^q$ fraction of eigenstates are delocalized and
$(1-\alpha_g^q)$ fraction of eigenstates are multifractal or localized
depending on whether $\sigma<1$ or $\sigma>1$. Hence $P_q$ phases for
$\sigma<1$ contain the delocalized-multifractal (DM) edges. $P_q$
phases for $\sigma>1$ contain the delocalized-localized (DL) edges,
also known as mobility edges. For the present numerical study we have
chosen some specific ($\lambda,\sigma$) values. For $\sigma=0.5$, we
consider $\lambda=0.1,0.5,1.0,2.0$ which correspond to the
delocalized, $P_1$, $P_2$ and $P_3$ phases (with DM edge)
respectively. For $\sigma=1.5$, we look at $\lambda=0.1,1.3,2.0,3.0$
which correspond to the delocalized, $P_1$, $P_2$ and $P_3$ phases
(with DL edge) respectively. For $\sigma=3.0$, we look at
$\lambda=0.1,1.7,2.1,2.5,5.0$ which correspond to the delocalized,
$P_1$, $P_2$, $P_3$ phases (with DL edge) and localized phases
respectively with $\sigma=3.0$ being essentially the short-range
limit. Next we discuss the nonequilibrium dynamics of free fermions in
the AAH and LRH models.

\section{Entanglement entropy} \label{sec3}
The study of out-of-equilibrium properties of disordered quantum
systems has been proved to be a very efficient tool to detect
delocalized and localized phases. The system is initially prepared in
a suitable state, and the properties of the time-evolved state are
tracked. Since a charge density wave (CDW) type of state (for fermions
at half-filling) is easily prepared in experiments involving
ultra-cold atoms, we consider a CDW state as the initial state in our
study. The initial state can be written as:
\begin{equation}
\ket{\Psi_{in}}=\prod\limits_{i=1}^{N/2} \hat{c}_{2i}^\dagger \ket{0}.
\label{cdw_state}
\end{equation}
We are mainly interested in the dynamics of entanglement entropy and
the out-of-time-order correlator which are of current interest for
integrable disordered quantum systems~\cite{mcginley2019slow}. In this
section we discuss the dynamics of entanglement entropy. OTOC will be
discussed in the following section. We will stick to the
quasiperiodicity parameter $\alpha_g=(\sqrt{5}-1)/2$ unless otherwise
mentioned.
\begin{figure}
\centering
\stackunder{\includegraphics[width=4.25cm,height=3.7cm]{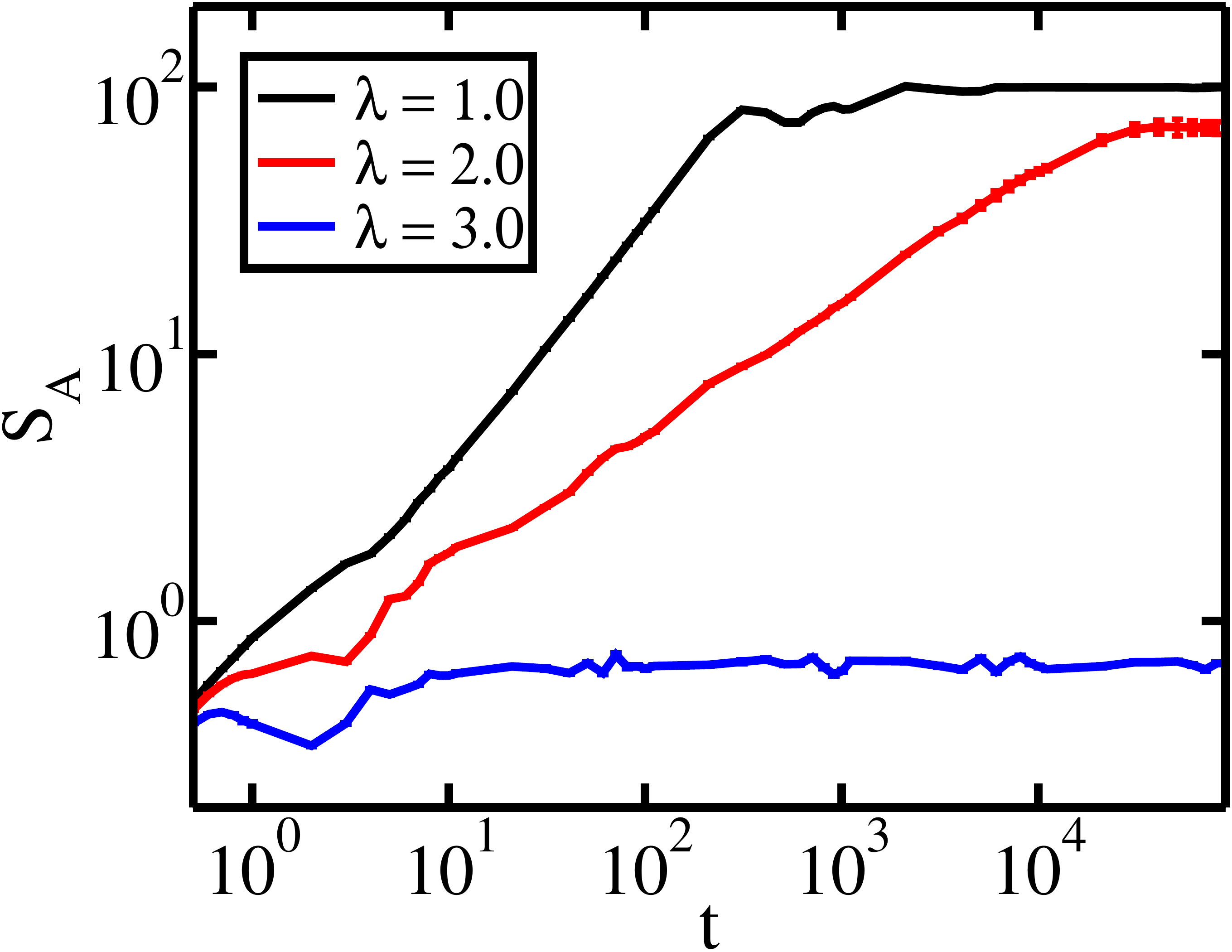}}{(a)}
\stackunder{\includegraphics[width=4.25cm,height=3.7cm]{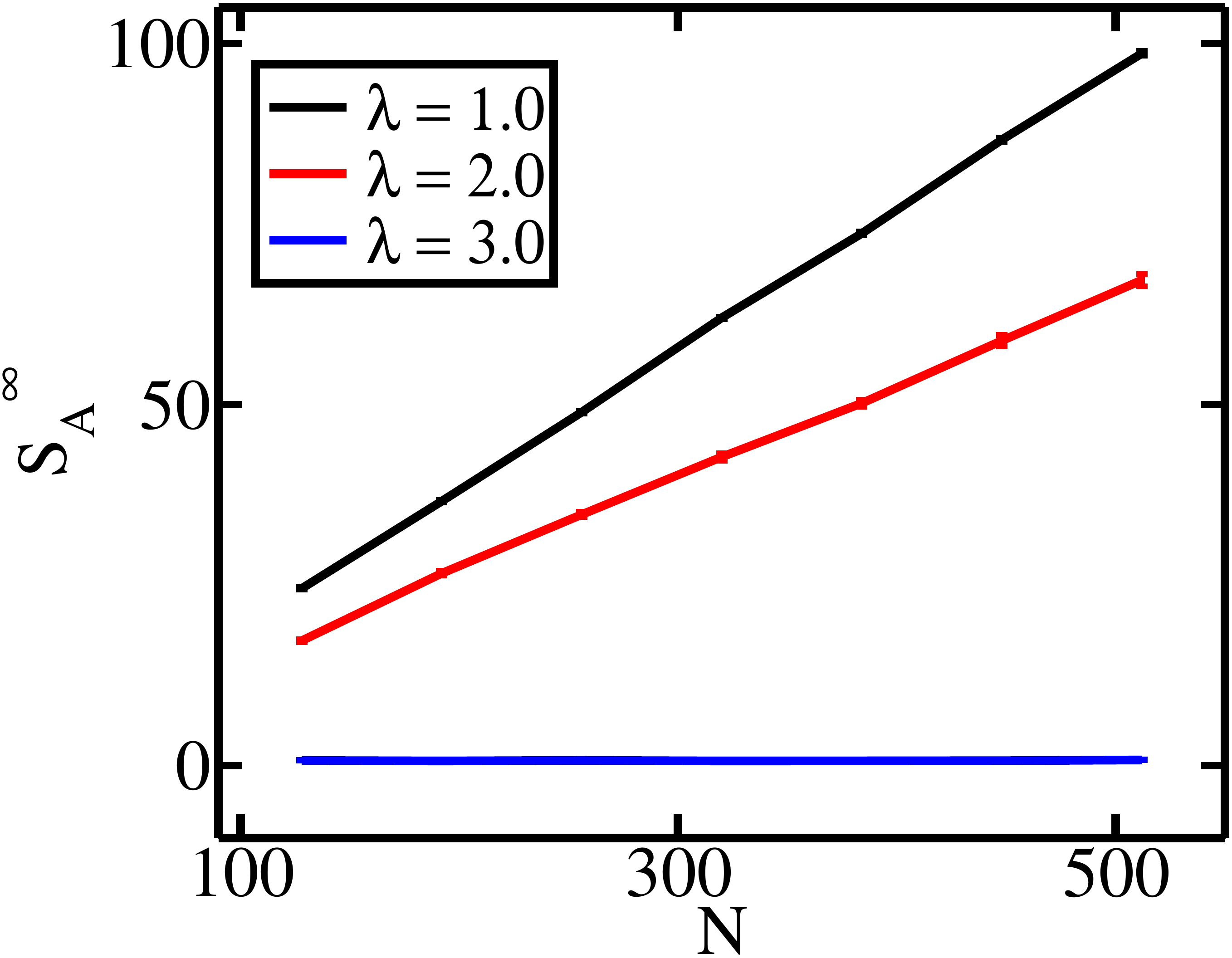}}{(b)}
\caption{Entanglement entropy in the AAH model. (a) The dynamics of
  the half-chain entanglement entropy $S_A$ with increasing values of
  $\lambda$ for free fermions at half-filling. Here system size
  $N=512$. (b) The system size $N$ dependence of the saturation value
  of the half-chain entanglement entropy $S_A^\infty$ of free fermions
  at half-filling for increasing values of $\lambda$. For all the
  plots, total number of $\theta_p$ realizations is $100$ with
  quasi-periodicity fixed to be $\alpha_g$.}
\label{eevst_aah}
\end{figure}
\begin{figure*}
  \centering
  \stackunder{\includegraphics[width=0.6\columnwidth,height=4.4cm]{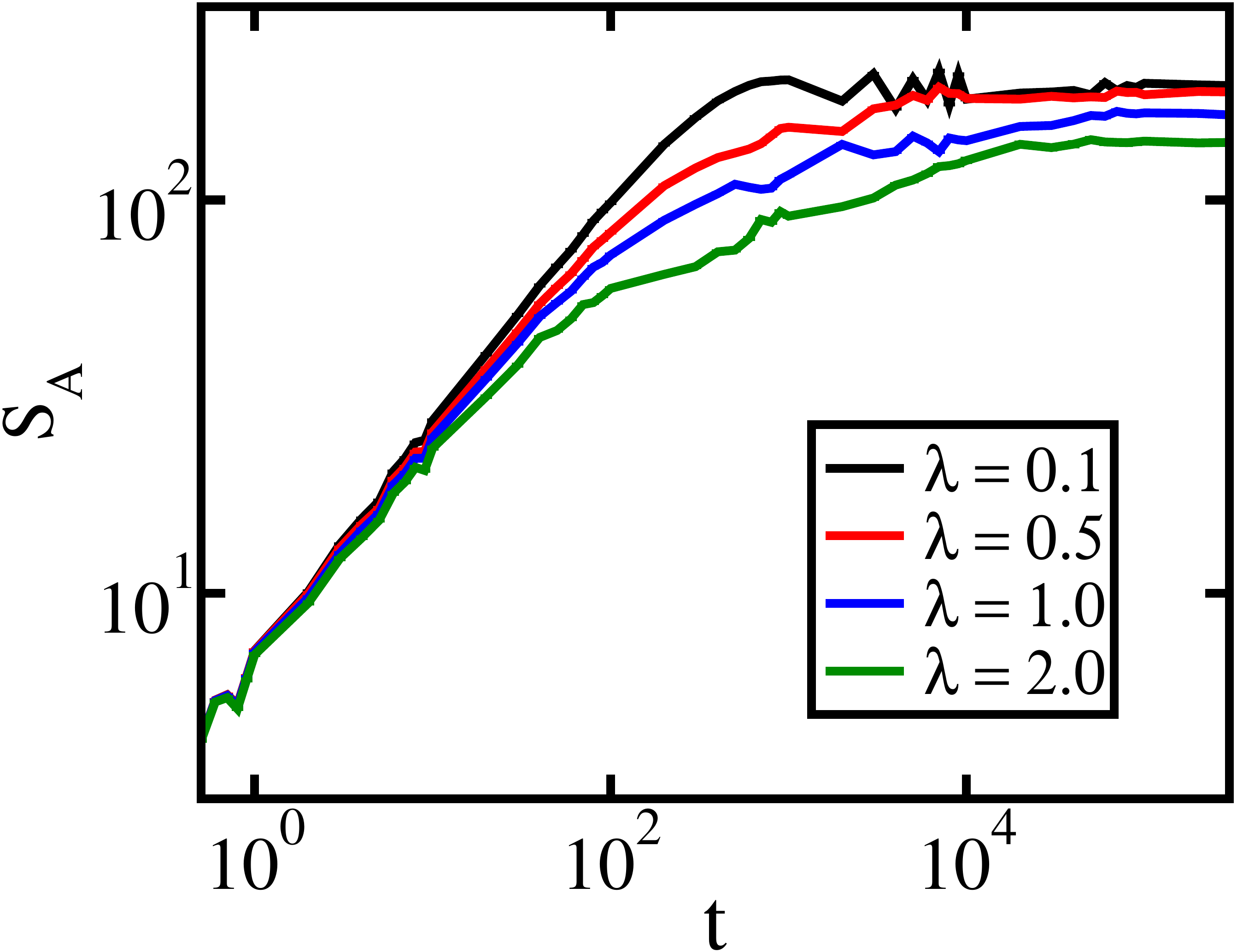}}{(a)}
   \stackunder{\includegraphics[width=0.6\columnwidth,height=4.4cm]{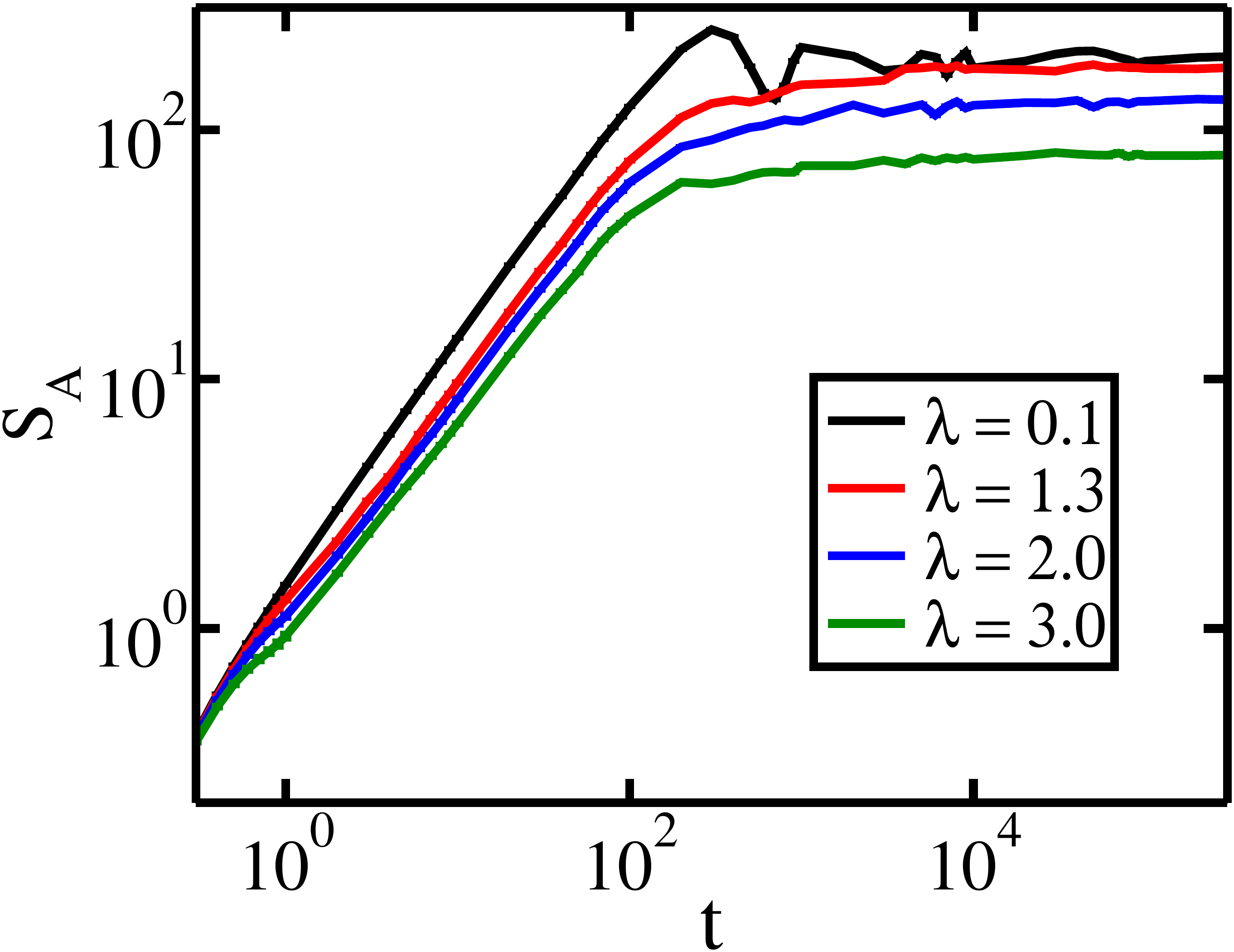}}{(b)}
   \stackunder{\includegraphics[width=0.6\columnwidth,height=4.4cm]{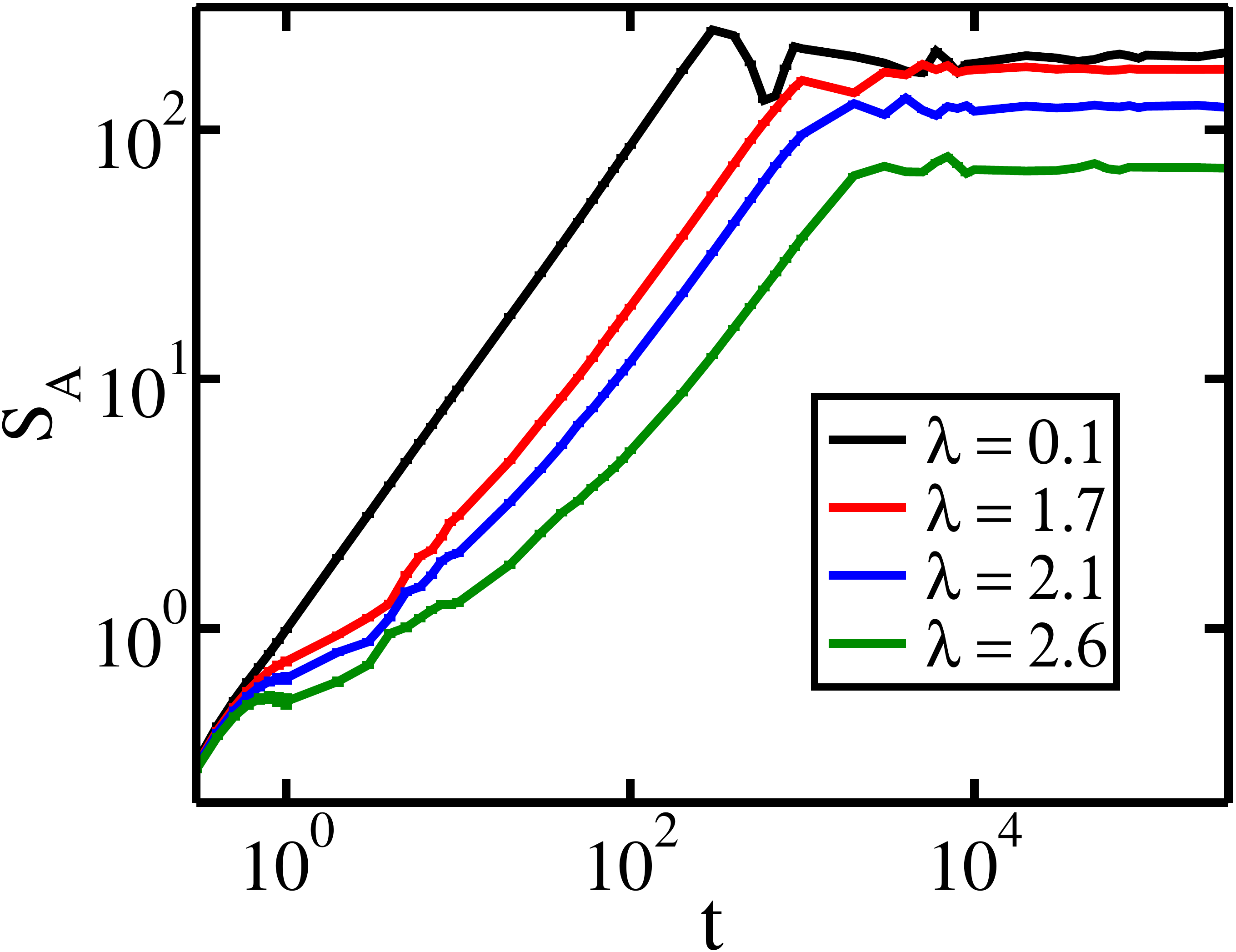}}{(c)}
  \caption{(a-c) The dynamics of the half-chain entanglement
    entropy $S_A$ with increasing values of $\lambda$ for free
    fermions at half-filling and for $\sigma=0.5,1.5$ and $3.0$
    respectively.  For all the plots system size $N=1024$.}
  \label{eevst_lrh}
\end{figure*}

When the overall state of the system is pure, entanglement entropy is
simply given by $S_A=-Tr(\rho_A \ln \rho_A)$ where $\rho_A$ is the
reduced density matrix of the subsystem A. We calculate the dynamics
of the half-chain entanglement entropy using free fermionic
techniques~\cite{peschel2003calculation,prb} that allow for the study
of significantly large system sizes. In the AAH model, the growth of
$S_A$ is ballistic in time in the delocalized phase $(\lambda=1)$ and
(almost) diffusive at the critical point $(\lambda=2)$ whereas there
is essentially no growth in the localized phase $(\lambda=3)$ as shown
in Fig.~\ref{eevst_aah}(a). These results are in agreement with those
of an earlier study of quench dynamics in the AAH
model~\cite{divakaran}. Fig.~\ref{eevst_aah}(b) shows that the
saturation value $S_A^\infty$ scales linearly with system sizes
$(S_A^\infty\propto N)$ at $\lambda=1$ and
$\lambda=2$, while $S_A^\infty\propto N^0$ for $\lambda=3$. Also we have
checked that these results remain independent of the choice of the
quasiperiodicity parameter $\alpha$.

\begin{figure}
\centering
\stackunder{\includegraphics[width=0.493\columnwidth,height=3.9cm]{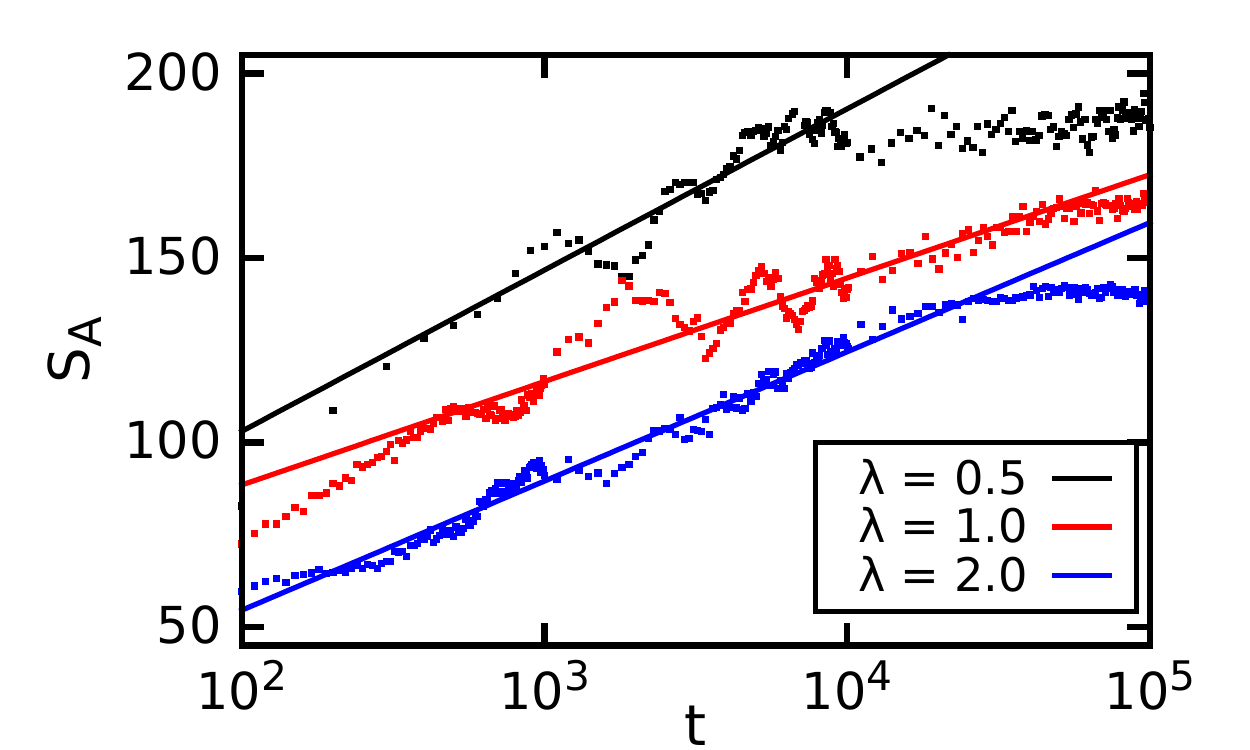}}{(a)}
\stackunder{\includegraphics[width=0.493\columnwidth,height=3.9cm]{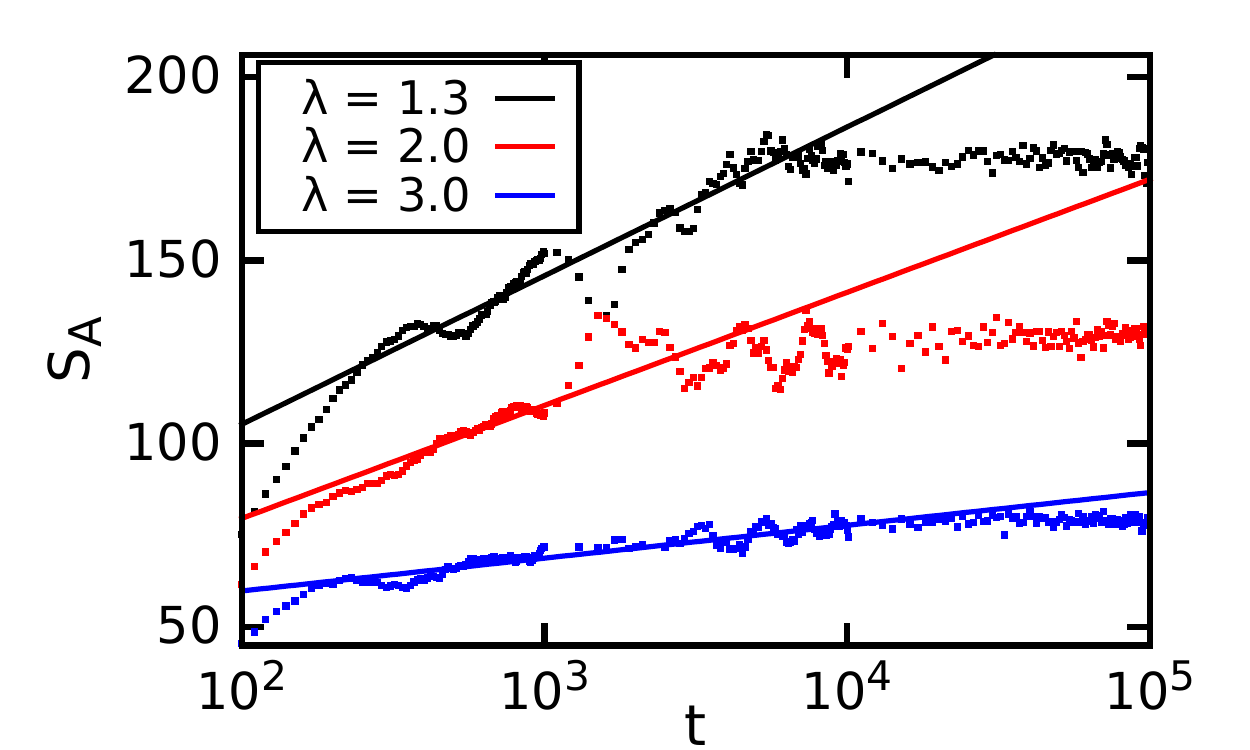}}{(b)}
\caption{(a) The secondary logarithmic growth of the half-chain
  entanglement entropy $S_A$ in the LRH model for $\sigma=0.5$ and
  $\lambda=0.5,1.0,2.0$ for which the best fits are $18.98\ln t +
  15.64$, $12.18\ln t + 32.38$ and $15.21\ln t - 15.54$
  respectively. (b) Similar plots for $\sigma=1.5$ and
  $\lambda=1.3,2.0,3.0$ for which best fits are $17.61\ln t + 24.15$,
  $13.39\ln t + 17.94$ and $3.89\ln t + 41.93$ respectively. The solid
  lines are best fits whereas the scattered points represent the
  corresponding data-points. The $x$-axis is in the log scale. For all
  the plots system size $N=1024$ and fermionic filling fraction is
  $1/2$.}
\label{entzoomed_lrh}
\end{figure}

The plots of $S_A$ as a function of time for the LRH model
are shown in Fig.~\ref{eevst_lrh}(a-c) for increasing values of
$\lambda$ and $\sigma=0.5,1.5$ and $3.0$ respectively. In the plots
for $\sigma=0.5$ and $\sigma=1.5$ each, $S_A$ shows two different
behaviors with time which can be noticed both in
Fig.~\ref{eevst_lrh}(a) and Fig.~\ref{eevst_lrh}(b). In
Fig.~\ref{eevst_lrh}(a) after the initial transient a power-law growth
is found followed by a secondary logarithmic growth (see
Fig.~\ref{entzoomed_lrh}(a)). The secondary growth appears presumably
due to the presence of the DM edge. It is to be noted that the
secondary growth is absent for $\lambda=0.1$ for which all the
eigenstates are delocalized. The primary growth in the dynamics of
$S_A$ can be fitted with a function $S_A(t)=c_1t^\beta+c_2$ to extract
the values of the power-law exponent $\beta$. For $\lambda=0.1$,
$\beta$ turns out to be $0.53$. For other values of
$\lambda=0.5,1.0,2.0$ which correspond to mixed phases with DM edges, $\beta=0.45,0.38$ and $0.31$ respectively.

In Fig.~\ref{eevst_lrh}(b) for $\sigma=1.5$ a primary power-law growth
and a subsequent secondary logarithmic (see
Fig.~\ref{entzoomed_lrh}(b)) growth is observed. For $\sigma=1.5$,
$\lambda=0.1$ corresponds to the delocalized phase whereas
$\lambda=1.3,2.0,3.0$ here correspond to mixed phases with DL
edges. For $\lambda=0.1,1.3,2.0$ and $3.0$, the power-law exponent
$\beta=0.89,0.82,0.80$ and $0.76$ respectively. The secondary growth
is again absent for $\lambda=0.1$ for which there is no DL edge. For
$\sigma=3.0$ the secondary growth is absent as seen from
Fig.~\ref{eevst_lrh}(c) since the LRH model approaches the short-range
AAH limit at this point. For $\lambda=0.1$ the growth of $S_A$ happens
ballistically as $\beta=1.0$ as in the delocalized phase of the
short-range AAH model. For $\lambda=1.7,2.1,2.6$ the system is in the
mixed phases with the DL edges. In the mixed phases the growth of
$S_A$ is initially less sensitive to the delocalized eigenstates due
to the short-rangeness of the system. After some time the delocalized
eigenstates start to dominate as indicated by the increasing change of
rate of $S_A$ in Fig.~\ref{eevst_lrh}(c). Right before reaching
saturation the power-law fit provides $\beta=0.84,0.82,0.79$ for
$\lambda=1.7,2.1,2.6$ respectively.  The secondary logarithmic growth
for $\sigma=0.5, 1.5$ are depicted in Fig.~\ref{entzoomed_lrh}(a,b)
respectively where the plots are fitted with the function
$S_A(t)=a_1\ln t + b_1$. Lots of intrinsic fluctuations can be seen in
the plots due to the quasiperiodicity in the system. The secondary
logarithmic growth tends to vanish in the short-range limit of hopping
as these are barely seen for $\sigma=3.0$ (see
Fig.~\ref{eevst_lrh}(c)).
Logarithmic growth of entanglement entropy has been seen recently in a few noninteracting randomly disordered systems~\cite{mcginley2019slow,hetterich2017noninteracting}. Logarithmic growth of entanglement in longrange interacting systems has also recently been addressed~\cite{lerose2020origin}.
The logarithmic behavior in quasiperiodically disordered long-ranged LRH model is attributed to the presence of mixed phases in the longrange regime since this feature is not found in the short-range regime or in absence of mixed phases.

We notice that the power-law exponent $\beta$ is larger for $\sigma>1$ as compared to $\sigma<1$. The counter-intuitive behavior of power-law exponent in the entanglement growth has been addressed earlier in a clean free fermionic long-range model~\cite{daley}.
It is noteworthy that the exponent $\beta$ changes very little with
$\lambda$ for $\sigma=1.5$ and $3.0$ for each of which
$(\lambda,\sigma)$ combinations correspond to the same $P_1, P_2$ and
$P_3$ phases with DL edges. This happens possibly because the
properties of the localized states barely vary in the different $P_q$
phases. On the other hand $\beta$ changes rapidly with $\lambda$ for
$\sigma=0.5$ in the presence of multifractal states the properties of
which may change significantly as one moves from $P_1$ to $P_2$ to
$P_3$ and so on. Another observation is that the late time dynamics
of $S_A$ slows down for $\sigma=1.5$ whereas it speeds up for
$\sigma=3.0$. This happens due to varying degrees of effectiveness of
the delocalized eigenstates in the presence of long-range and
short-range hopping.  In a particular $P_q$ phase (with DM or DL
edges) the values of all the exponents discussed here barely change
with $\lambda$ for a fixed value of $\sigma$. Similar results have
been discussed in a recent work~\cite{modak2020many}.  Also we have
checked that the qualitative behaviors of all the $S_A$ plots and the
values of the exponent $\beta$ change very little if, instead of
$\alpha_g$, one uses $\alpha_s$ or $\alpha_b$ for an initial half
filled CDW state. However, the exponents associated with the secondary
$S_A$ growth may change significantly as this part of the dynamics is
dominated by the multifractal or the localized single particle
eigenstates, the fraction of which depends on the choice of the
quasiperiodicity parameter in a particular $P_q$ $(q=1,2,3...)$
phase.
\begin{figure}
\centering
	\stackunder{\includegraphics[width=0.493\columnwidth,height=3.9cm]{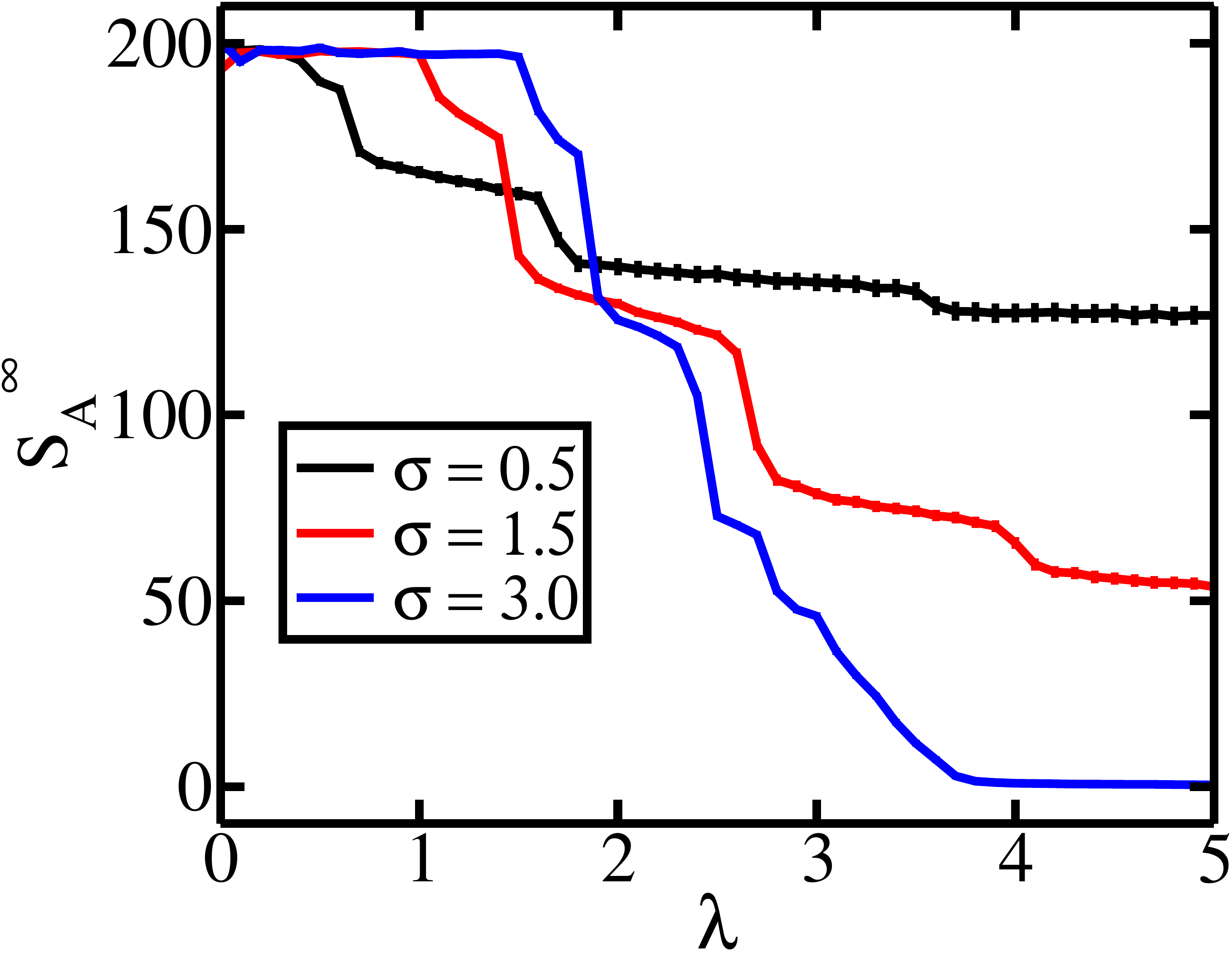}}{(a)}
		\stackunder{\includegraphics[width=0.493\columnwidth,height=3.9cm]{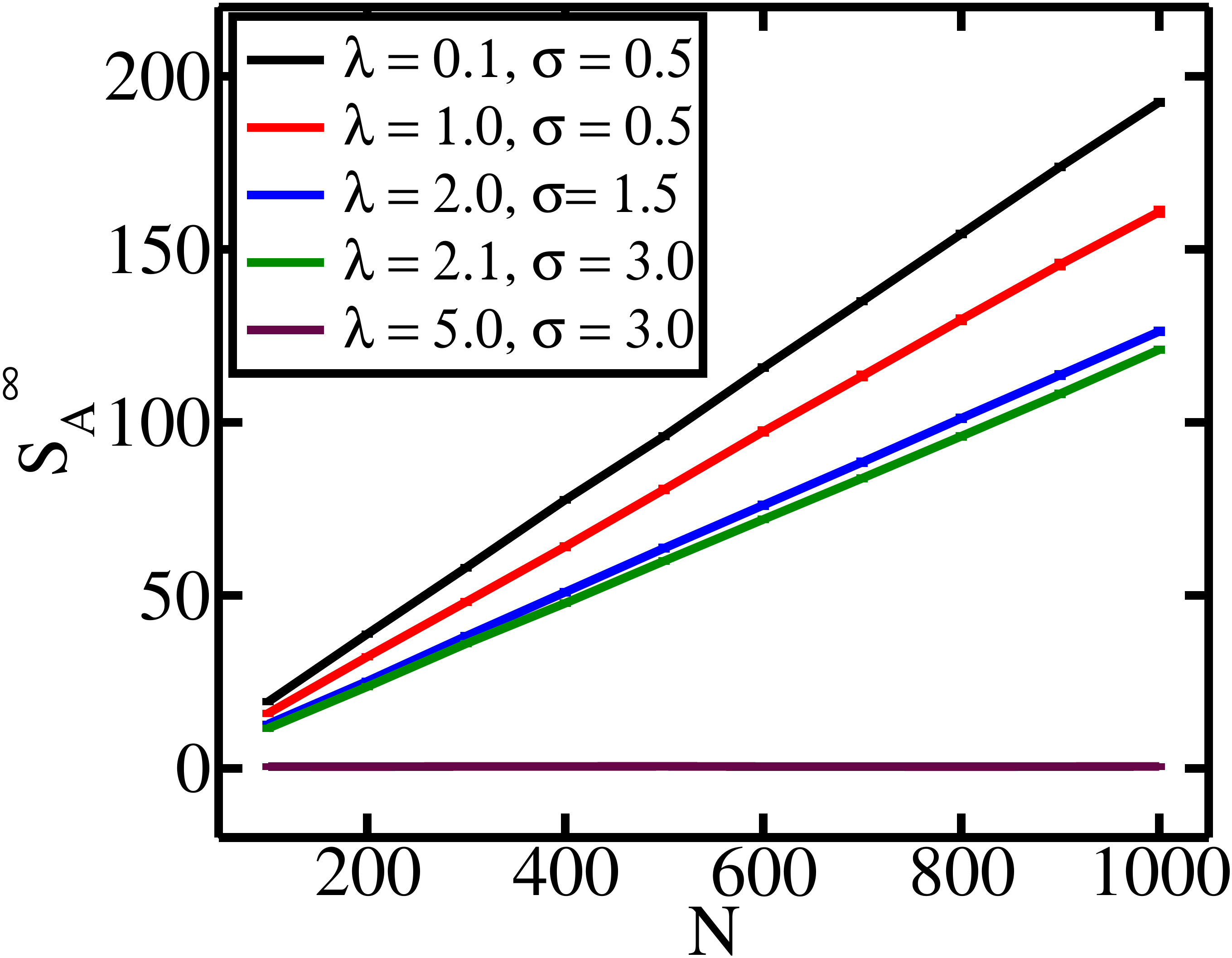}}{(b)}
	\caption{
	(a) The saturation value of the half-system $S_A$ as
	          a function of $\lambda$ for $\sigma=0.5,1.5$ and $3.0$
	          respectively for fermions at half-filling.  (b) The system
	          size $N$ dependence of the saturation value of the
	          half-chain entanglement entropy $S_A^\infty$ of free
	          fermions at half-filling for different combinations of
	          $\lambda$ and $\sigma$.
	}
	\label{eescaling_lrh}
\end{figure}

The saturation value of entanglement entropy $S_A^\infty$ turns out to
be a useful quantity. Fig.~\ref{eescaling_lrh}(a) shows $S_A^\infty$
as a function of $\lambda$ for $\sigma=0.5,1.5,3.0$. The steps
appearing in the plots denote the transitions from the delocalized-to
$P_1$-to-$P_2$-to-$P_3$ etc. phases. The $P_q$ phases have a fraction
of eigenstates that are multifractal for $\sigma=0.5$ and a fraction
of eigenstates that are localized for $\sigma=1.5,3.0$. Hence
$S_A^\infty$ is much lower for $\sigma=1.5,3.0$ than for $\sigma=0.5$
in these phases. Also we have looked at the system size $N$ dependence
of $S_A^\infty$ in these phases as shown in
Fig.~\ref{eescaling_lrh}(b). The combinations of ($\lambda$,$\sigma$)
are chosen in such a way that the system is in the delocalized phase
for $(0.1,0.5)$; $P_2$ phase with DM edge for $(1.0,0.5)$; $P_2$ phase
with DL edge for $(2.0,1.5)$, and $(2.1,3.0)$; and localized phase for
$(5.0,3.0)$. For the delocalized and mixed phases with DM or DL edges
$S_A^\infty\propto N$. In the localized phase $S_A^\infty\propto N^0$,
which is obtained effectively in the short-range AAH limit. Typically
in a sufficiently long-ranged regime one can obtain algebraic
localization such as seen in the random long-range hopping
model~\cite{prb}. In the random long-range hopping model an algebraic
localization dominated phase is found for $1<\sigma<2$ for which
$S_A^\infty$ varies sub-linearly with $N$~\cite{prb}.

\section{Out-of-time-order correlator} \label{sec4}
Out-of-time-order correlators (OTOC) are good observables to capture
chaos or information scrambling in quantum systems. The majority of
studies looking at OTOC have been in the context of localization
transitions in interacting
systems~\cite{chen2017out,fan2017out,lewis2019unifying,swingle2017slow,huang2017out}. However,
OTOC has been barely~\cite{riddell2019out,riddell2020out} addressed in
the literature in relation to the localization transition in
disordered noninteracting (quadratic) Hamiltonians. Our goal here is
to investigate OTOC as a distinguisher for the various phases found in
the AAH and LRH models.  In this work we choose the two
unitary-and-Hermitian operators $\hat{\sigma}_i^z$ and
$\hat{\sigma}_j^z$ at a distance $x=|i-j|$. The function $F(x,t)$ in
Eq.~\ref{otoc2} is then given by
\begin{eqnarray}
F(x,t)=\langle \hat{\sigma}_i^z(t) \hat{\sigma}_j^z(0) \hat{\sigma}_i^z(t) \hat{\sigma}_j^z(0) \rangle .
\label{otoc3}
\end{eqnarray}
We keep the position of the time evolved operator fixed at $i=N/2$. By
varying $j$ we study the scrambling of quantum information over the
lattice as a function of time. The initial state is fixed as the
product state of half-filled fermions defined in
Eq.~\ref{cdw_state}. For free fermions one can use the Jordan-Wigner
transformation $\hat{\sigma}_j^z=2\hat{n}_j-1$ to simplify the
expression of $F(x,t)$~\cite{riddell2019out}. We elaborate on this
ahead.

\subsection{Formalism}\label{app_otoc}
Here we provide a brief description of the formalism in relation to
OTOC which is used in this work.  Let us consider a generic quadratic
Hamiltonian:
\begin{equation} 
\hat{H}_{free} = \sum\limits_{i,j} H_{ij} \hat{c}_i^{\dagger} \hat{c}_j,  
\end{equation}
where $H_{ij}$'s are the elements of a Hermitian matrix $H$ and
$\hat{c}_i^{\dagger}$'s ($\hat{c}_i$'s) are fermionic creation
(annihilation) operators obeying the following anti-commutation
relations :
\begin{equation}
\{ \mathit{\hat{c}_i^{\dagger}}\textsf{,}\hspace{1mm}\mathit{\hat{c}_j} \} = \delta_{ij} \textsf{;}\hspace{4mm} \{\mathit{\hat{c}_i^{\dagger}}\textsf{,}\hspace{1mm}\mathit{\hat{c}_j^{\dagger}} \}\hspace{2mm} = \hspace{2mm} \{ \mathit{\hat{c}_i}\textsf{,}\hspace{1mm}\mathit{\hat{c}_j} \} \hspace{2mm} = \hspace{2mm}0.
\end{equation} 
Using the eigenvectors of the coupling matrix $H$, we can define new fermionic operators that diagonalize the Hamiltonian. If $A_{jk}$ represent the coefficients of 
the eigenvectors of the matrix $H$, we introduce the fermionic operators:
\begin{equation}
\mathit{\hat{d}_k^{\dagger}} = \sum\limits_j A_{jk}^\ast \mathit{\hat{c}_j^{\dagger}},\hspace{4mm} 
\mathit{\hat{d}_k} = \sum\limits_j A_{jk} \mathit{\hat{c}_j}
\end{equation}
that transform the Hamiltonian into a diagonal form:
\begin{equation}
\hat{H}_{free} = \sum\limits_{k} \epsilon_k \mathit{\hat{d}_k^{\dagger}}\mathit{\hat{d}_k}.
\end{equation}
Here $\mathit{\hat{d}_k^{\dagger}}$ ($\mathit{\hat{d}_k}$) creates (annihilates) a particle with energy $\epsilon_k$ and obeys similar anti-commutation relations as $\hat{c_i}$'s:
\begin{equation}
\{ \mathit{\hat{d}_k^{\dagger}}\textsf{,}\mathit{\hat{d}_l}\} = \delta_{kl}, \hspace{2mm}
\{ \mathit{\hat{d}_k}\textsf{,}\mathit{\hat{d}_l}\} = 0, \hspace{2mm}
\{ \mathit{\hat{d}_k^{\dagger}}\textsf{,}\mathit{\hat{d}_l^{\dagger}}\} = 0.
\end{equation}
Using the Heisenberg equation for operators\textsf{,} the time-evolved operators $\mathit{\hat{d}_k^{\dagger}}(t)$ and $\mathit{\hat{d}_k}(t)$ can be found.
\begin{eqnarray}
\frac{d}{dt}\mathit{\hat{d}_k} = \imath \left[\hat{H}_{free}\textsf{,} \mathit{\hat{d}_k} \right]=-\imath\epsilon_k \mathit{\hat{d}}_k,
\end{eqnarray}
which leads to
\begin{equation}
\mathit{\hat{d}_k} (t) = e^{-\imath\epsilon_k t} \mathit{\hat{d}_k}(t=0)
\end{equation}
and hence $\mathit{\hat{d}_k}^{\dagger} (t) = e^{\imath\epsilon_k t} \mathit{\hat{d}_k}^{\dagger}(t=0)$.
Using the relations: 
\begin{equation}
\mathit{\hat{c}_j^{\dagger}}(t) = \sum\limits_k A_{jk}\mathit{\hat{d}_k^{\dagger}}(t), \hspace{2mm}
\mathit{\hat{c}_j}(t) = \sum\limits_k A^*_{jk}\mathit{\hat{d}_k}(t)
\end{equation}
one finds the following anti-commutation relations between creation and annihilation operators at different times in position space. 
\begin{align}
&&\{\hat{c}_i^{\dagger} (t), \hat{c}_j\} = \sum\limits_{k} e^{\imath \epsilon_kt} A_{ik}^\ast A_{jk} = a_{ij}(t)\nonumber\\
&&\{\hat{c}_i(t), \hat{c}_j^{\dagger}\} = \sum\limits_{k} e^{-\imath \epsilon_kt} A_{ik}A_{jk}^\ast = {a}_{ij}^\ast(t)
\end{align}
along with $\{\hat{c}_i^{\dagger} (t), \hat{c}_j^{\dagger}\} = \{\hat{c}_i (t), \hat{c}_j\} = 0$, which are trivially satisfied.
Here the parantheses used to denote time are dropped from the operators for $t=0$. This convention is used further in the paper.

In this work we consider an initial product state of the form
\begin{equation}
\ket{\Psi} = \prod\limits_{j\in S}\hspace{1mm}\mathit{\hat{c}_j^{\dagger}} \ket{0}
\end{equation}
where $j$ refers to the index of the site which is occupied. Let $S$
be the set consisting of site indices of sites which are
occupied. The initial occupation matrix in position space is then given by
\begin{equation}
<\mathit{\hat{c}_i^{\dagger}}\mathit{\hat{c}_j}> = \bigg\{ {1 \hspace{4mm} \textsf{if} \hspace{1mm}i=j\hspace{2mm} \forall\hspace{2mm} i\in S \atop 0 \hspace{12mm}\textsf{otherwise} }.
\end{equation}
Using the Jordan-Wigner transformation $\hat{\sigma}_i^z=2\hat{n}_i -
1$ with $\hat{n}_i=\hat{c}_i^\dagger\hat{c}_i$ in Eq.~\ref{otoc3}
we have:
\begin{eqnarray}
F(x\textsf{,} t) &=& 16\langle\hat{n}_i(t)\hat{n}_j\hat{n}_i(t)\hat{n}_j\rangle 
+ 4\langle\hat{n}_j\hat{n}_i(t)\rangle -4\langle\hat{n}_i(t)\hat{n}_j\rangle
\nonumber\\ && 
-8\langle\hat{n}_i(t)\hat{n}_j\hat{n}_i(t)\rangle-8\langle\hat{n}_j\hat{n}_i(t)\hat{n}_j\rangle   + 1 .
\label{otocF}
\end{eqnarray} 
In this work we have kept $i=N/2$ where $N$ is the number of sites in the lattice and calculated $F(x,t)$ by varying $j$.
For the case $j \in S$ such that $\mathit{\hat{c}_j^{\dagger}}\ket{\Psi} = 0$, Eq.~\ref{otocF} can be written as~\cite{riddell2019out} 
\begin{equation}
F(x\textsf{,} t) = 8\vert a_{ij}\vert^2\left<\hat{n}_i(t)\right> - 8\vert a_{ij}\vert^2 + 1 .
\end{equation}
For $j\notin S$,  $\mathit{\hat{c}_j}\ket{\Psi} = 0$ which leads to
\begin{equation}
F(x\textsf{,} t) = 1 - 8\vert a_{ij}\vert^2\left<\hat{n}_i(t)\right>.
\end{equation}

\subsection{Results} \label{results}
We now discuss the OTOC-related results for the AAH and LRH models.

\begin{figure}
  \centering
  \stackunder{\includegraphics[width=4.25cm,height=3.7cm]{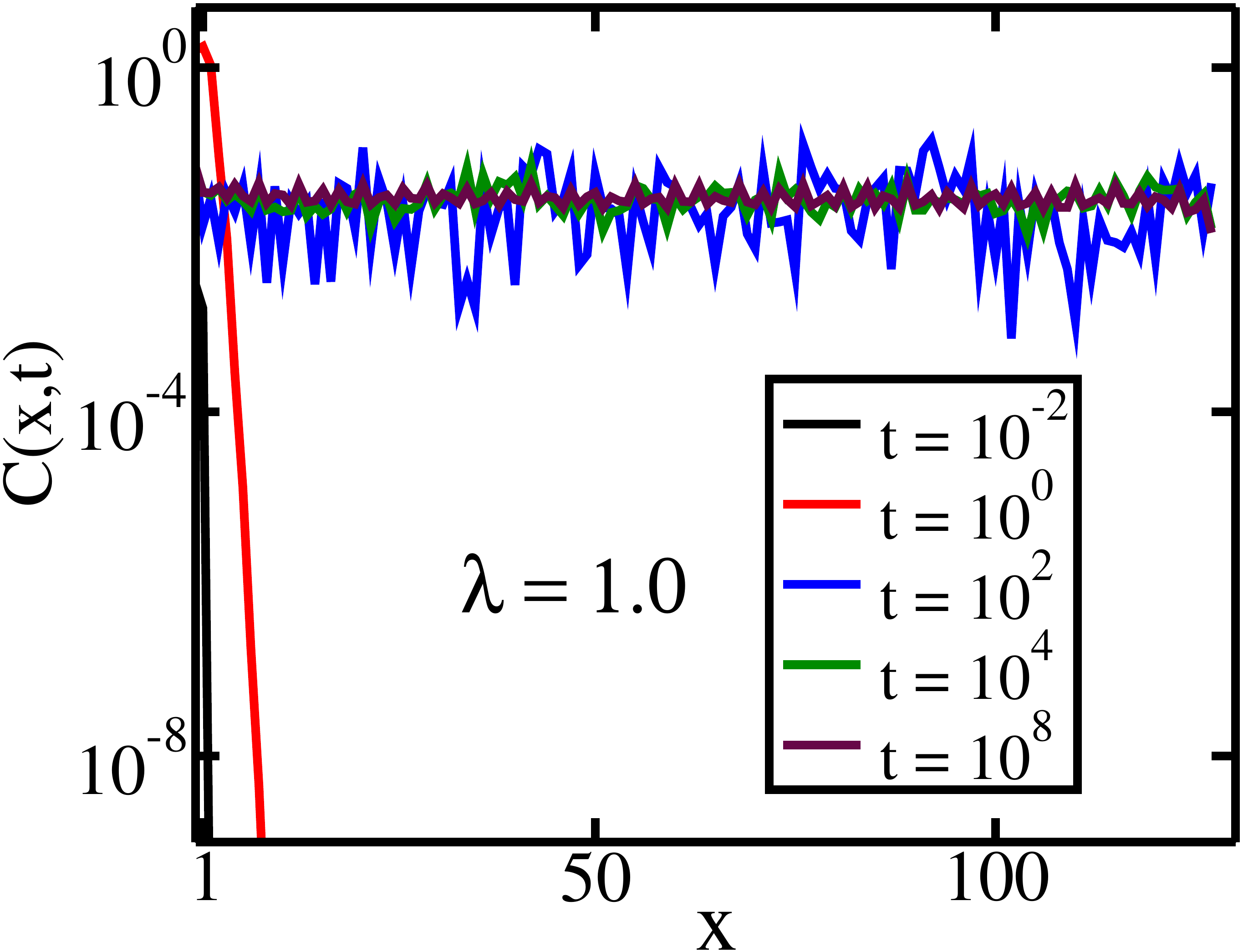}}{(a)}
    \stackunder{\includegraphics[width=4.25cm,height=3.7cm]{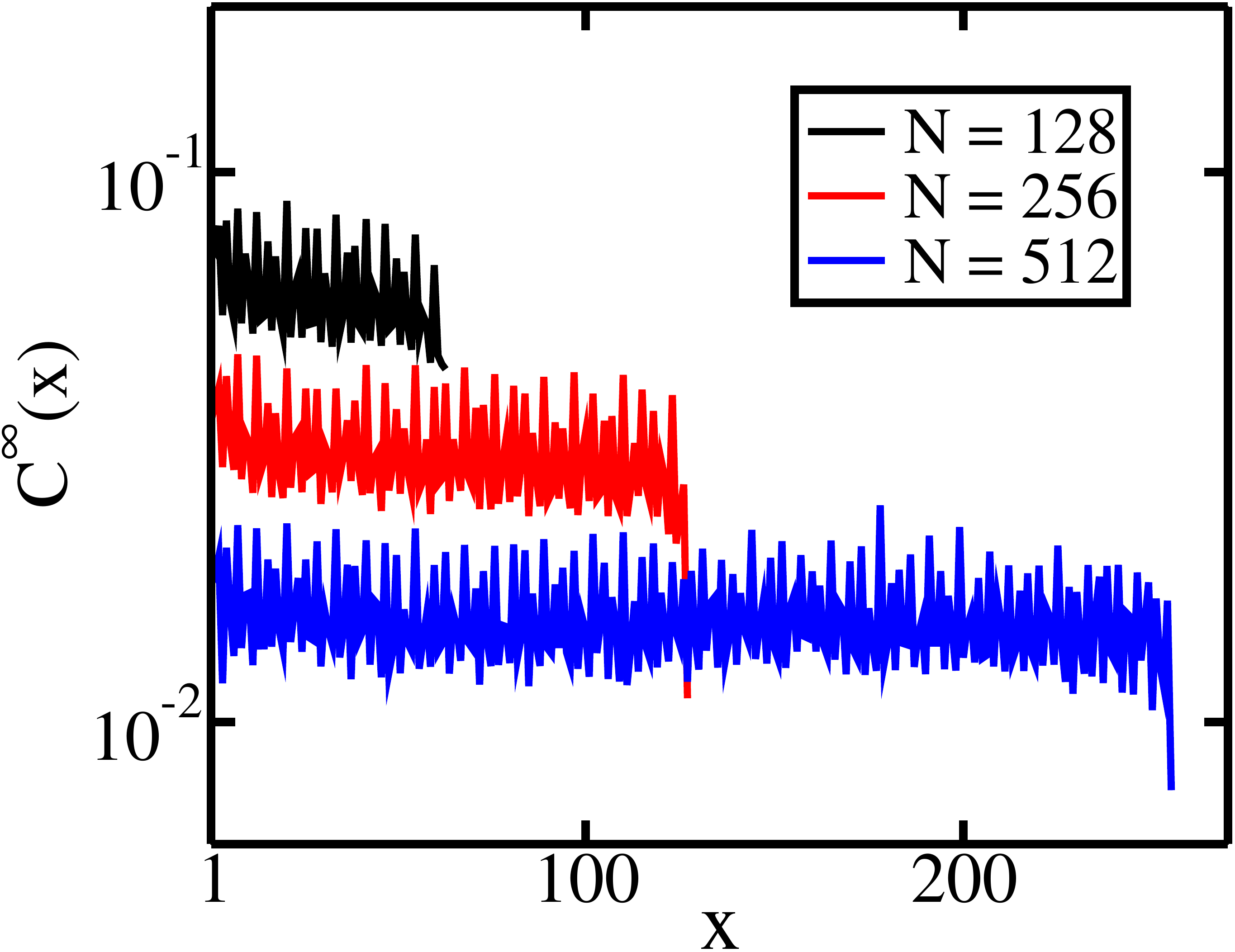}}{(d)}
    \stackunder{\includegraphics[width=4.25cm,height=3.7cm]{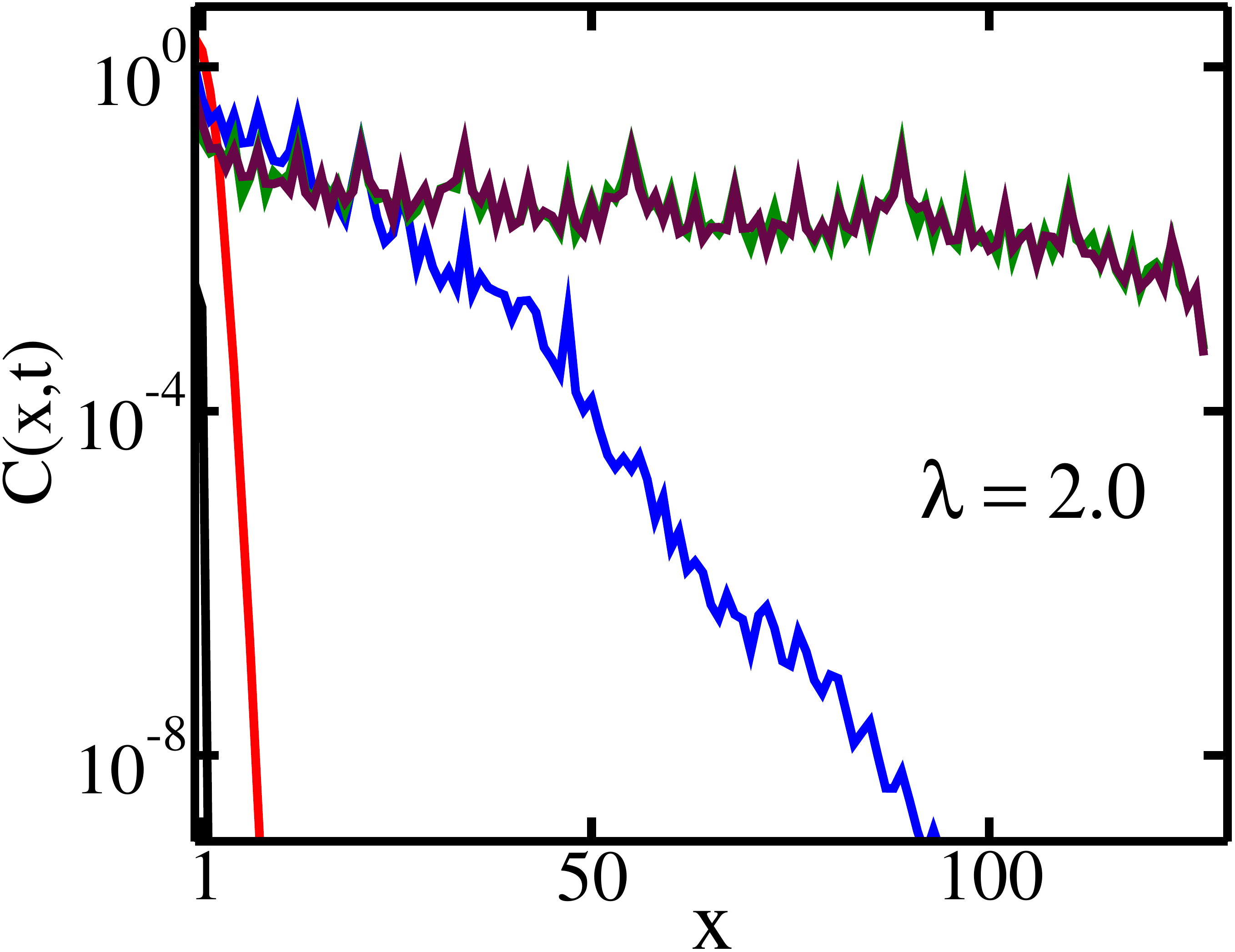}}{(b)}
    \stackunder{\includegraphics[width=4.25cm,height=3.7cm]{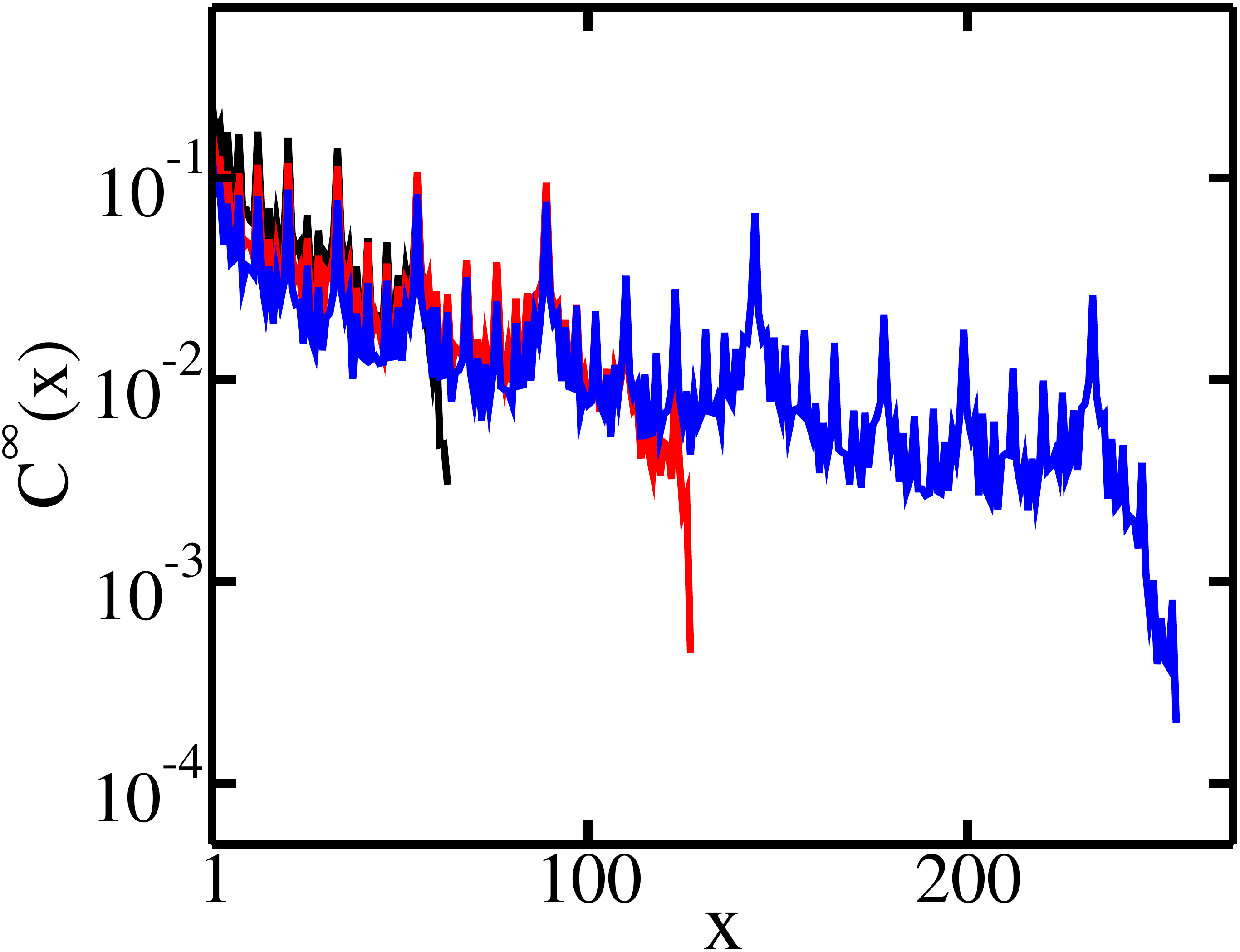}}{(e)}
    \stackunder{\includegraphics[width=4.25cm,height=3.7cm]{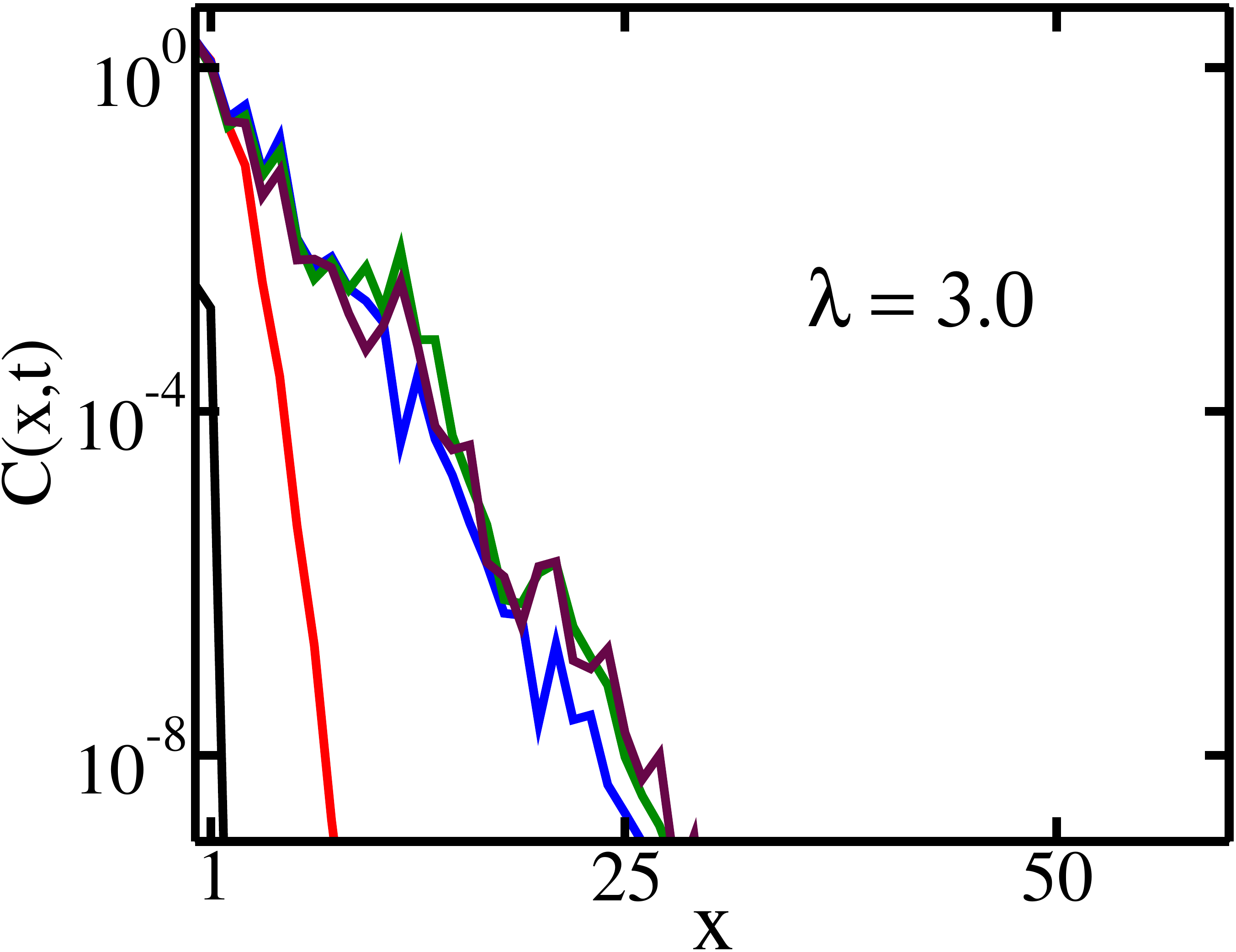}}{(c)}
    \stackunder{\includegraphics[width=4.25cm,height=3.7cm]{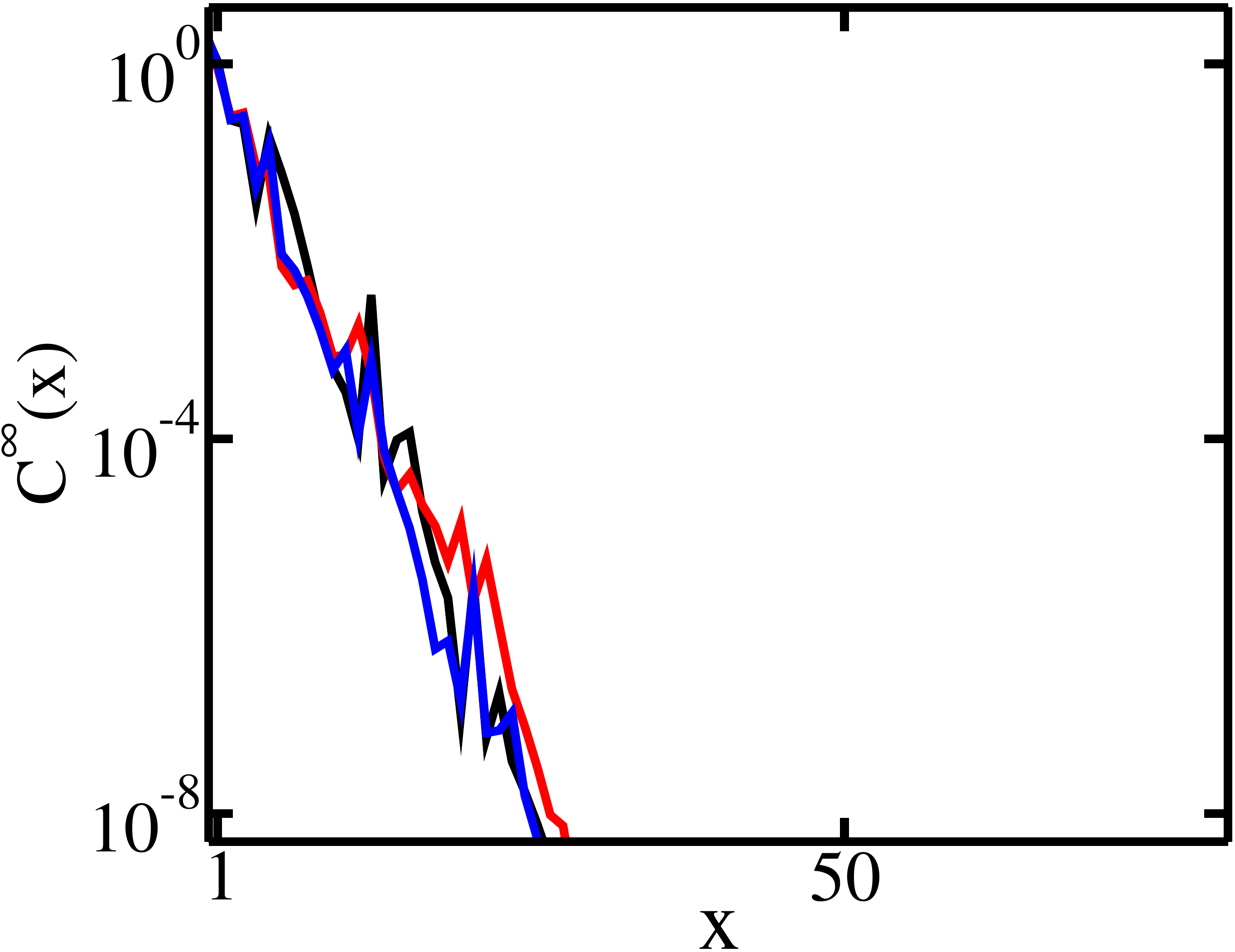}}{(f)}
  \caption{OTOC in the AAH model. (a-c) OTOC $C(x,t)$ as a function of
    distance $x$ at different instants $t$ for $\lambda=1.0,2.0$ and
    $3.0$ respectively. System size $N=256$. The plot legend shown in
    (a) also applies to (b) and (c). (d-f) Saturation value $C^\infty(x)$
    as a function of distance $x$ for increasing system sizes $N$ and
    for $\lambda=1.0,2.0$ and $3.0$ respectively. The plot legend
    shown in (d) also applies to (e) and (f). For all the plots, total
    number of $\theta_p$ realizations is $500$.}
  \label{otocprof_aah}
\end{figure}

{\it AAH model}: First we calculate $C(x,t)$ in the AAH model. The
profiles of $C(x,t)$ in position space for increasing instants of time
are shown in Fig.~\ref{otocprof_aah}(a-c) for $\lambda=1.0,2.0$ and
$3.0$ respectively. At $t=0$ $C(x)$ is zero for all $x$ because
$F(x,0)$ reduces to the squares of Pauli matrices yielding unity in
Eq.~\ref{otoc3}. Then $C(x)$ starts developing for small values of the
distance $x$ due to the non-commutation of the matrices
$\hat{\sigma}_i^z(t)$ and $\hat{\sigma}_j^z(0)$ for small $x$ at early
times. During this period of time $C(x)$ attains high values for small
$x$ while the maximum value of $C(x)$ happens to be at $x=1$. This is
shown in Fig.~\ref{otocprof_aah}(a) for $\lambda=1$. Then $C(x,t)$
starts decreasing for small $x$ whereas it keeps growing for large
values of $x$ due to the spreading of non-commutativity among Pauli
matrices. In the long run $C(x,t)$ shows a uniform dependence on $x$ for
$\lambda=1$ (see Fig.~\ref{otocprof_aah}(a)) when $S_A$ also reaches
saturation.
For the critical point $\lambda=2$ the initial dynamics of $C(x,t)$
shown in Fig.~\ref{otocprof_aah}(b) is similar to that for
$\lambda=1$. But in the long-time limit $C(x,t)$ shows a non-uniform
dependence on $x$ with occasionally large fluctuations especially at
$x=21,34,55$ etc. which are terms in the Fibonacci sequence of the
`golden mean'~\cite{fibonacci}. These large fluctuations appear
possibly due to the multifractal nature of the eigenstates. In
Fig.~\ref{otocprof_aah}(c) for the localized phase at $\lambda=3$,
$C(x,t)$ grows for small $x$ at early times while the subsequent decay
is absent in the dynamics. Eventually in the long-time limit $C(x)$
drops exponentially with $x$ i.e. $C(x)\sim
e^{-x/{\xi_{OTOC}}}$~\cite{riddell2019out} such that $C(x)\neq0$ for
$x<\xi_{OTOC}$ but is zero for large $x$. $\xi_{OTOC}$ decreases with
$\lambda$ in the localized phase.

Also we analyze the system size $N$-dependence of the spatial profile
of $C^\infty(x)$ in the long-time limit as shown in
Fig.~\ref{otocprof_aah}(d-f) for $\lambda=1,2,3$ respectively. For
$\lambda=1$, $C^\infty\propto1/N$.  This can be explained by looking
at the long-time behavior of $|a_{ij}(t)|^2$ defined in
Section~\ref{app_otoc}. $\lim\limits_{T\rightarrow\infty}
\frac{1}{T}\int\limits_{0}^{T} dt |a_{ij}(t)|^2=\sum\limits_{k}
|A_{ik}|^2 |A_{jk}|^2$, which scales with $1/N$ as $A_{ik}\propto
1/\sqrt{N}$ in the delocalized phase. At the critical point
$\lambda=2$, $C^\infty$ depends on $x$ and shows a sub-linearly decreasing
dependence with $N$ except on the points where large fluctuations are observed due to
the multifractal nature of the eigenstates. At these special points the $N$-dependence is not regular. The number of these large
fluctuations increases with $N$. However, in the localized phase for
$\lambda=3$, $C^\infty\propto N^0$ for $x<\xi_{OTOC}$ and is in any
case zero for large $x$.

\begin{figure}
  \centering
  \stackunder{\includegraphics[width=4.25cm,height=3.7cm]{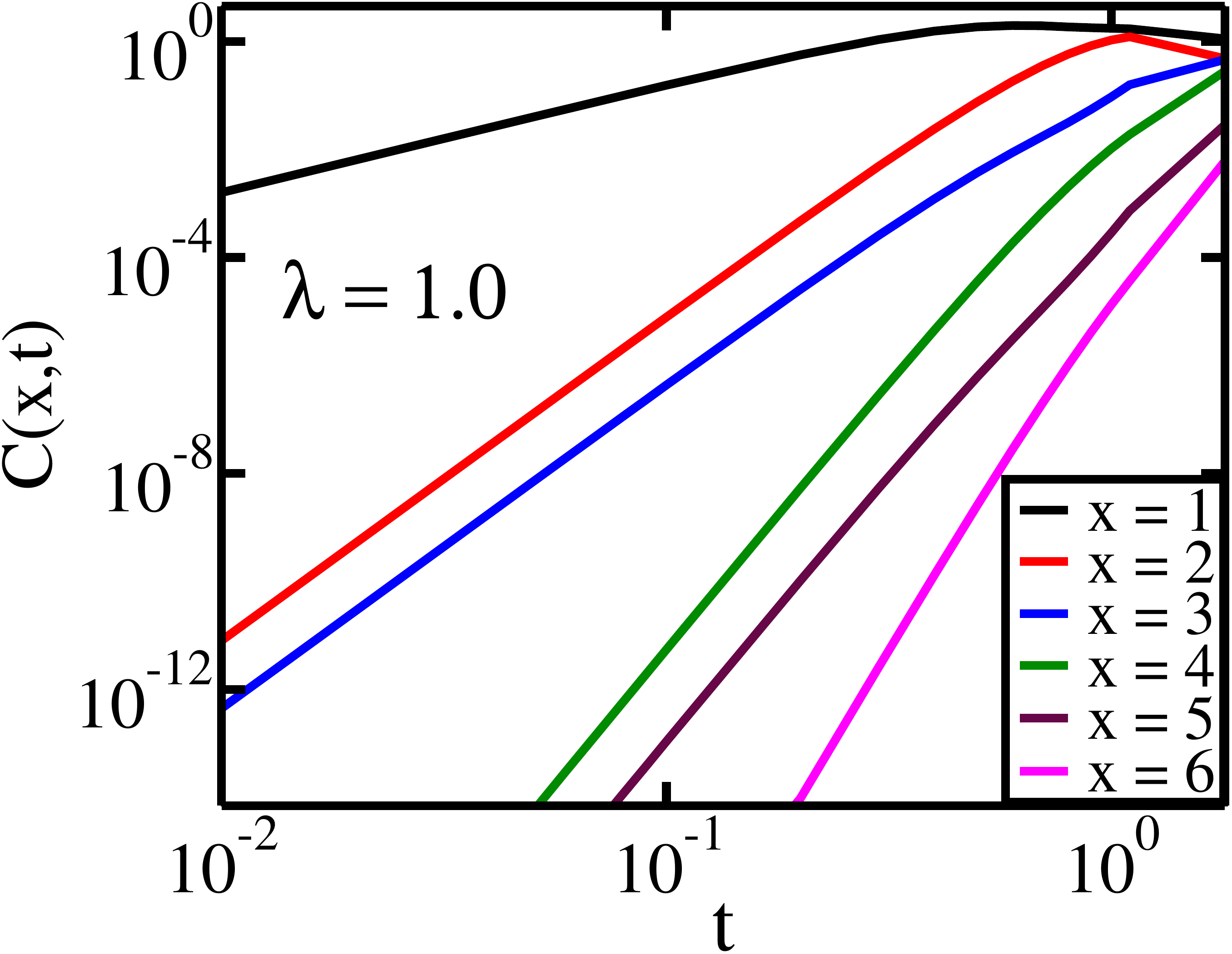}}{(a)}
    \stackunder{\includegraphics[width=4.25cm,height=3.7cm]{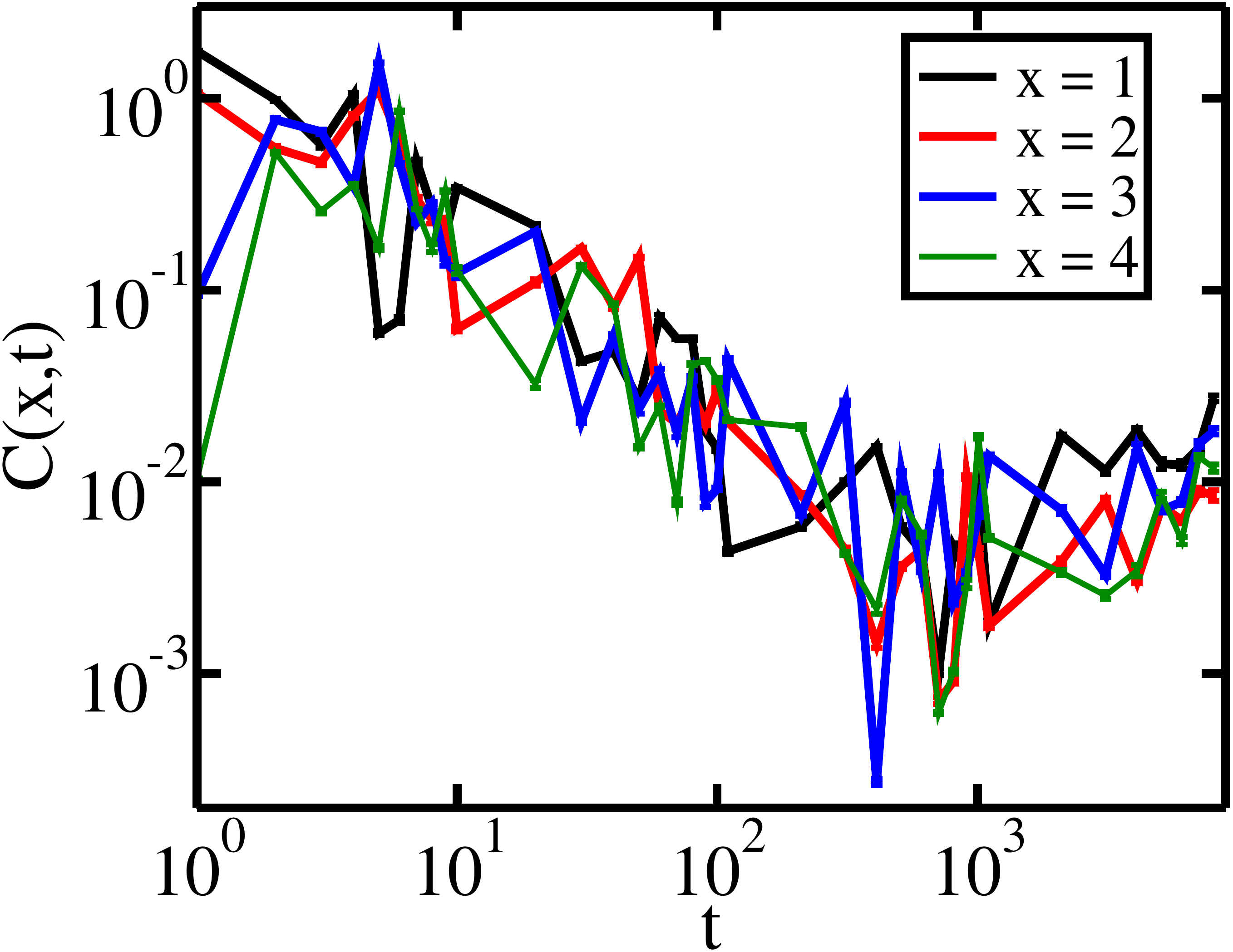}}{(d)}
    \stackunder{\includegraphics[width=4.25cm,height=3.7cm]{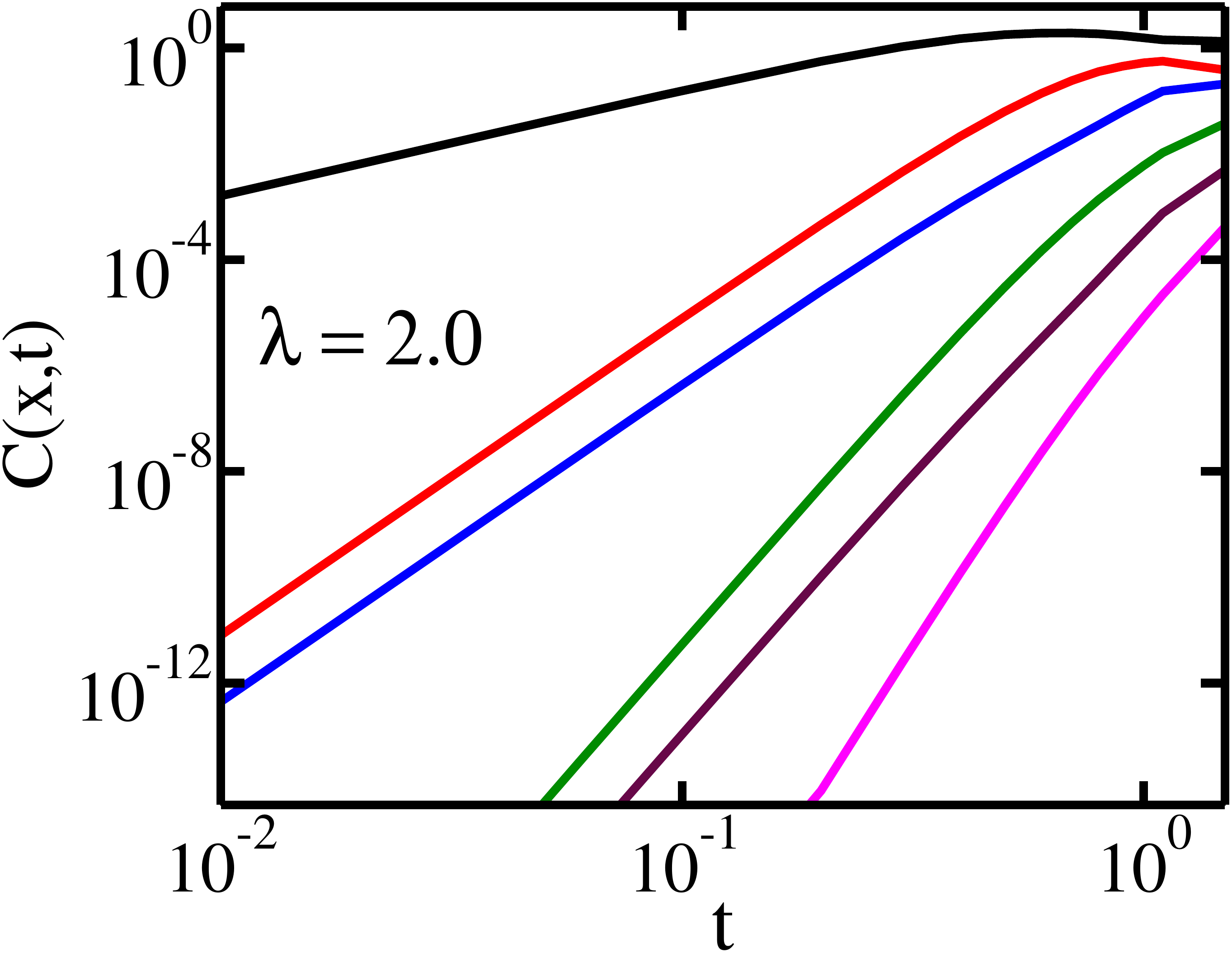}}{(b)}
    \stackunder{\includegraphics[width=4.25cm,height=3.7cm]{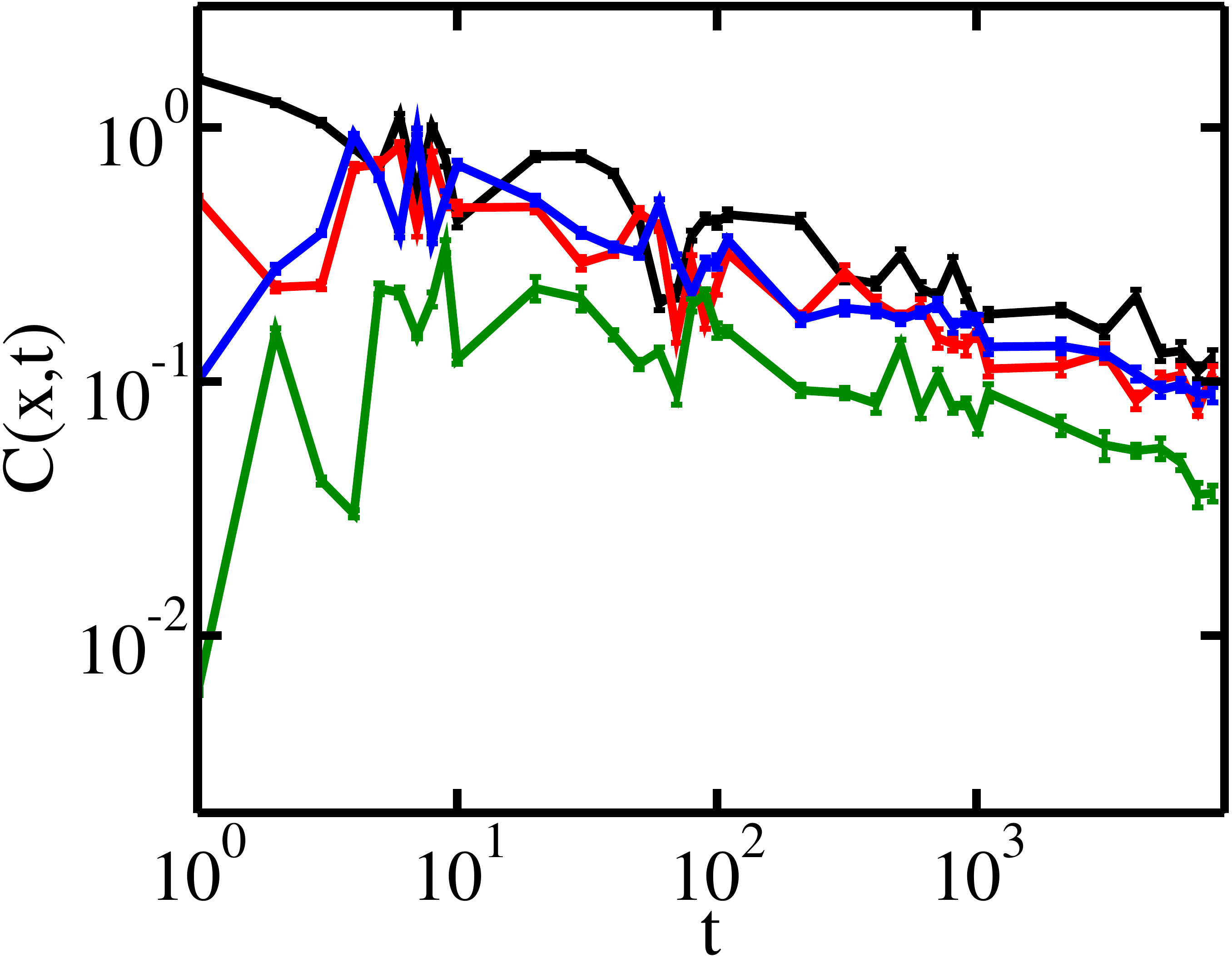}}{(e)}
    \stackunder{\includegraphics[width=4.25cm,height=3.7cm]{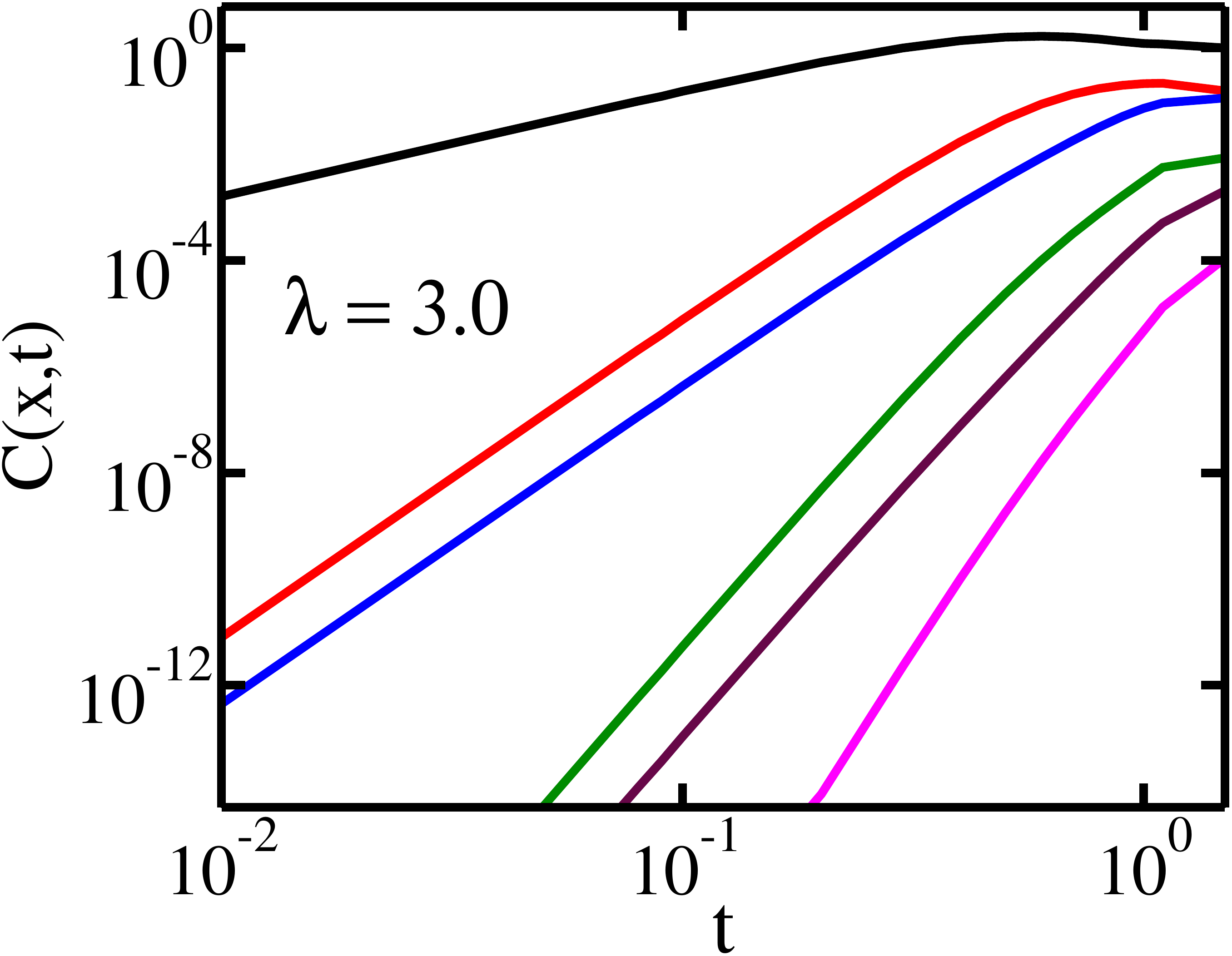}}{(c)}
    \stackunder{\includegraphics[width=4.25cm,height=3.7cm]{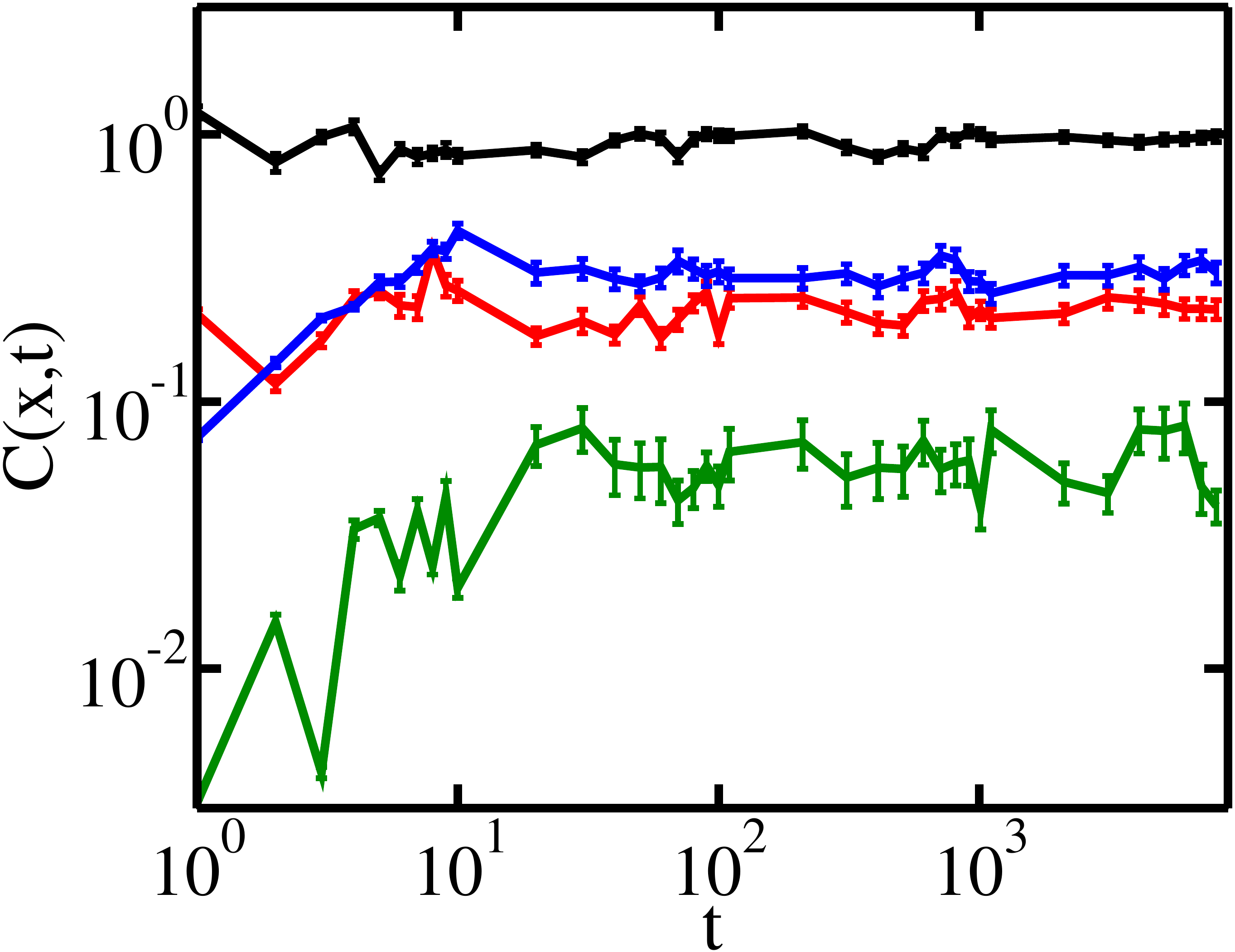}}{(f)}
  \caption{Time dynamics of OTOC in the AAH model for increasing
    values of $x$. (a-c) $C(x,t)$ vs $t$ plots for early times for
    $\lambda=1.0,2.0$ and $3.0$ respectively. The plot legend shown in
        (a) also applies to (b) and (c).  (d-f) $C(x,t)$ vs $t$
    plots for late times for $\lambda=1.0,2.0$ and $3.0$ respectively.
    The plot legend shown in (d) also applies to (e) and (f).
    For all the plots, system size $N=1024$ and total number of
    $\theta_p$ realizations is $500$.}
  \label{otocearly_aah}
\end{figure}  

The early-time growth of OTOC in the AAH model is shown in
Fig.~\ref{otocearly_aah}(a-c) for small values of $x$ and for
$\lambda=1,2,3$ respectively. For all values of $\lambda$ we notice
that $C(x,t)\sim t^{2x}$ $\forall$ odd $x$ and $C(x,t)\sim t^{2(x+1)}$
$\forall$ even $x$, which is also found in translationally invariant
models~\cite{lin2018out}. 
This can be understood by writing the Heisenberg 
time evolution of $\hat{W}(t)$ using Hausdorff-Baker-Campbel (HBC) formula
\begin{eqnarray}
e^{it\hat{H}} \hat{W} e^{-it\hat{H}} = \sum\limits_{m=0}^{\infty} \frac{(it)^m}{m!} \hat{L}^m(\hat{W}) ,
\label{hbc}
\end{eqnarray}
where $\hat{L}(\hat{W})=[\hat{H},\hat{W}]$ and $\hat{W}=\hat{\sigma}^z_{L/2}$. The power-law growth obtained in the early-time dynamics is controlled by the term with smallest $m$ such that $[\hat{L}^m,\hat{\sigma}^z_{L/2+x}])]\neq 0$. For short-range AAH Hamiltonian it is clear that this happens when $m=x$ leading to $C(x,t)\sim t^{2x}$~\cite{lin2018out}. For $x=2,4,6,..$ one includes the next leading term which gives $C(x,t)\sim t^{2(x+1)}$~\cite{riddell2019out}. 
This shows that the quasiperiodic disorder
does not play any important role in the initial dynamics. However, in
the long-time limit OTOC is found to decay as $1/t^\gamma$ with time
and the power-law exponent $\gamma$ depends on $\lambda$ as shown in
Fig.~\ref{otocearly_aah}(d-f) for $\lambda=1,2,3$ respectively. We
find that the values of $\gamma\approx1.0,0.3,0.0$ for $\lambda=1,2,3$
respectively which correspond to the delocalized, critical and
localized phases respectively. The $t^{-1}$ decay in the delocalized
phase is also seen in a clean system~\cite{lin2018out}.
The extended (ergodic or nonergodic) states are responsible for the correlation wavefront to reach a particular distant site in the lattice (leading to OTOC growth) and then proceed further (leading to OTOC decay) until OTOC reaches a saturation.
Although the decay rate is expectedly less in presence of (nonergodic) multifractal phase in comparison to (ergodic) delocalized phase.
A lot of
intrinsic fluctuations are found in these plots due to the presence of
quasiperiodic disorder. We also note that in the late time dynamics
for a fixed value of $\lambda$ the value of $\gamma$ does not depend
on $x$ unlike the early-time growth.

\begin{figure}
\centering
\stackunder{\includegraphics[width=4.25cm,height=3.7cm]{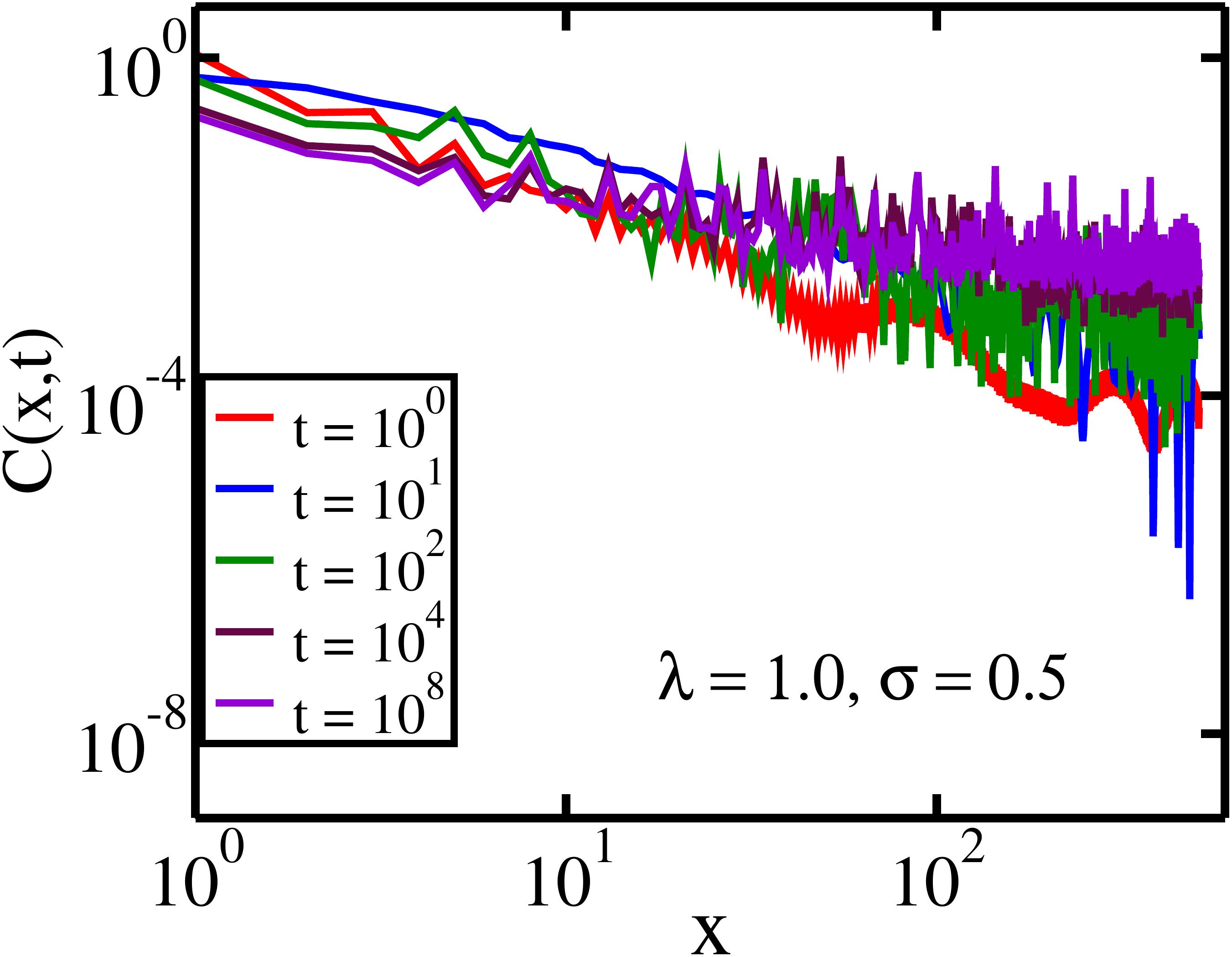}}{(a)}
\stackunder{\includegraphics[width=4.25cm,height=3.7cm]{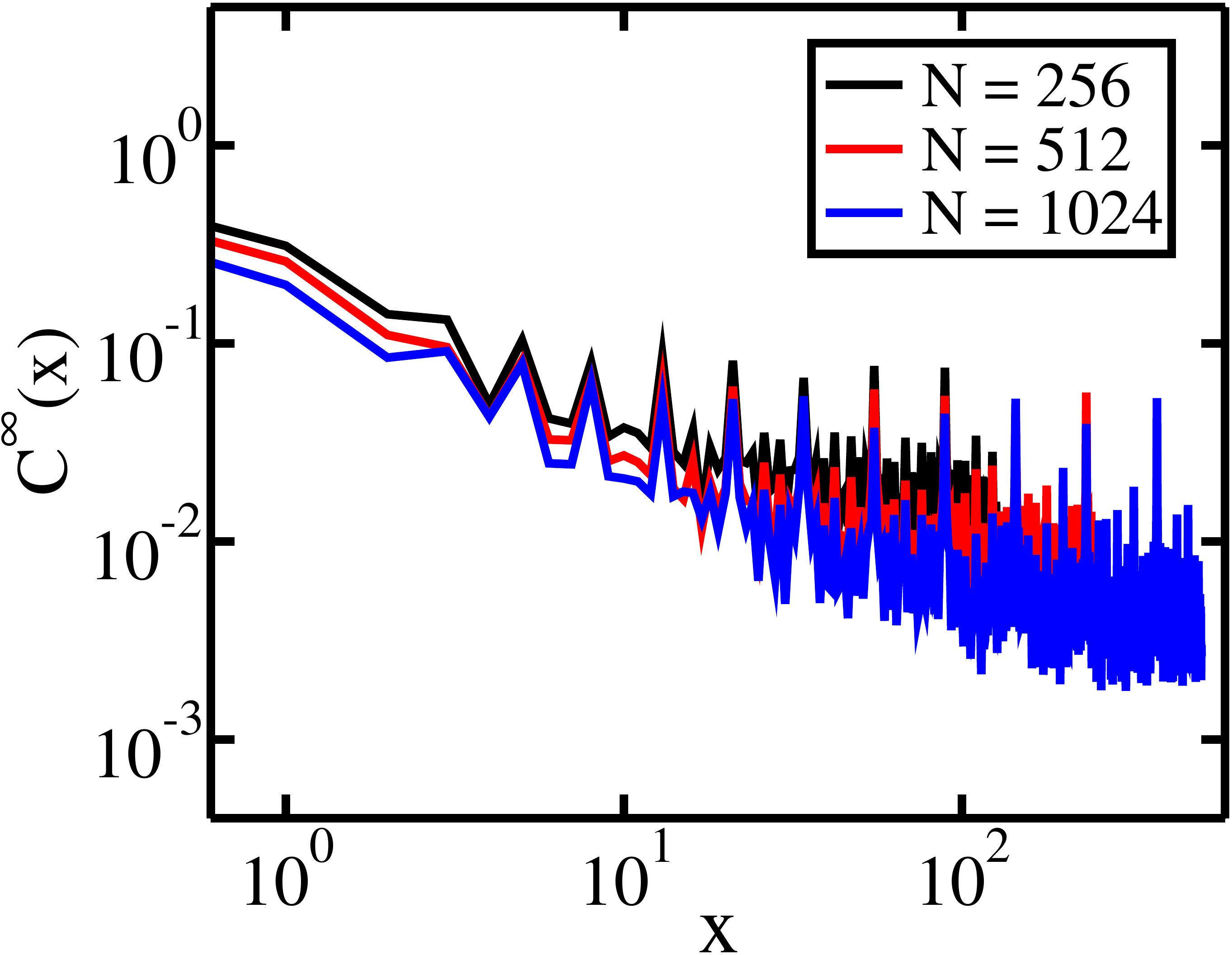}}{(e)}
\stackunder{\includegraphics[width=4.25cm,height=3.7cm]{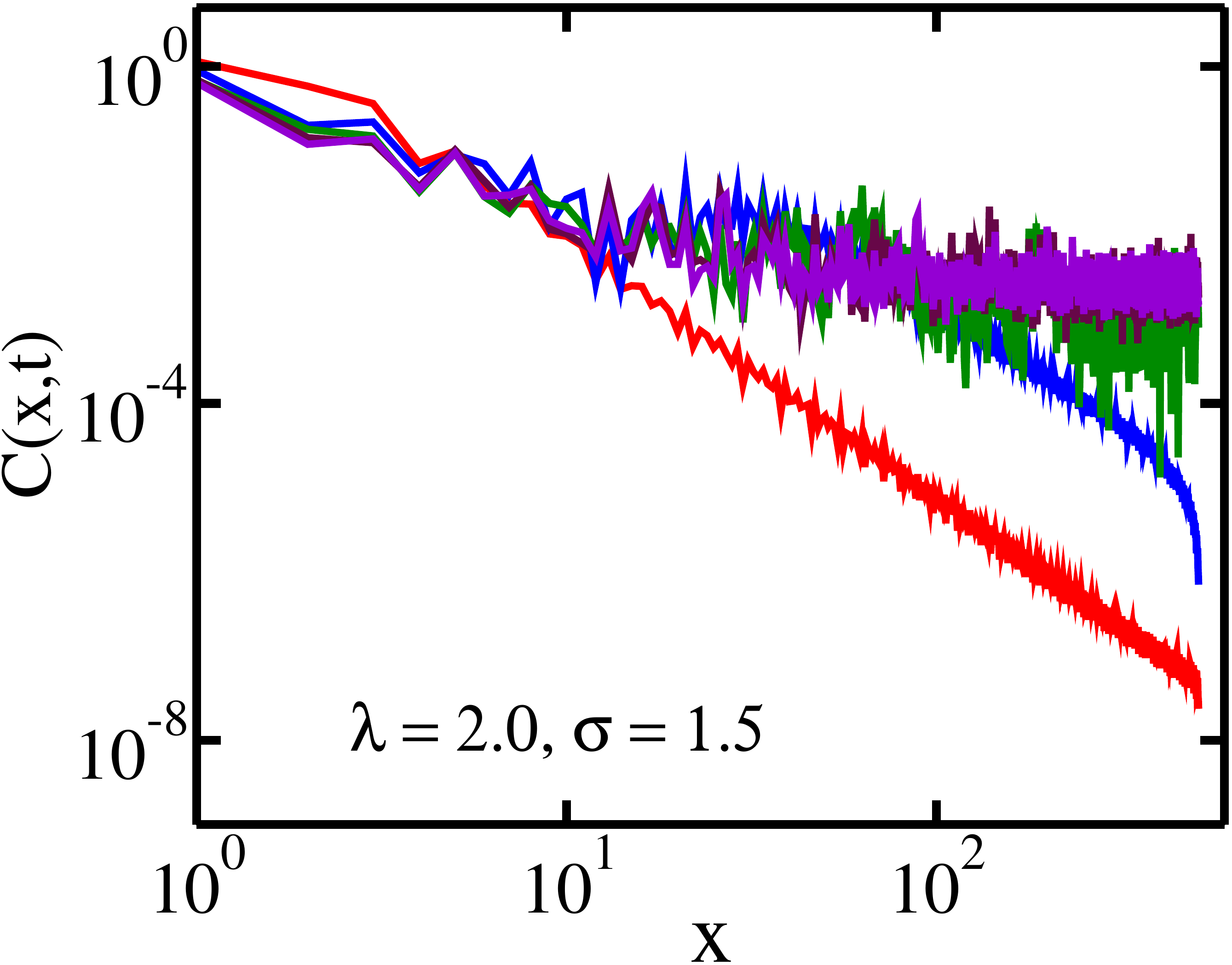}}{(b)}
\stackunder{\includegraphics[width=4.25cm,height=3.7cm]{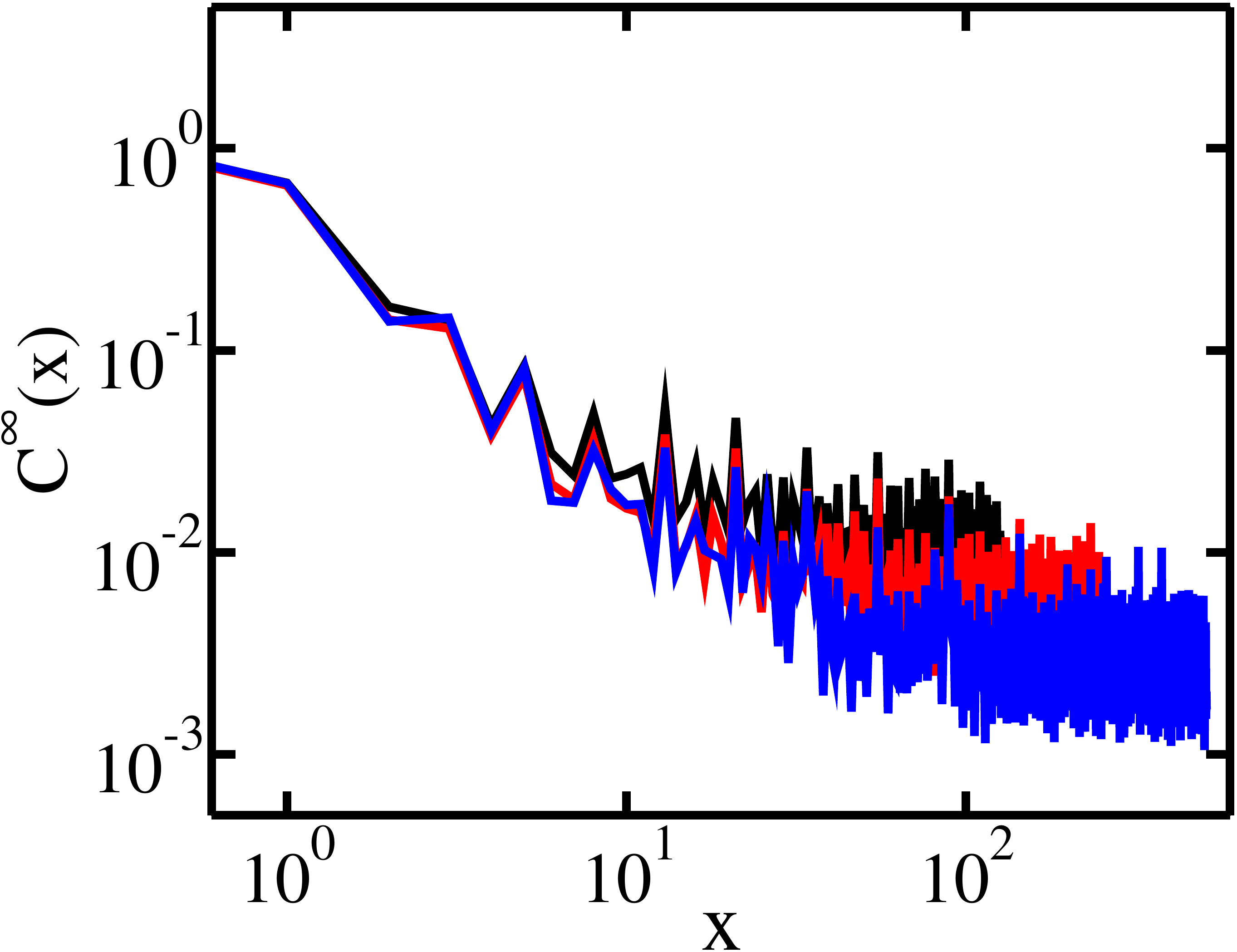}}{(f)}
\stackunder{\includegraphics[width=4.25cm,height=3.7cm]{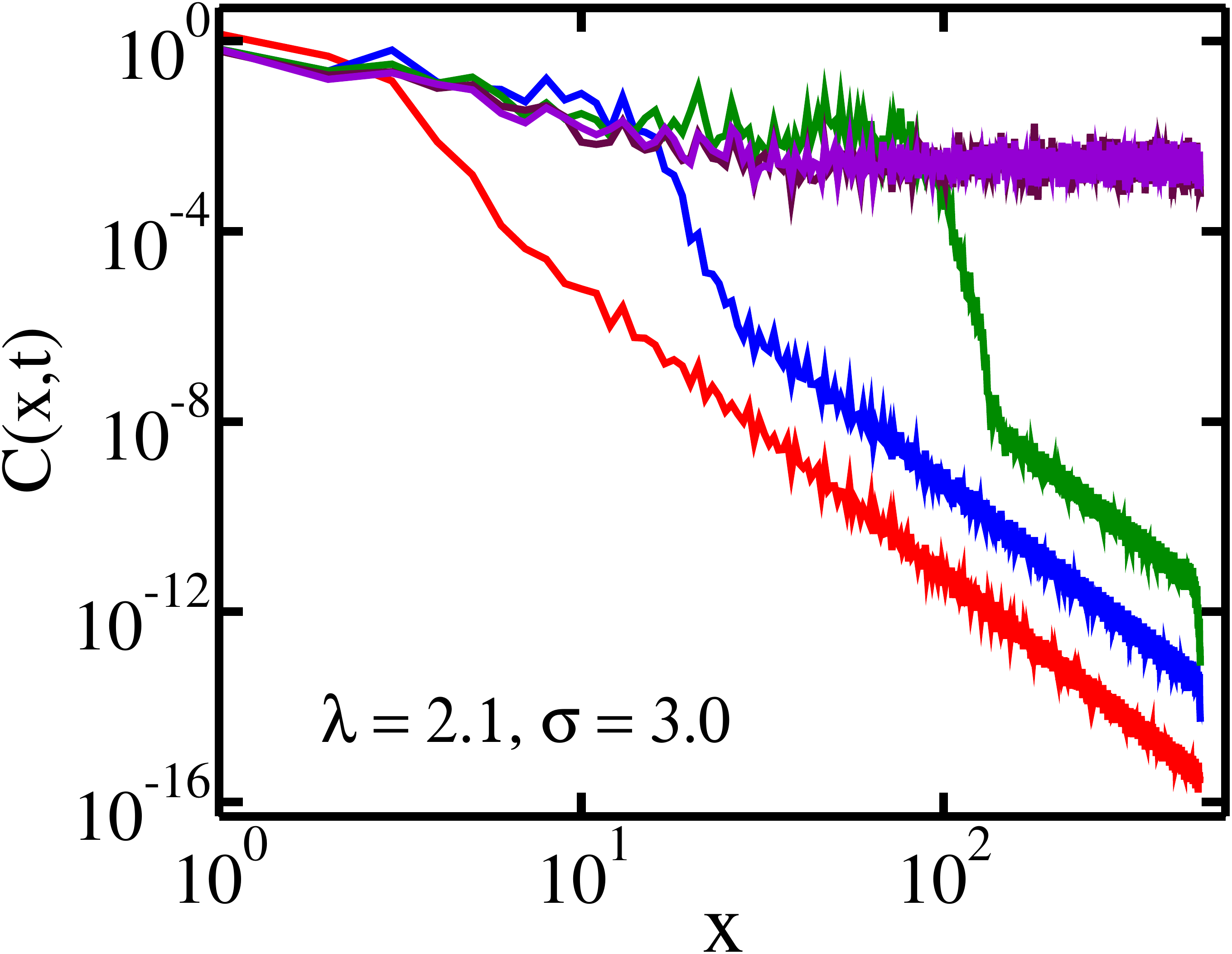}}{(c)}
\stackunder{\includegraphics[width=4.25cm,height=3.7cm]{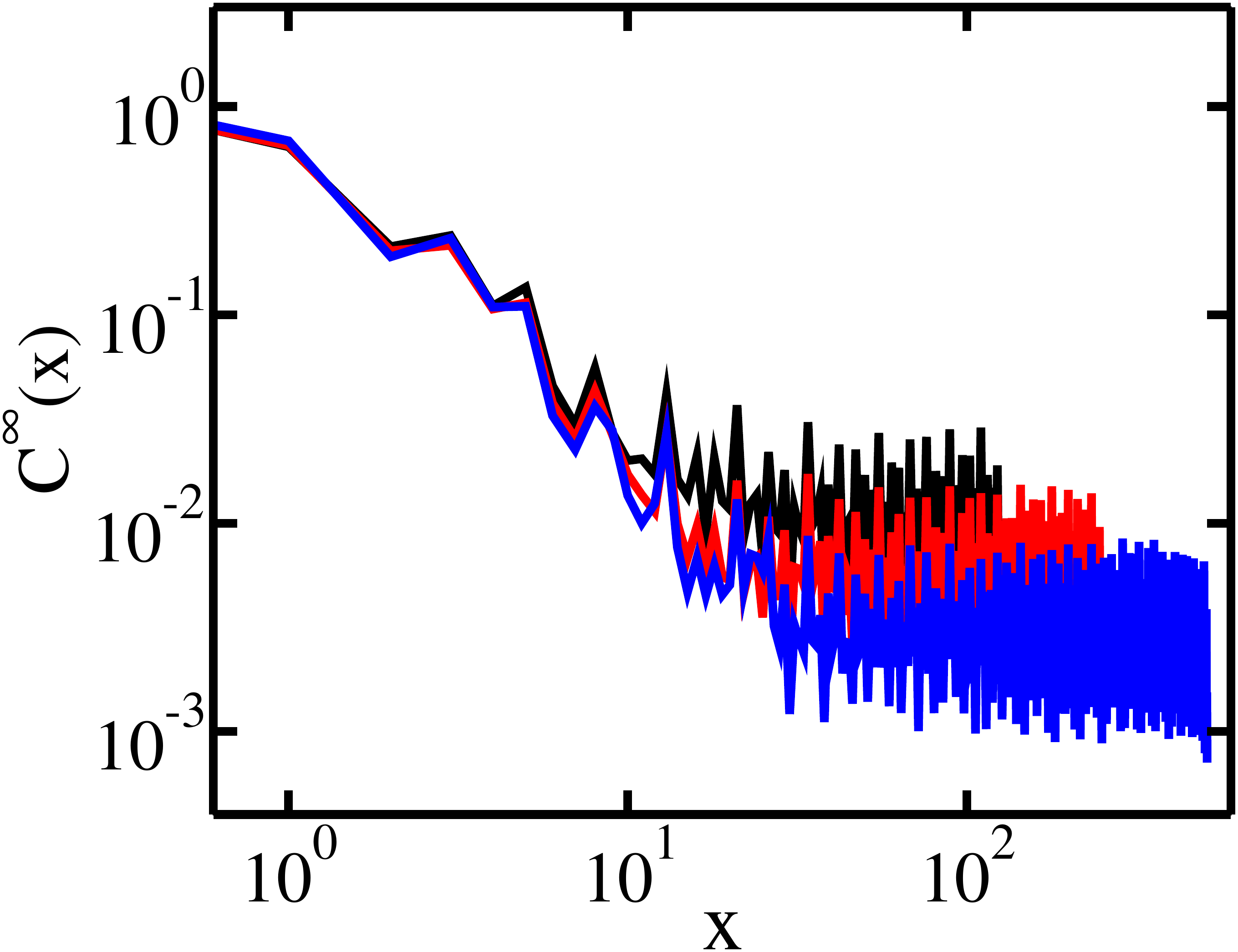}}{(g)}
\stackunder{\includegraphics[width=4.25cm,height=3.7cm]{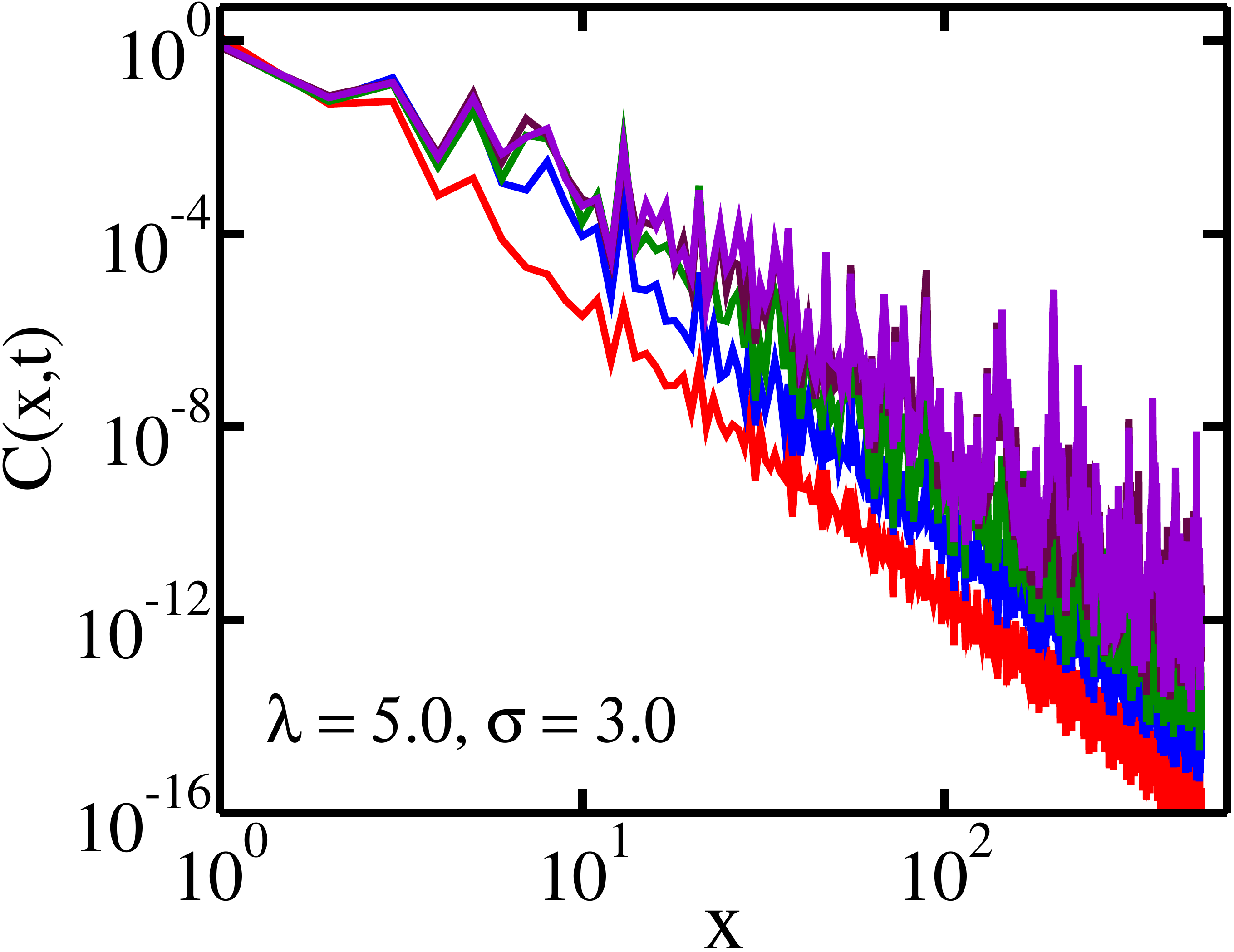}}{(d)}
\stackunder{\includegraphics[width=4.25cm,height=3.7cm]{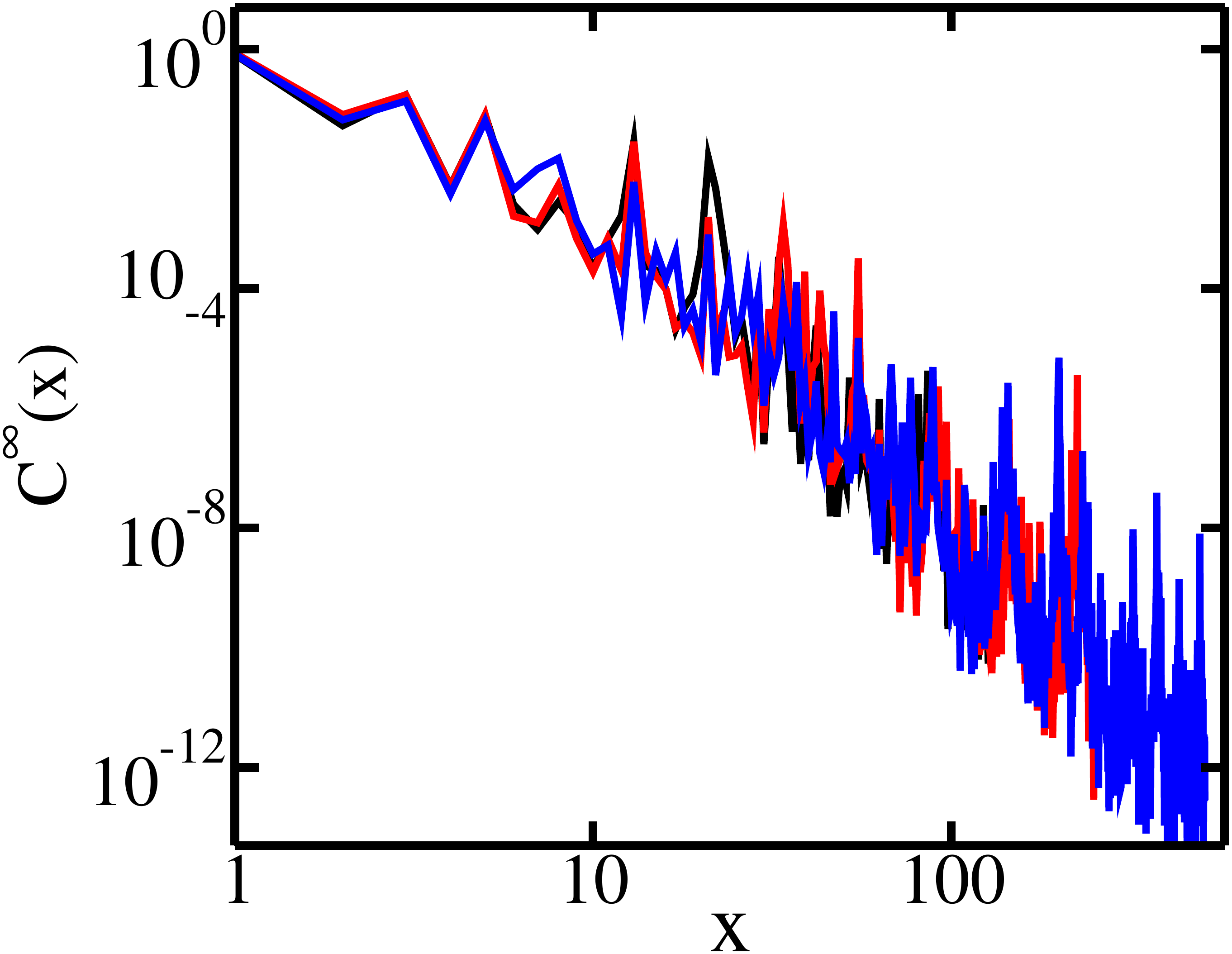}}{(h)}
\caption{OTOC in the LRH model. (a-d) OTOC $C(x,t)$ as a function of
  distance $x$ at different instants $t$ for
  $(\lambda=1.0,\sigma=0.5)$, $(\lambda=2.0,\sigma=1.5)$,
  $(\lambda=2.1,\sigma=3.0)$ and $(\lambda=5.0,\sigma=3.0)$
  respectively. System size $N=1024$ The plot legend shown in
  (a) also applies to (b), (c) and (d). (e-h) Saturation value
  $C^\infty(x)$ as a function of distance $x$ for increasing system sizes
  $N$ and for $(\lambda=1.0,\sigma=0.5)$, $(\lambda=2.0,\sigma=1.5)$,
  $(\lambda=2.1,\sigma=3.0)$ and $(\lambda=5.0,\sigma=3.0)$
  respectively. The plot legend shown in figure (e) also applies to
  figures (f), (g) and (h). For all the plots, total
    number of $\theta_p$ realizations is $500$.}
\label{otocprof_lrh}
\end{figure}

{\it LRH model}: The spatial distribution of OTOC for the LRH model of
fermions at half-filling is shown in Fig.~\ref{otocprof_lrh}. We have
chosen the combination of parameters $(\lambda,\sigma)$ in such a way
that the system is in four different types of phases: (i) $P_2$ phase
with DM edge ($\sigma=0.5$), (ii) $P_2$ phase with DL edge where the
hopping is relatively long-range ($\sigma=1.5$), (iii) $P_2$ phase
with DL edge where the hopping is short-range ($\sigma=3.0$) and (iv)
the localized phase. The spatial profiles of $C(x,t)$ for each of
the above kinds of parameter combinations are shown in
Fig.~\ref{otocprof_lrh}(a-d) respectively.  For early times $C(x,t)$
shows $1/x^{2\sigma}$ dependence for all the choices of parameters.
\begin{figure}
\centering
\stackunder{\includegraphics[width=4.25cm,height=3.7cm]{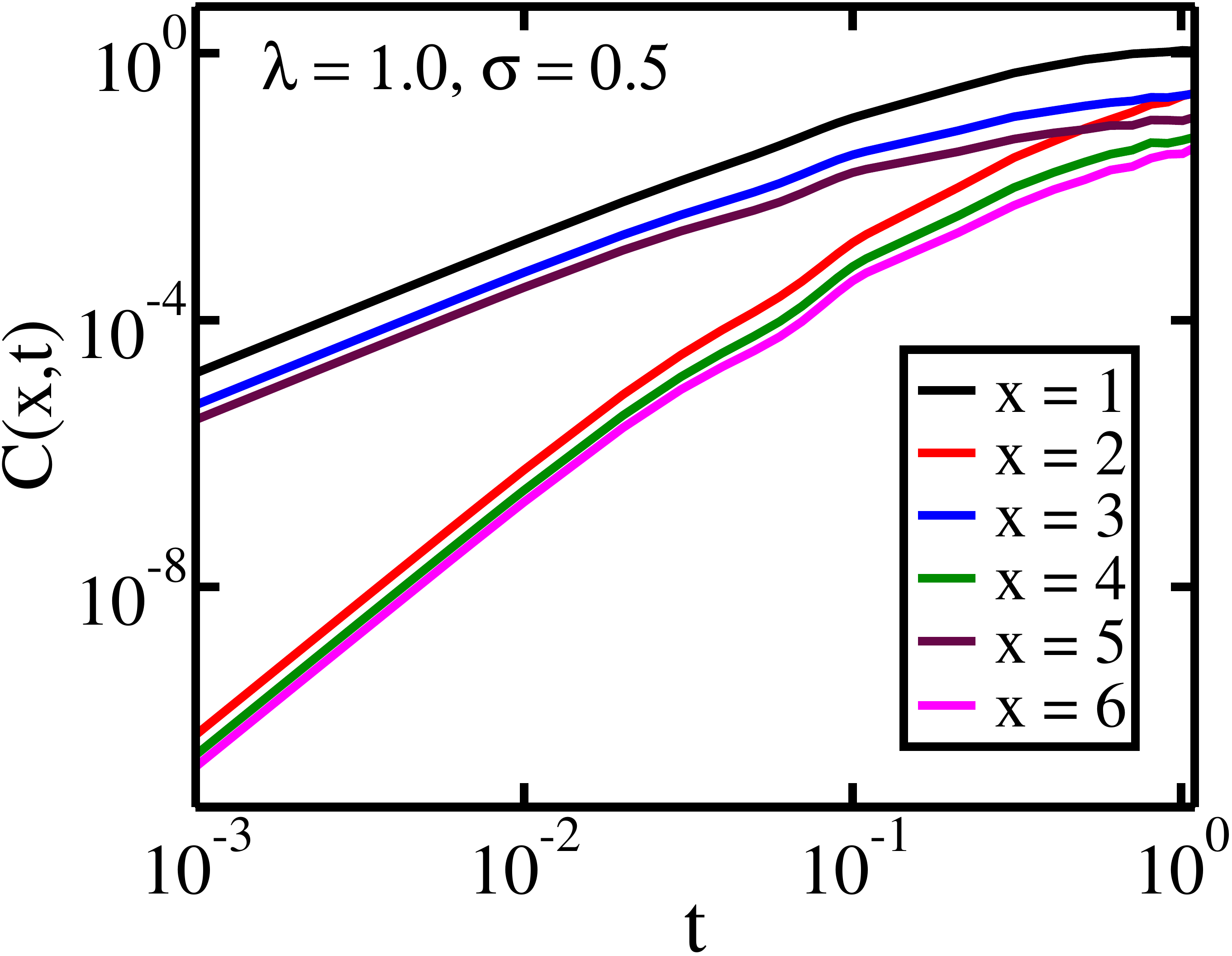}}{(a)}
\stackunder{\includegraphics[width=4.25cm,height=3.7cm]{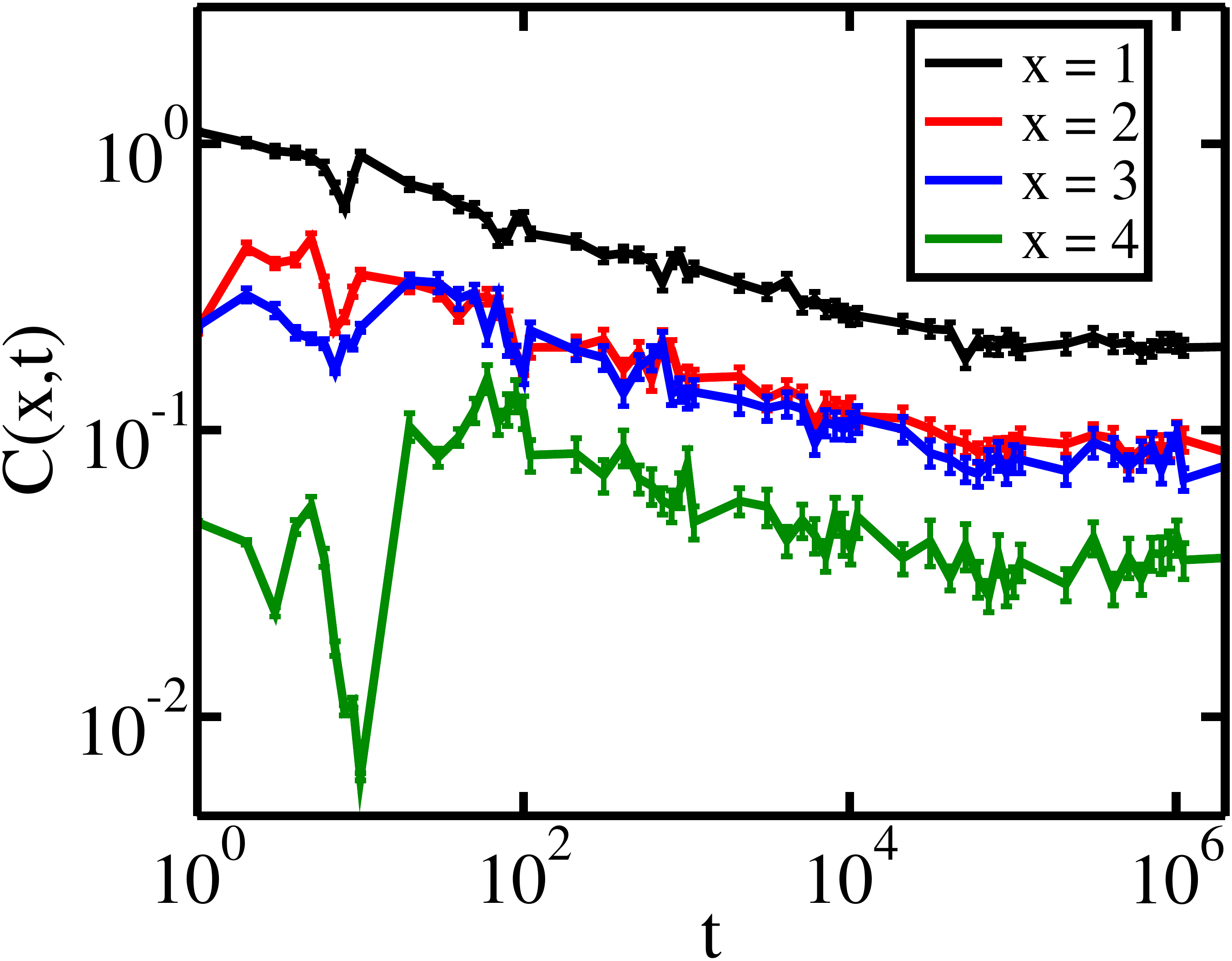}}{(e)}
\stackunder{\includegraphics[width=4.25cm,height=3.7cm]{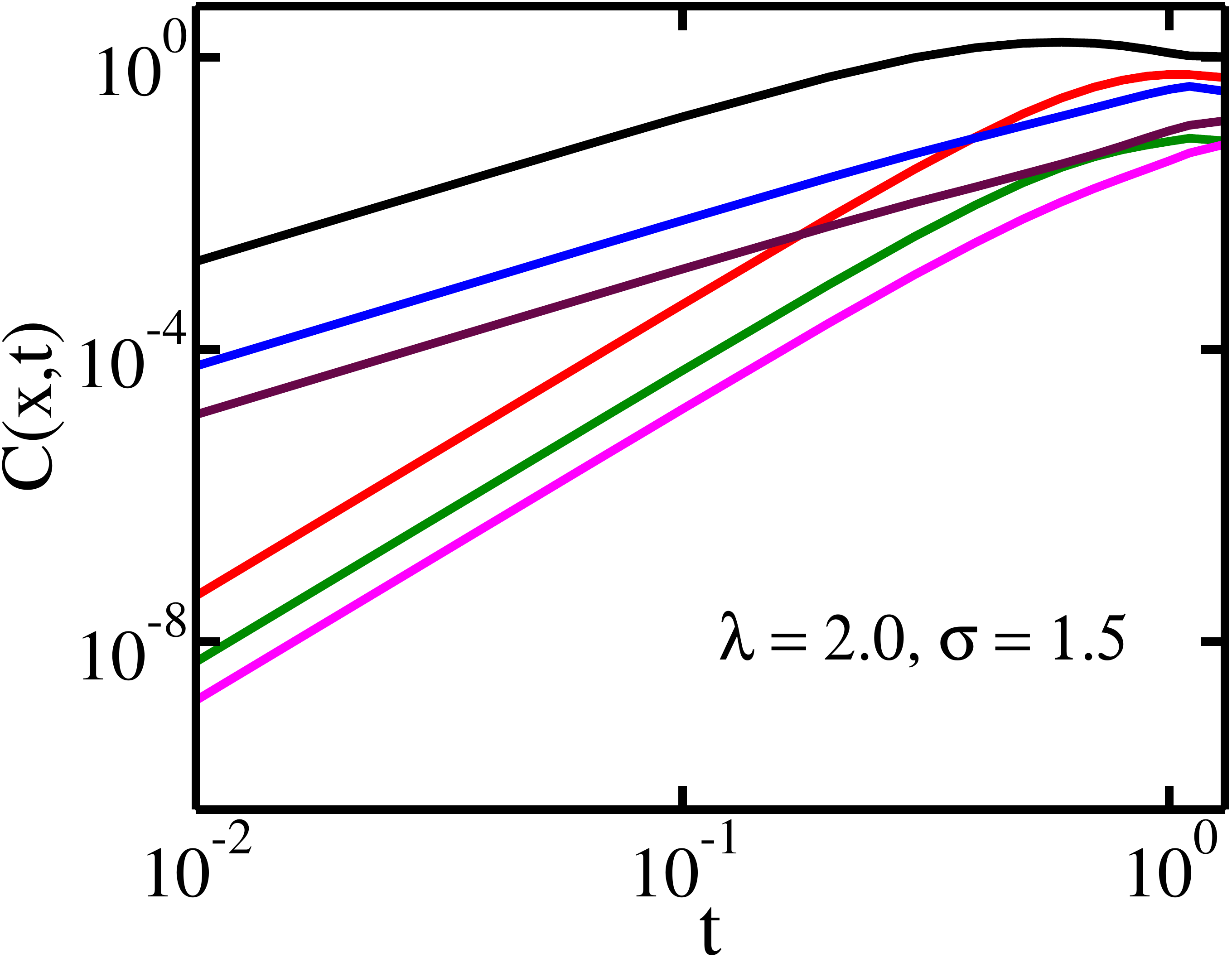}}{(b)}
\stackunder{\includegraphics[width=4.25cm,height=3.7cm]{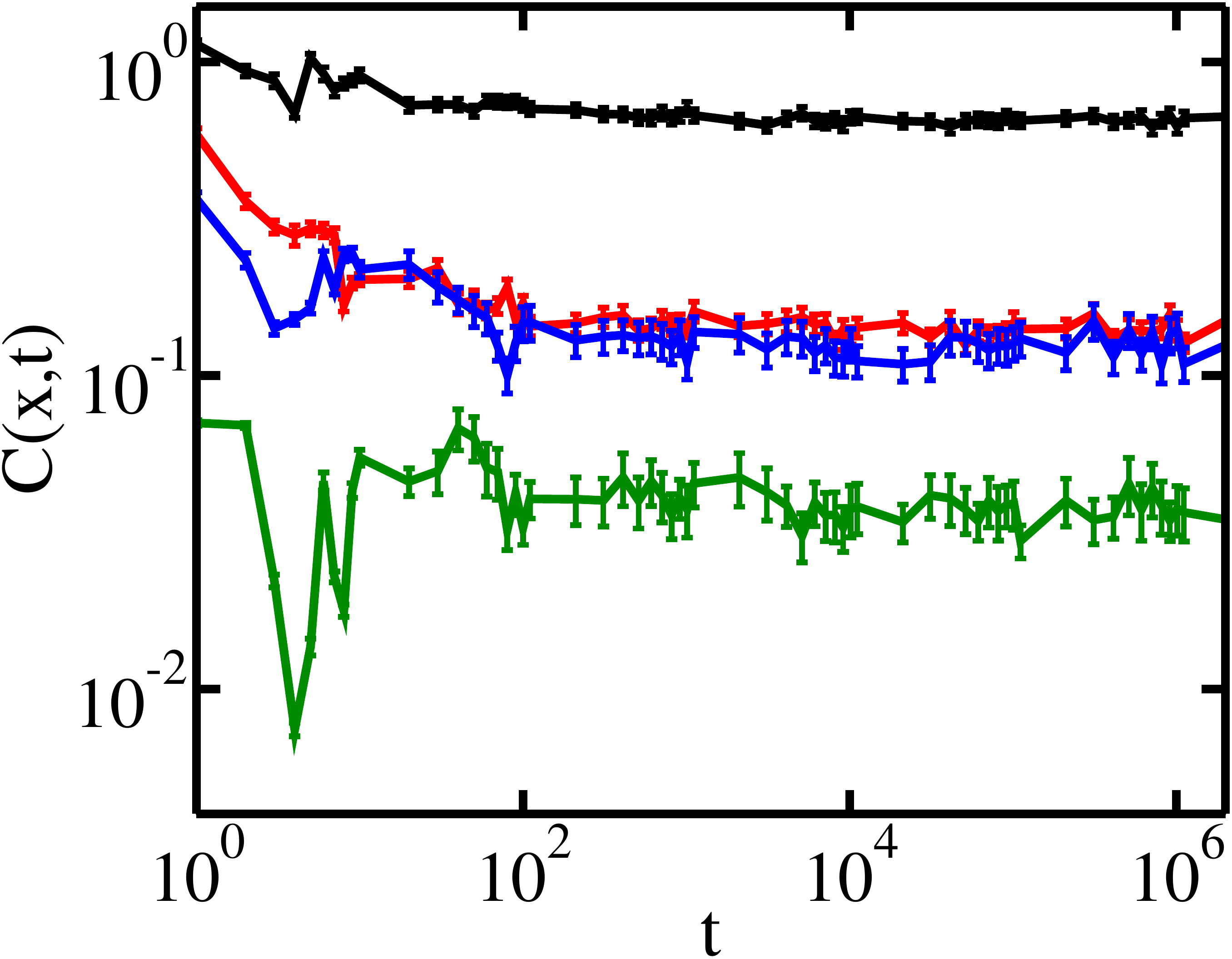}}{(f)}
\stackunder{\includegraphics[width=4.25cm,height=3.7cm]{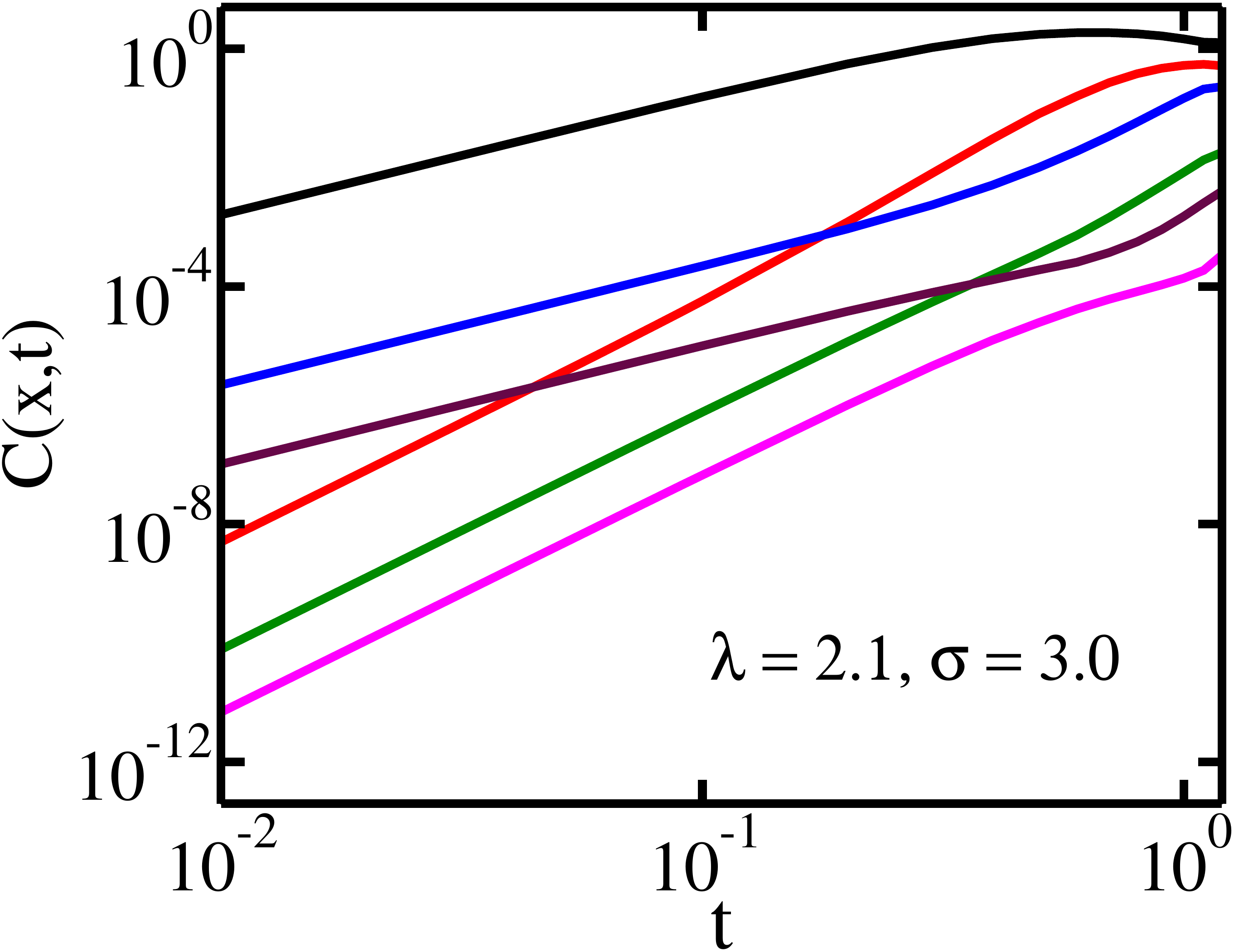}}{(c)}
\stackunder{\includegraphics[width=4.25cm,height=3.7cm]{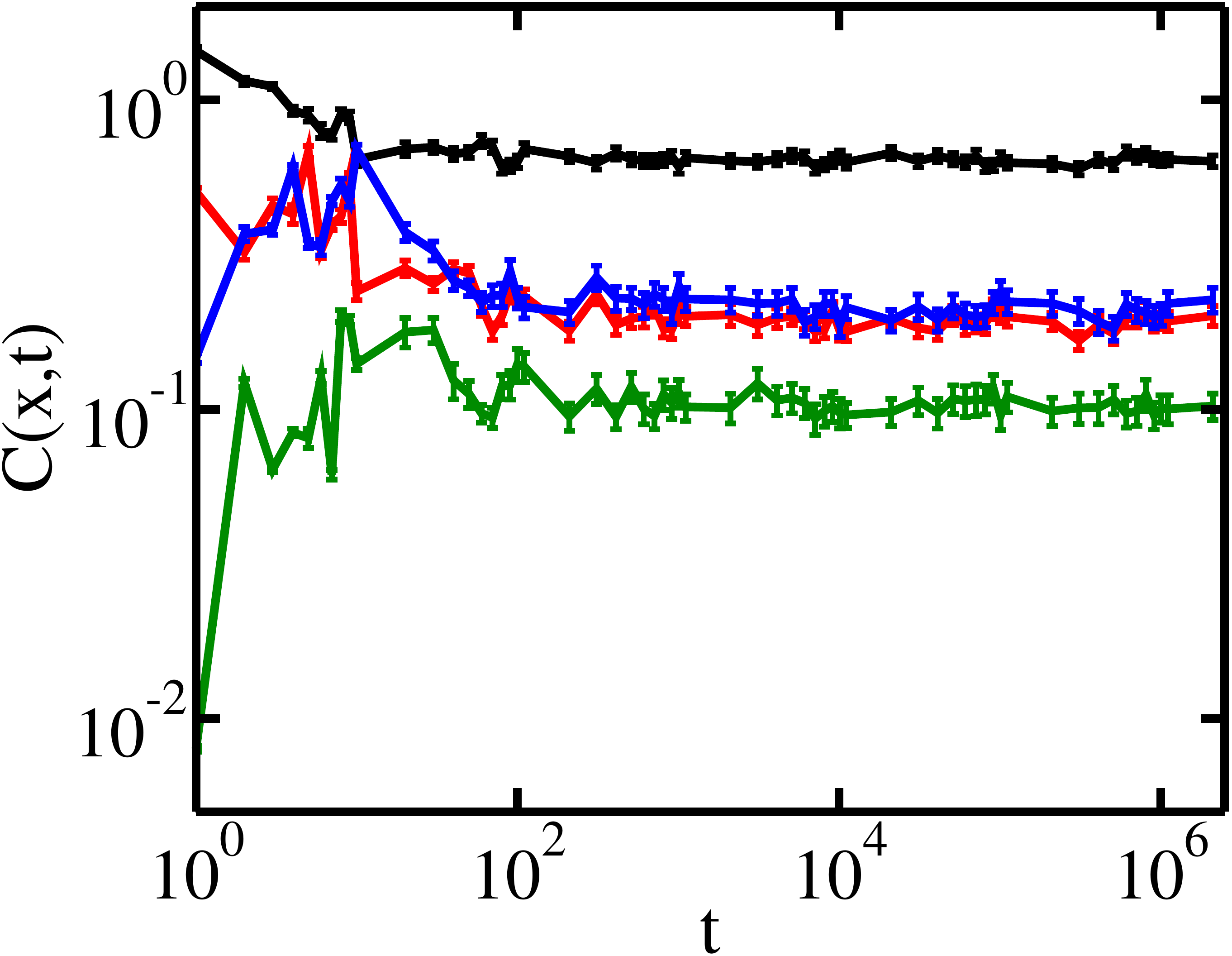}}{(g)}
\stackunder{\includegraphics[width=4.25cm,height=3.7cm]{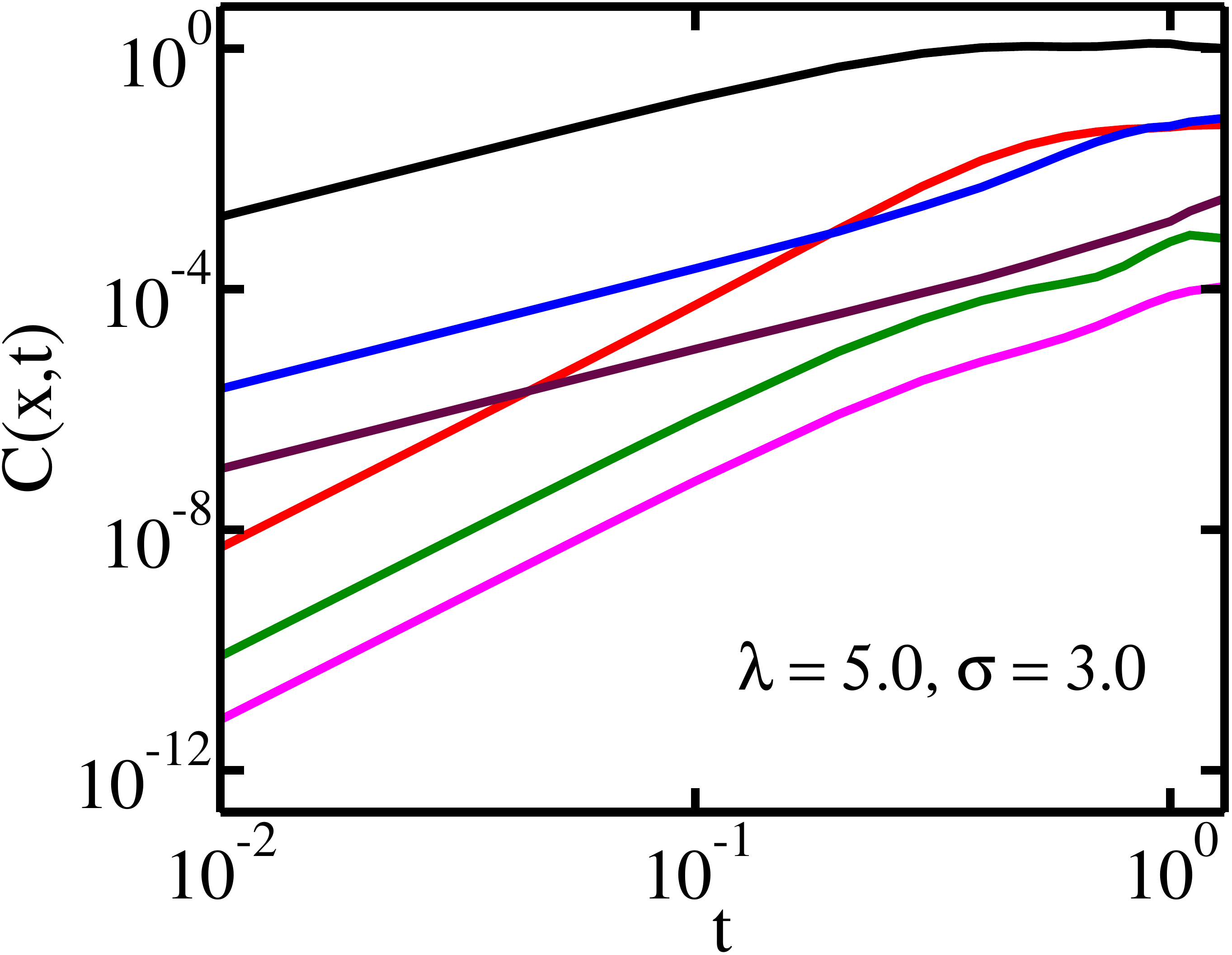}}{(d)}
\stackunder{\includegraphics[width=4.25cm,height=3.7cm]{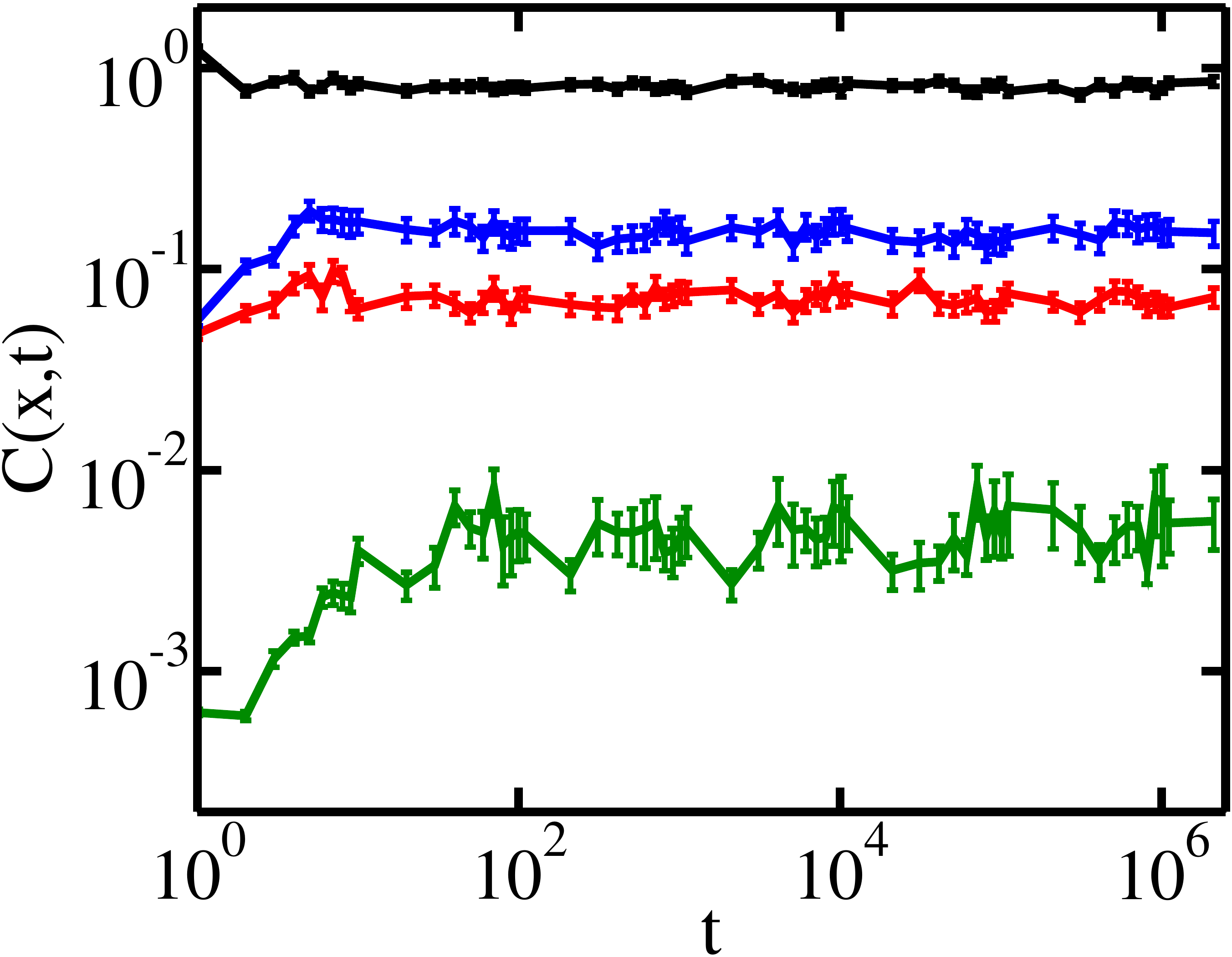}}{(h)}
\caption{Time dynamics of OTOC in the LRH model for increasing values
  of $x$. (a-d) $C(x,t)$ vs $t$ plots for early times for
  $(\lambda=1.0,\sigma=0.5)$, $(\lambda=2.0,\sigma=1.5)$,
  $(\lambda=2.1,\sigma=3.0)$ and $(\lambda=5.0,\sigma=3.0)$
  respectively. The plot legend shown in (a) also applies to
  (b), (c) and (d).  (e-h) $C(x,t)$ vs $t$ plots for late times for
  $(\lambda=1.0,\sigma=0.5)$, $(\lambda=2.0,\sigma=1.5)$,
  $(\lambda=2.1,\sigma=3.0)$ and $(\lambda=5.0,\sigma=3.0)$
  respectively. The plot legend shown in figure (e) also applies to
  figures (f), (g) and (h).
  For all the plots, system size $N=1024$ and total
  number of $\theta_p$ realizations is $500$.}
\label{otocdyn_lrh}
\end{figure}

In the mixed phases we see that in the long time limit $C(x,t)$ follows
$1/x^{\delta}$ (power-law) behavior for small $x$ and almost
$x$-independent behavior for large $x$. In Fig.~\ref{otocprof_lrh}(a)
$\delta\approx1.0$ for small $x$ whereas occasional large fluctuations
can be seen for large values of $x$ which are terms in the Fibonacci
sequence of the `golden mean'. The occasional large fluctuations are
signatures of the multifractal states similar to the AAH model at the
transition point. In Fig.~\ref{otocprof_lrh}(b) for intermediate times
$(t\sim10)$ $C(x,t)$ shows $1/x^{1.5}$ dependence for small $x$ and
$1/x^{2\sigma}$ dependence for large $x$. However, in the long time
limit $x$-independent behavior of $C(x,t)$ is seen for large $x$ along
with the $1/x^{1.5}$ dependence for small $x$. Here
$\delta\approx1.5$.  In Fig.~\ref{otocprof_lrh}(c) for intermediate
times $C(x,t)$ shows $1/x^{1.5}$ dependence for small $x$ and
$1/x^{2\sigma}$ dependence for large $x$. A sharp boundary can be seen
between these two behaviors, which is a characteristic signature of
the short-range regime~\cite{lieb}. In the long time limit $C(x,t)$ does
not depend on $x$ for large $x$ whereas it continues to show the
$1/x^{1.5}$ dependence for small $x$ corresponding to
$\delta\approx1.5$ once again. In Fig.~\ref{otocprof_lrh}(d)
corresponding to the localized phase the spatial profile $C(x,t)$ continues
to be $1/x^{2\sigma}$ $\forall x$. We do not a see a `mixed' behavior
in this case as the system is unambiguously in the localized phase.
We notice that the value of $\delta\approx1.0$ for all the mixed
phases with DM edges ($\sigma<1$) while $\delta\approx 1.5$ for the
mixed phases with DL edges ($\sigma>1$). It is noticeable that the value of $\delta$ is larger in the presence of localized states than that in the presence of multifractal states. 

In Fig.~\ref{otocprof_lrh}(e-h) we show the system size dependence of
the spatial profile of OTOC in the long time limit $C^\infty(x)$
corresponding to phases described in Fig.~\ref{otocprof_lrh}(a-d)
respectively. Since we have already mentioned earlier that the
calculation of $C^\infty(x)$ involves each eigenstate in the spectrum,
in the mixed phases ($P_2$ phase with DL/DM edge), in
Fig.~\ref{otocprof_lrh}(e-g) $C^\infty(x)$ for large $x$ decreases
with the system size $N$ although its functional dependence on $N$ is
not very clear due to fluctuations. In Fig.~\ref{otocprof_lrh}(e) we
see the occasional large fluctuations due to the presence of
multifractal states increase and become more prominent with $N$. In
Fig.~\ref{otocprof_lrh}(f) and Fig.~\ref{otocprof_lrh}(g) the large
fluctuations are not seen due to the absence of multifractal states.
In the presence of DM edge $C(x)$ depends on $N$ for small $x$ whereas
in the presence of DL edge $C(x)$ remains invariant with the change in
$N$ for small $x$.  In Fig.~\ref{otocprof_lrh}(h) the spatial profile
of $C^\infty(x)$ becomes system size independent which is
characteristic of a localized phase. We note that in the presence of
localized states the peak of the profile of $C^\infty(x)$ has a
higher value (Fig.~\ref{otocprof_lrh}(f-h)) than in the absence of
localized states (Fig.~\ref{otocprof_lrh}(e)).

The early-time growth of OTOC in the LRH model is shown for small $x$
in Fig.~\ref{otocdyn_lrh} for the $P_2$ phase with DM edge in
Fig.~\ref{otocdyn_lrh}(a), $P_2$ phase with DL edge in
Fig.~\ref{otocdyn_lrh}(b-c) and the localized phase in
Fig.~\ref{otocdyn_lrh}(d). Independent of the values of $\lambda$ and
$\sigma$, we find that $C(x,t)\sim t^2$ $\forall$ odd $x$ and
$C(x,t)\sim t^4$ $\forall$ even $x$ as also found for the
translationally invariant long-range hopping model (we have
checked). Unlike the short-range AAH model here OTOC does not have a
power-law behaviour with $x$ and the growth here is in fact largely
$x$-independent. This can be again understood from Eq.~\ref{hbc}. Since LRH Hamiltonian is long-ranged $\hat{\sigma}^z_{L/2}$ and $\hat{\sigma}^z_{L/2+x}$ immediately gets connected for smallest $m=1$ which gives $C(x,t)\sim t^2$~\cite{lin2018out_2}. For even $x$ $C(x,t)\sim t^4$ by including next term as the leading order. For the same phases of the LRH model, the late-time
decay of OTOC is shown in Fig.~\ref{otocdyn_lrh}(e-h). From
Fig.~\ref{otocdyn_lrh}(e) we see that in the $P_2$ phase with DM edge
the power-law decay exponent $\gamma=0.15$ $\forall$ $x$. In
Fig.~\ref{otocdyn_lrh}(f-g) we find that in the $P_2$ phase with DL
edge the decay exponent gets smaller, and is difficult to determine. A
power-law decay is found due to the presence of the delocalized states
in the phase. However it is smaller as compared to that in
Fig.~\ref{otocdyn_lrh}(e) due to the presence of localized states
instead of (extended nonergodic) multifractal states. In
Fig.~\ref{otocdyn_lrh}(h) we do not see any decay of OTOC after the
early-time growth due to the absence of delocalized or multifractal
states.
All the dynamical behaviors shown by both the entanglement entropy and OTOC can be seen more clearly as one increases the system sizes.

We would also like to mention that we have checked that for a clean
(undisordered) system in the presence of long-range hopping, in the
long-time limit $C(x,t)$ decays as $1/t$ independent of $x$ and the
long-range parameter $\sigma$. The values of $C^\infty$ are also
independent of $x$ and $\sigma$ as the phases are delocalized for all
$\sigma$. On the other hand in the LRH model, the power-law decay
exponent $\gamma$ is much lower than that in the clean system. In the
LRH model $\gamma$ depends on the values of $\sigma$. The values of
$C^\infty$ depend on $x$ as the phases are (nonergodic) mixed or
localized. However, the values of $\gamma$ and $C^\infty$ change very
little with the fraction of delocalized states present in the mixed
phases, especially in the presence of the DM edge.
 
\section{Conclusion} \label{conclusion}
To conclude we study the nonequilibrium dynamics of entanglement
entropy and out-of-time-order correlator of noninteracting fermions at
half-filling starting from a product state to distinguish different
phases hosted by the quasiperiodic Aubry-Andr\'e-Harper model with
long-range hopping. Apart from the delocalized and localized phases,
the model also shows mixed phases which consist of delocalized and
multifractal or localized states. In the nearest neighbor hopping
limit due to the restoration of self-duality the model hosts
delocalized, multifractal and localized phases. When the hopping is
sufficiently long-ranged a secondary logarithmic behavior in the
entanglement entropy is seen in the mixed phases whereas the primary
behavior is a power-law growth which can be different in different
phases. 
The saturation value of entanglement entropy in the delocalized, multifractal and mixed phases depends linearly on system
size whereas in the localized phase (in the short-range regime) it is
independent of system size.
The secondary growth is a unique feature that we expect to see in the long-ranged mixed phases of other models as this feature seems to be absent in the short-range regime. Although the logarithmic behavior in our case is surprising and it may not necessarily be logarithmic in nature for other cases.

In early-time dynamics OTOC shows very different behavior in the
presence of nearest neighbor hopping and long-range hopping, like is
seen also in clean systems. The late-time decay rate of OTOC is
different in the delocalized and multifractal phases of the nearest
neighbor AAH model whereas the localized phase of the same model shows
no such decay. In the long-time limit the spatial profile of OTOC is
independent, dependent (with large fluctuations) and exponentially
dependent on space in the delocalized, multifractal and localized
phases respectively. Also the profile decreases linearly and
sub-linearly with system size in the delocalized and multifractal phases respectively
whereas it is independent of system size in the localized phase. In
the multifractal phase, large fluctuations are observed at the special
points which are related to the Fibonacci sequence of the
quasiperiodicity parameter. In the long-range Harper model, the
late-time power-law decay is present in the mixed phases due to the presence of extended states although the
power-law decay exponent is smaller compared to the inverse power-law
behavior found in the delocalized phase of the (clean) system. The
power-law exponent barely changes with the change in the fraction of
delocalized states in the mixed phases showing the dominace of the nonergodic states in the dynamics. Among the mixed phases the presence of localized states supresses the late-time decay even more than that of multifractal states. The localized phase of
this model does not show any such decay due to absence of extended states. The dynamics of the spatial profile of OTOC in the mixed
phase in the short-range limit reveals a sharp boundary which is typical of longrange models~\cite{daley}. The spatial
profile of OTOC in the long-time limit in the mixed phases shows a
mixed behavior: power-law dependence for small distance (nonergodic behavior) and no
dependence for large distance (ergodic behavior). In the mixed phases containing multifractal states the profile shows large fluctuations at
special points for large distance similar to the critical point of the
AAH model. In the localized phase the spatial dependence of OTOC is a
power-law one for all distances and is also independent of system
size. Also in the mixed phases the spatial profile shows different system-size dependences for small and large distances which is expected.
One may expect to see these behaviors in the mixed phases of other long-range (Harper-like) models.

Entanglement entropy and OTOC are two quantities that are of great
interest in dynamical studies of quantum systems with the second one
being easier to be implemented in experiments. Very recently, a surge
of interest in the community~\cite{chen2020detecting,daug2019detection,lewis2020detecting,nie2020experimental}
has been seen in the experimental detection of quantum phase
transitions using OTOC.  At this point our work provides the temporal
and spatial features of OTOC to detect a host of different quantum
phases which can potentially be implemented in the ongoing
experiments.  Also there are possibilities of studying the temperature
dependence of OTOC in the longrange Harper model using a thermal state
which one can address in the future.
\section*{Acknowledgements}
NR would like to acknowledge University Grants Commision (UGC), India
for providing a PhD fellowship and thanks Kamanpreet Singh Manoor for
fruitful discussions on OTOC. A.S acknowledges financial support from
SERB via the grant (File Number: CRG/2019/003447), and from DST via
the DST-INSPIRE Faculty Award [DST/INSPIRE/04/2014/002461].

\bibliography{refs}

\begin{thebibliography}{74}%
\makeatletter
\providecommand \@ifxundefined [1]{%
 \@ifx{#1\undefined}
}%
\providecommand \@ifnum [1]{%
 \ifnum #1\expandafter \@firstoftwo
 \else \expandafter \@secondoftwo
 \fi
}%
\providecommand \@ifx [1]{%
 \ifx #1\expandafter \@firstoftwo
 \else \expandafter \@secondoftwo
 \fi
}%
\providecommand \natexlab [1]{#1}%
\providecommand \enquote  [1]{``#1''}%
\providecommand \bibnamefont  [1]{#1}%
\providecommand \bibfnamefont [1]{#1}%
\providecommand \citenamefont [1]{#1}%
\providecommand \href@noop [0]{\@secondoftwo}%
\providecommand \href [0]{\begingroup \@sanitize@url \@href}%
\providecommand \@href[1]{\@@startlink{#1}\@@href}%
\providecommand \@@href[1]{\endgroup#1\@@endlink}%
\providecommand \@sanitize@url [0]{\catcode `\\12\catcode `\$12\catcode
  `\&12\catcode `\#12\catcode `\^12\catcode `\_12\catcode `\%12\relax}%
\providecommand \@@startlink[1]{}%
\providecommand \@@endlink[0]{}%
\providecommand \url  [0]{\begingroup\@sanitize@url \@url }%
\providecommand \@url [1]{\endgroup\@href {#1}{\urlprefix }}%
\providecommand \urlprefix  [0]{URL }%
\providecommand \Eprint [0]{\href }%
\providecommand \doibase [0]{http://dx.doi.org/}%
\providecommand \selectlanguage [0]{\@gobble}%
\providecommand \bibinfo  [0]{\@secondoftwo}%
\providecommand \bibfield  [0]{\@secondoftwo}%
\providecommand \translation [1]{[#1]}%
\providecommand \BibitemOpen [0]{}%
\providecommand \bibitemStop [0]{}%
\providecommand \bibitemNoStop [0]{.\EOS\space}%
\providecommand \EOS [0]{\spacefactor3000\relax}%
\providecommand \BibitemShut  [1]{\csname bibitem#1\endcsname}%
\let\auto@bib@innerbib\@empty
\bibitem [{\citenamefont {Laflorencie}(2016)}]{laflorencie2016quantum}%
  \BibitemOpen
  \bibfield  {author} {\bibinfo {author} {\bibfnamefont {N.}~\bibnamefont
  {Laflorencie}},\ }\href@noop {} {\bibfield  {journal} {\bibinfo  {journal}
  {Physics Reports}\ }\textbf {\bibinfo {volume} {646}},\ \bibinfo {pages} {1}
  (\bibinfo {year} {2016})}\BibitemShut {NoStop}%
\bibitem [{\citenamefont {Eisert}\ \emph {et~al.}(2010)\citenamefont {Eisert},
  \citenamefont {Cramer},\ and\ \citenamefont {Plenio}}]{eisert}%
  \BibitemOpen
  \bibfield  {author} {\bibinfo {author} {\bibfnamefont {J.}~\bibnamefont
  {Eisert}}, \bibinfo {author} {\bibfnamefont {M.}~\bibnamefont {Cramer}}, \
  and\ \bibinfo {author} {\bibfnamefont {M.~B.}\ \bibnamefont {Plenio}},\
  }\href {\doibase 10.1103/RevModPhys.82.277} {\bibfield  {journal} {\bibinfo
  {journal} {Rev. Mod. Phys.}\ }\textbf {\bibinfo {volume} {82}},\ \bibinfo
  {pages} {277} (\bibinfo {year} {2010})}\BibitemShut {NoStop}%
\bibitem [{\citenamefont {Vidal}\ \emph {et~al.}(2003)\citenamefont {Vidal},
  \citenamefont {Latorre}, \citenamefont {Rico},\ and\ \citenamefont
  {Kitaev}}]{vidal_entanglement}%
  \BibitemOpen
  \bibfield  {author} {\bibinfo {author} {\bibfnamefont {G.}~\bibnamefont
  {Vidal}}, \bibinfo {author} {\bibfnamefont {J.~I.}\ \bibnamefont {Latorre}},
  \bibinfo {author} {\bibfnamefont {E.}~\bibnamefont {Rico}}, \ and\ \bibinfo
  {author} {\bibfnamefont {A.}~\bibnamefont {Kitaev}},\ }\href {\doibase
  10.1103/PhysRevLett.90.227902} {\bibfield  {journal} {\bibinfo  {journal}
  {Phys. Rev. Lett.}\ }\textbf {\bibinfo {volume} {90}},\ \bibinfo {pages}
  {227902} (\bibinfo {year} {2003})}\BibitemShut {NoStop}%
\bibitem [{\citenamefont {Serbyn}\ \emph {et~al.}(2013)\citenamefont {Serbyn},
  \citenamefont {Papi{\'c}},\ and\ \citenamefont
  {Abanin}}]{serbyn2013universal}%
  \BibitemOpen
  \bibfield  {author} {\bibinfo {author} {\bibfnamefont {M.}~\bibnamefont
  {Serbyn}}, \bibinfo {author} {\bibfnamefont {Z.}~\bibnamefont {Papi{\'c}}}, \
  and\ \bibinfo {author} {\bibfnamefont {D.~A.}\ \bibnamefont {Abanin}},\
  }\href@noop {} {\bibfield  {journal} {\bibinfo  {journal} {Physical review
  letters}\ }\textbf {\bibinfo {volume} {110}},\ \bibinfo {pages} {260601}
  (\bibinfo {year} {2013})}\BibitemShut {NoStop}%
\bibitem [{\citenamefont {Alet}\ and\ \citenamefont
  {Laflorencie}(2018)}]{alet}%
  \BibitemOpen
  \bibfield  {author} {\bibinfo {author} {\bibfnamefont {F.}~\bibnamefont
  {Alet}}\ and\ \bibinfo {author} {\bibfnamefont {N.}~\bibnamefont
  {Laflorencie}},\ }\href {\doibase https://doi.org/10.1016/j.crhy.2018.03.003}
  {\bibfield  {journal} {\bibinfo  {journal} {Comptes Rendus Physique}\
  }\textbf {\bibinfo {volume} {19}},\ \bibinfo {pages} {498 } (\bibinfo {year}
  {2018})},\ \bibinfo {note} {quantum simulation / Simulation
  quantique}\BibitemShut {NoStop}%
\bibitem [{\citenamefont {Abanin}\ \emph {et~al.}(2019)\citenamefont {Abanin},
  \citenamefont {Altman}, \citenamefont {Bloch},\ and\ \citenamefont
  {Serbyn}}]{abanin2019colloquium}%
  \BibitemOpen
  \bibfield  {author} {\bibinfo {author} {\bibfnamefont {D.~A.}\ \bibnamefont
  {Abanin}}, \bibinfo {author} {\bibfnamefont {E.}~\bibnamefont {Altman}},
  \bibinfo {author} {\bibfnamefont {I.}~\bibnamefont {Bloch}}, \ and\ \bibinfo
  {author} {\bibfnamefont {M.}~\bibnamefont {Serbyn}},\ }\href@noop {}
  {\bibfield  {journal} {\bibinfo  {journal} {Reviews of Modern Physics}\
  }\textbf {\bibinfo {volume} {91}},\ \bibinfo {pages} {021001} (\bibinfo
  {year} {2019})}\BibitemShut {NoStop}%
\bibitem [{\citenamefont {Hashimoto}\ \emph {et~al.}(2017)\citenamefont
  {Hashimoto}, \citenamefont {Murata},\ and\ \citenamefont
  {Yoshii}}]{hashimoto2017out}%
  \BibitemOpen
  \bibfield  {author} {\bibinfo {author} {\bibfnamefont {K.}~\bibnamefont
  {Hashimoto}}, \bibinfo {author} {\bibfnamefont {K.}~\bibnamefont {Murata}}, \
  and\ \bibinfo {author} {\bibfnamefont {R.}~\bibnamefont {Yoshii}},\
  }\href@noop {} {\bibfield  {journal} {\bibinfo  {journal} {Journal of High
  Energy Physics}\ }\textbf {\bibinfo {volume} {2017}},\ \bibinfo {pages} {138}
  (\bibinfo {year} {2017})}\BibitemShut {NoStop}%
\bibitem [{\citenamefont {Shenker}\ and\ \citenamefont
  {Stanford}(2015)}]{shenker2015stringy}%
  \BibitemOpen
  \bibfield  {author} {\bibinfo {author} {\bibfnamefont {S.~H.}\ \bibnamefont
  {Shenker}}\ and\ \bibinfo {author} {\bibfnamefont {D.}~\bibnamefont
  {Stanford}},\ }\href@noop {} {\bibfield  {journal} {\bibinfo  {journal}
  {Journal of High Energy Physics}\ }\textbf {\bibinfo {volume} {2015}},\
  \bibinfo {pages} {132} (\bibinfo {year} {2015})}\BibitemShut {NoStop}%
\bibitem [{\citenamefont {Maldacena}\ \emph {et~al.}(2016)\citenamefont
  {Maldacena}, \citenamefont {Shenker},\ and\ \citenamefont
  {Stanford}}]{maldacena2016bound}%
  \BibitemOpen
  \bibfield  {author} {\bibinfo {author} {\bibfnamefont {J.}~\bibnamefont
  {Maldacena}}, \bibinfo {author} {\bibfnamefont {S.~H.}\ \bibnamefont
  {Shenker}}, \ and\ \bibinfo {author} {\bibfnamefont {D.}~\bibnamefont
  {Stanford}},\ }\href@noop {} {\bibfield  {journal} {\bibinfo  {journal}
  {Journal of High Energy Physics}\ }\textbf {\bibinfo {volume} {2016}},\
  \bibinfo {pages} {106} (\bibinfo {year} {2016})}\BibitemShut {NoStop}%
\bibitem [{\citenamefont {Roberts}\ and\ \citenamefont
  {Swingle}(2016)}]{roberts2016lieb}%
  \BibitemOpen
  \bibfield  {author} {\bibinfo {author} {\bibfnamefont {D.~A.}\ \bibnamefont
  {Roberts}}\ and\ \bibinfo {author} {\bibfnamefont {B.}~\bibnamefont
  {Swingle}},\ }\href@noop {} {\bibfield  {journal} {\bibinfo  {journal}
  {Physical review letters}\ }\textbf {\bibinfo {volume} {117}},\ \bibinfo
  {pages} {091602} (\bibinfo {year} {2016})}\BibitemShut {NoStop}%
\bibitem [{\citenamefont {Roberts}\ and\ \citenamefont
  {Stanford}(2014)}]{roberts2014two}%
  \BibitemOpen
  \bibfield  {author} {\bibinfo {author} {\bibfnamefont {D.~A.}\ \bibnamefont
  {Roberts}}\ and\ \bibinfo {author} {\bibfnamefont {D.}~\bibnamefont
  {Stanford}},\ }\href@noop {} {\bibfield  {journal} {\bibinfo  {journal}
  {arXiv preprint arXiv:1412.5123}\ } (\bibinfo {year} {2014})}\BibitemShut
  {NoStop}%
\bibitem [{\citenamefont {Swingle}(2018)}]{swingle2018unscrambling}%
  \BibitemOpen
  \bibfield  {author} {\bibinfo {author} {\bibfnamefont {B.}~\bibnamefont
  {Swingle}},\ }\href@noop {} {\bibfield  {journal} {\bibinfo  {journal}
  {Nature Physics}\ }\textbf {\bibinfo {volume} {14}},\ \bibinfo {pages} {988}
  (\bibinfo {year} {2018})}\BibitemShut {NoStop}%
\bibitem [{\citenamefont {Swingle}\ and\ \citenamefont
  {Chowdhury}(2017)}]{swingle2017slow}%
  \BibitemOpen
  \bibfield  {author} {\bibinfo {author} {\bibfnamefont {B.}~\bibnamefont
  {Swingle}}\ and\ \bibinfo {author} {\bibfnamefont {D.}~\bibnamefont
  {Chowdhury}},\ }\href@noop {} {\bibfield  {journal} {\bibinfo  {journal}
  {Physical Review B}\ }\textbf {\bibinfo {volume} {95}},\ \bibinfo {pages}
  {060201} (\bibinfo {year} {2017})}\BibitemShut {NoStop}%
\bibitem [{\citenamefont {Rozenbaum}\ \emph {et~al.}(2017)\citenamefont
  {Rozenbaum}, \citenamefont {Ganeshan},\ and\ \citenamefont
  {Galitski}}]{rozenbaum2017lyapunov}%
  \BibitemOpen
  \bibfield  {author} {\bibinfo {author} {\bibfnamefont {E.~B.}\ \bibnamefont
  {Rozenbaum}}, \bibinfo {author} {\bibfnamefont {S.}~\bibnamefont {Ganeshan}},
  \ and\ \bibinfo {author} {\bibfnamefont {V.}~\bibnamefont {Galitski}},\
  }\href@noop {} {\bibfield  {journal} {\bibinfo  {journal} {Physical review
  letters}\ }\textbf {\bibinfo {volume} {118}},\ \bibinfo {pages} {086801}
  (\bibinfo {year} {2017})}\BibitemShut {NoStop}%
\bibitem [{\citenamefont {Chen}\ \emph {et~al.}(2017)\citenamefont {Chen},
  \citenamefont {Zhou}, \citenamefont {Huse},\ and\ \citenamefont
  {Fradkin}}]{chen2017out}%
  \BibitemOpen
  \bibfield  {author} {\bibinfo {author} {\bibfnamefont {X.}~\bibnamefont
  {Chen}}, \bibinfo {author} {\bibfnamefont {T.}~\bibnamefont {Zhou}}, \bibinfo
  {author} {\bibfnamefont {D.~A.}\ \bibnamefont {Huse}}, \ and\ \bibinfo
  {author} {\bibfnamefont {E.}~\bibnamefont {Fradkin}},\ }\href@noop {}
  {\bibfield  {journal} {\bibinfo  {journal} {Annalen der Physik}\ }\textbf
  {\bibinfo {volume} {529}},\ \bibinfo {pages} {1600332} (\bibinfo {year}
  {2017})}\BibitemShut {NoStop}%
\bibitem [{\citenamefont {Fan}\ \emph {et~al.}(2017)\citenamefont {Fan},
  \citenamefont {Zhang}, \citenamefont {Shen},\ and\ \citenamefont
  {Zhai}}]{fan2017out}%
  \BibitemOpen
  \bibfield  {author} {\bibinfo {author} {\bibfnamefont {R.}~\bibnamefont
  {Fan}}, \bibinfo {author} {\bibfnamefont {P.}~\bibnamefont {Zhang}}, \bibinfo
  {author} {\bibfnamefont {H.}~\bibnamefont {Shen}}, \ and\ \bibinfo {author}
  {\bibfnamefont {H.}~\bibnamefont {Zhai}},\ }\href@noop {} {\bibfield
  {journal} {\bibinfo  {journal} {Science bulletin}\ }\textbf {\bibinfo
  {volume} {62}},\ \bibinfo {pages} {707} (\bibinfo {year} {2017})}\BibitemShut
  {NoStop}%
\bibitem [{\citenamefont {Lewis-Swan}\ \emph {et~al.}(2019)\citenamefont
  {Lewis-Swan}, \citenamefont {Safavi-Naini}, \citenamefont {Bollinger},\ and\
  \citenamefont {Rey}}]{lewis2019unifying}%
  \BibitemOpen
  \bibfield  {author} {\bibinfo {author} {\bibfnamefont {R.}~\bibnamefont
  {Lewis-Swan}}, \bibinfo {author} {\bibfnamefont {A.}~\bibnamefont
  {Safavi-Naini}}, \bibinfo {author} {\bibfnamefont {J.~J.}\ \bibnamefont
  {Bollinger}}, \ and\ \bibinfo {author} {\bibfnamefont {A.~M.}\ \bibnamefont
  {Rey}},\ }\href@noop {} {\bibfield  {journal} {\bibinfo  {journal} {Nature
  communications}\ }\textbf {\bibinfo {volume} {10}},\ \bibinfo {pages} {1}
  (\bibinfo {year} {2019})}\BibitemShut {NoStop}%
\bibitem [{\citenamefont {G{\"a}rttner}\ \emph {et~al.}(2018)\citenamefont
  {G{\"a}rttner}, \citenamefont {Hauke},\ and\ \citenamefont
  {Rey}}]{garttner2018relating}%
  \BibitemOpen
  \bibfield  {author} {\bibinfo {author} {\bibfnamefont {M.}~\bibnamefont
  {G{\"a}rttner}}, \bibinfo {author} {\bibfnamefont {P.}~\bibnamefont {Hauke}},
  \ and\ \bibinfo {author} {\bibfnamefont {A.~M.}\ \bibnamefont {Rey}},\
  }\href@noop {} {\bibfield  {journal} {\bibinfo  {journal} {Physical review
  letters}\ }\textbf {\bibinfo {volume} {120}},\ \bibinfo {pages} {040402}
  (\bibinfo {year} {2018})}\BibitemShut {NoStop}%
\bibitem [{\citenamefont {Halpern}\ \emph {et~al.}(2019)\citenamefont
  {Halpern}, \citenamefont {Bartolotta},\ and\ \citenamefont
  {Pollack}}]{halpern2019entropic}%
  \BibitemOpen
  \bibfield  {author} {\bibinfo {author} {\bibfnamefont {N.~Y.}\ \bibnamefont
  {Halpern}}, \bibinfo {author} {\bibfnamefont {A.}~\bibnamefont {Bartolotta}},
  \ and\ \bibinfo {author} {\bibfnamefont {J.}~\bibnamefont {Pollack}},\
  }\href@noop {} {\bibfield  {journal} {\bibinfo  {journal} {Communications
  Physics}\ }\textbf {\bibinfo {volume} {2}},\ \bibinfo {pages} {1} (\bibinfo
  {year} {2019})}\BibitemShut {NoStop}%
\bibitem [{\citenamefont {Larkin}\ and\ \citenamefont
  {Ovchinnikov}(1969)}]{larkin1969quasiclassical}%
  \BibitemOpen
  \bibfield  {author} {\bibinfo {author} {\bibfnamefont {A.}~\bibnamefont
  {Larkin}}\ and\ \bibinfo {author} {\bibfnamefont {Y.~N.}\ \bibnamefont
  {Ovchinnikov}},\ }\href@noop {} {\bibfield  {journal} {\bibinfo  {journal}
  {Sov Phys JETP}\ }\textbf {\bibinfo {volume} {28}},\ \bibinfo {pages} {1200}
  (\bibinfo {year} {1969})}\BibitemShut {NoStop}%
\bibitem [{\citenamefont {Aleiner}\ and\ \citenamefont
  {Larkin}(1996)}]{aleiner1996divergence}%
  \BibitemOpen
  \bibfield  {author} {\bibinfo {author} {\bibfnamefont {I.}~\bibnamefont
  {Aleiner}}\ and\ \bibinfo {author} {\bibfnamefont {A.}~\bibnamefont
  {Larkin}},\ }\href@noop {} {\bibfield  {journal} {\bibinfo  {journal}
  {Physical Review B}\ }\textbf {\bibinfo {volume} {54}},\ \bibinfo {pages}
  {14423} (\bibinfo {year} {1996})}\BibitemShut {NoStop}%
\bibitem [{\citenamefont {Li}\ \emph {et~al.}(2017)\citenamefont {Li},
  \citenamefont {Fan}, \citenamefont {Wang}, \citenamefont {Ye}, \citenamefont
  {Zeng}, \citenamefont {Zhai}, \citenamefont {Peng},\ and\ \citenamefont
  {Du}}]{li2017measuring}%
  \BibitemOpen
  \bibfield  {author} {\bibinfo {author} {\bibfnamefont {J.}~\bibnamefont
  {Li}}, \bibinfo {author} {\bibfnamefont {R.}~\bibnamefont {Fan}}, \bibinfo
  {author} {\bibfnamefont {H.}~\bibnamefont {Wang}}, \bibinfo {author}
  {\bibfnamefont {B.}~\bibnamefont {Ye}}, \bibinfo {author} {\bibfnamefont
  {B.}~\bibnamefont {Zeng}}, \bibinfo {author} {\bibfnamefont {H.}~\bibnamefont
  {Zhai}}, \bibinfo {author} {\bibfnamefont {X.}~\bibnamefont {Peng}}, \ and\
  \bibinfo {author} {\bibfnamefont {J.}~\bibnamefont {Du}},\ }\href@noop {}
  {\bibfield  {journal} {\bibinfo  {journal} {Physical Review X}\ }\textbf
  {\bibinfo {volume} {7}},\ \bibinfo {pages} {031011} (\bibinfo {year}
  {2017})}\BibitemShut {NoStop}%
\bibitem [{\citenamefont {Wei}\ \emph {et~al.}(2018)\citenamefont {Wei},
  \citenamefont {Ramanathan},\ and\ \citenamefont
  {Cappellaro}}]{wei2018exploring}%
  \BibitemOpen
  \bibfield  {author} {\bibinfo {author} {\bibfnamefont {K.~X.}\ \bibnamefont
  {Wei}}, \bibinfo {author} {\bibfnamefont {C.}~\bibnamefont {Ramanathan}}, \
  and\ \bibinfo {author} {\bibfnamefont {P.}~\bibnamefont {Cappellaro}},\
  }\href@noop {} {\bibfield  {journal} {\bibinfo  {journal} {Physical review
  letters}\ }\textbf {\bibinfo {volume} {120}},\ \bibinfo {pages} {070501}
  (\bibinfo {year} {2018})}\BibitemShut {NoStop}%
\bibitem [{\citenamefont {Niknam}\ \emph {et~al.}(2020)\citenamefont {Niknam},
  \citenamefont {Santos},\ and\ \citenamefont {Cory}}]{niknam2020sensitivity}%
  \BibitemOpen
  \bibfield  {author} {\bibinfo {author} {\bibfnamefont {M.}~\bibnamefont
  {Niknam}}, \bibinfo {author} {\bibfnamefont {L.~F.}\ \bibnamefont {Santos}},
  \ and\ \bibinfo {author} {\bibfnamefont {D.~G.}\ \bibnamefont {Cory}},\
  }\href@noop {} {\bibfield  {journal} {\bibinfo  {journal} {Physical Review
  Research}\ }\textbf {\bibinfo {volume} {2}},\ \bibinfo {pages} {013200}
  (\bibinfo {year} {2020})}\BibitemShut {NoStop}%
\bibitem [{\citenamefont {G{\"a}rttner}\ \emph {et~al.}(2017)\citenamefont
  {G{\"a}rttner}, \citenamefont {Bohnet}, \citenamefont {Safavi-Naini},
  \citenamefont {Wall}, \citenamefont {Bollinger},\ and\ \citenamefont
  {Rey}}]{garttner2017measuring}%
  \BibitemOpen
  \bibfield  {author} {\bibinfo {author} {\bibfnamefont {M.}~\bibnamefont
  {G{\"a}rttner}}, \bibinfo {author} {\bibfnamefont {J.~G.}\ \bibnamefont
  {Bohnet}}, \bibinfo {author} {\bibfnamefont {A.}~\bibnamefont
  {Safavi-Naini}}, \bibinfo {author} {\bibfnamefont {M.~L.}\ \bibnamefont
  {Wall}}, \bibinfo {author} {\bibfnamefont {J.~J.}\ \bibnamefont {Bollinger}},
  \ and\ \bibinfo {author} {\bibfnamefont {A.~M.}\ \bibnamefont {Rey}},\
  }\href@noop {} {\bibfield  {journal} {\bibinfo  {journal} {Nature Physics}\
  }\textbf {\bibinfo {volume} {13}},\ \bibinfo {pages} {781} (\bibinfo {year}
  {2017})}\BibitemShut {NoStop}%
\bibitem [{\citenamefont {Landsman}\ \emph {et~al.}(2019)\citenamefont
  {Landsman}, \citenamefont {Figgatt}, \citenamefont {Schuster}, \citenamefont
  {Linke}, \citenamefont {Yoshida}, \citenamefont {Yao},\ and\ \citenamefont
  {Monroe}}]{landsman2019verified}%
  \BibitemOpen
  \bibfield  {author} {\bibinfo {author} {\bibfnamefont {K.~A.}\ \bibnamefont
  {Landsman}}, \bibinfo {author} {\bibfnamefont {C.}~\bibnamefont {Figgatt}},
  \bibinfo {author} {\bibfnamefont {T.}~\bibnamefont {Schuster}}, \bibinfo
  {author} {\bibfnamefont {N.~M.}\ \bibnamefont {Linke}}, \bibinfo {author}
  {\bibfnamefont {B.}~\bibnamefont {Yoshida}}, \bibinfo {author} {\bibfnamefont
  {N.~Y.}\ \bibnamefont {Yao}}, \ and\ \bibinfo {author} {\bibfnamefont
  {C.}~\bibnamefont {Monroe}},\ }\href@noop {} {\bibfield  {journal} {\bibinfo
  {journal} {Nature}\ }\textbf {\bibinfo {volume} {567}},\ \bibinfo {pages}
  {61} (\bibinfo {year} {2019})}\BibitemShut {NoStop}%
\bibitem [{\citenamefont {Kohmoto}\ \emph {et~al.}(1987)\citenamefont
  {Kohmoto}, \citenamefont {Sutherland},\ and\ \citenamefont
  {Tang}}]{kohmoto_rev}%
  \BibitemOpen
  \bibfield  {author} {\bibinfo {author} {\bibfnamefont {M.}~\bibnamefont
  {Kohmoto}}, \bibinfo {author} {\bibfnamefont {B.}~\bibnamefont {Sutherland}},
  \ and\ \bibinfo {author} {\bibfnamefont {C.}~\bibnamefont {Tang}},\ }\href
  {\doibase 10.1103/PhysRevB.35.1020} {\bibfield  {journal} {\bibinfo
  {journal} {Phys. Rev. B}\ }\textbf {\bibinfo {volume} {35}},\ \bibinfo
  {pages} {1020} (\bibinfo {year} {1987})}\BibitemShut {NoStop}%
\bibitem [{\citenamefont {Kohmoto}(1983)}]{kohmoto2}%
  \BibitemOpen
  \bibfield  {author} {\bibinfo {author} {\bibfnamefont {M.}~\bibnamefont
  {Kohmoto}},\ }\href {\doibase 10.1103/PhysRevLett.51.1198} {\bibfield
  {journal} {\bibinfo  {journal} {Phys. Rev. Lett.}\ }\textbf {\bibinfo
  {volume} {51}},\ \bibinfo {pages} {1198} (\bibinfo {year}
  {1983})}\BibitemShut {NoStop}%
\bibitem [{\citenamefont {Tang}\ and\ \citenamefont
  {Kohmoto}(1986)}]{kohmoto3}%
  \BibitemOpen
  \bibfield  {author} {\bibinfo {author} {\bibfnamefont {C.}~\bibnamefont
  {Tang}}\ and\ \bibinfo {author} {\bibfnamefont {M.}~\bibnamefont {Kohmoto}},\
  }\href {\doibase 10.1103/PhysRevB.34.2041} {\bibfield  {journal} {\bibinfo
  {journal} {Phys. Rev. B}\ }\textbf {\bibinfo {volume} {34}},\ \bibinfo
  {pages} {2041} (\bibinfo {year} {1986})}\BibitemShut {NoStop}%
\bibitem [{\citenamefont {Goldman}\ and\ \citenamefont {Kelton}(1993)}]{rmp}%
  \BibitemOpen
  \bibfield  {author} {\bibinfo {author} {\bibfnamefont {A.~I.}\ \bibnamefont
  {Goldman}}\ and\ \bibinfo {author} {\bibfnamefont {R.~F.}\ \bibnamefont
  {Kelton}},\ }\href {\doibase 10.1103/RevModPhys.65.213} {\bibfield  {journal}
  {\bibinfo  {journal} {Rev. Mod. Phys.}\ }\textbf {\bibinfo {volume} {65}},\
  \bibinfo {pages} {213} (\bibinfo {year} {1993})}\BibitemShut {NoStop}%
\bibitem [{\citenamefont {Aubry}\ and\ \citenamefont {Andr\'e}(1980)}]{aubry}%
  \BibitemOpen
  \bibfield  {author} {\bibinfo {author} {\bibfnamefont {S.}~\bibnamefont
  {Aubry}}\ and\ \bibinfo {author} {\bibfnamefont {G.}~\bibnamefont
  {Andr\'e}},\ }\href@noop {} {\bibfield  {journal} {\bibinfo  {journal} {Ann.
  Israel Phys. Soc}\ }\textbf {\bibinfo {volume} {3}},\ \bibinfo {pages} {18}
  (\bibinfo {year} {1980})}\BibitemShut {NoStop}%
\bibitem [{\citenamefont {Harper}(1955)}]{harper}%
  \BibitemOpen
  \bibfield  {author} {\bibinfo {author} {\bibfnamefont {P.~G.}\ \bibnamefont
  {Harper}},\ }\href@noop {} {\bibfield  {journal} {\bibinfo  {journal} {Proc.
  Phys. Soc. A}\ }\textbf {\bibinfo {volume} {68}},\ \bibinfo {pages} {874}
  (\bibinfo {year} {1955})}\BibitemShut {NoStop}%
\bibitem [{\citenamefont {Roati}\ \emph {et~al.}(2008)\citenamefont {Roati},
  \citenamefont {D’Errico}, \citenamefont {Fallani}, \citenamefont {Fattori},
  \citenamefont {Fort}, \citenamefont {Zaccanti}, \citenamefont {Modugno},
  \citenamefont {Modugno},\ and\ \citenamefont {Inguscio}}]{roati2008anderson}%
  \BibitemOpen
  \bibfield  {author} {\bibinfo {author} {\bibfnamefont {G.}~\bibnamefont
  {Roati}}, \bibinfo {author} {\bibfnamefont {C.}~\bibnamefont {D’Errico}},
  \bibinfo {author} {\bibfnamefont {L.}~\bibnamefont {Fallani}}, \bibinfo
  {author} {\bibfnamefont {M.}~\bibnamefont {Fattori}}, \bibinfo {author}
  {\bibfnamefont {C.}~\bibnamefont {Fort}}, \bibinfo {author} {\bibfnamefont
  {M.}~\bibnamefont {Zaccanti}}, \bibinfo {author} {\bibfnamefont
  {G.}~\bibnamefont {Modugno}}, \bibinfo {author} {\bibfnamefont
  {M.}~\bibnamefont {Modugno}}, \ and\ \bibinfo {author} {\bibfnamefont
  {M.}~\bibnamefont {Inguscio}},\ }\href@noop {} {\bibfield  {journal}
  {\bibinfo  {journal} {Nature}\ }\textbf {\bibinfo {volume} {453}},\ \bibinfo
  {pages} {895} (\bibinfo {year} {2008})}\BibitemShut {NoStop}%
\bibitem [{\citenamefont {Lahini}\ \emph {et~al.}(2009)\citenamefont {Lahini},
  \citenamefont {Pugatch}, \citenamefont {Pozzi}, \citenamefont {Sorel},
  \citenamefont {Morandotti}, \citenamefont {Davidson},\ and\ \citenamefont
  {Silberberg}}]{lahini}%
  \BibitemOpen
  \bibfield  {author} {\bibinfo {author} {\bibfnamefont {Y.}~\bibnamefont
  {Lahini}}, \bibinfo {author} {\bibfnamefont {R.}~\bibnamefont {Pugatch}},
  \bibinfo {author} {\bibfnamefont {F.}~\bibnamefont {Pozzi}}, \bibinfo
  {author} {\bibfnamefont {M.}~\bibnamefont {Sorel}}, \bibinfo {author}
  {\bibfnamefont {R.}~\bibnamefont {Morandotti}}, \bibinfo {author}
  {\bibfnamefont {N.}~\bibnamefont {Davidson}}, \ and\ \bibinfo {author}
  {\bibfnamefont {Y.}~\bibnamefont {Silberberg}},\ }\href {\doibase
  10.1103/PhysRevLett.103.013901} {\bibfield  {journal} {\bibinfo  {journal}
  {Phys. Rev. Lett.}\ }\textbf {\bibinfo {volume} {103}},\ \bibinfo {pages}
  {013901} (\bibinfo {year} {2009})}\BibitemShut {NoStop}%
\bibitem [{\citenamefont {Lye}\ \emph {et~al.}(2005)\citenamefont {Lye},
  \citenamefont {Fallani}, \citenamefont {Modugno}, \citenamefont {Wiersma},
  \citenamefont {Fort},\ and\ \citenamefont {Inguscio}}]{lye2005bose}%
  \BibitemOpen
  \bibfield  {author} {\bibinfo {author} {\bibfnamefont {J.}~\bibnamefont
  {Lye}}, \bibinfo {author} {\bibfnamefont {L.}~\bibnamefont {Fallani}},
  \bibinfo {author} {\bibfnamefont {M.}~\bibnamefont {Modugno}}, \bibinfo
  {author} {\bibfnamefont {D.}~\bibnamefont {Wiersma}}, \bibinfo {author}
  {\bibfnamefont {C.}~\bibnamefont {Fort}}, \ and\ \bibinfo {author}
  {\bibfnamefont {M.}~\bibnamefont {Inguscio}},\ }\href@noop {} {\bibfield
  {journal} {\bibinfo  {journal} {Physical review letters}\ }\textbf {\bibinfo
  {volume} {95}},\ \bibinfo {pages} {070401} (\bibinfo {year}
  {2005})}\BibitemShut {NoStop}%
\bibitem [{\citenamefont {Lucioni}\ \emph {et~al.}(2011)\citenamefont
  {Lucioni}, \citenamefont {Deissler}, \citenamefont {Tanzi}, \citenamefont
  {Roati}, \citenamefont {Zaccanti}, \citenamefont {Modugno}, \citenamefont
  {Larcher}, \citenamefont {Dalfovo}, \citenamefont {Inguscio},\ and\
  \citenamefont {Modugno}}]{lucioni2011observation}%
  \BibitemOpen
  \bibfield  {author} {\bibinfo {author} {\bibfnamefont {E.}~\bibnamefont
  {Lucioni}}, \bibinfo {author} {\bibfnamefont {B.}~\bibnamefont {Deissler}},
  \bibinfo {author} {\bibfnamefont {L.}~\bibnamefont {Tanzi}}, \bibinfo
  {author} {\bibfnamefont {G.}~\bibnamefont {Roati}}, \bibinfo {author}
  {\bibfnamefont {M.}~\bibnamefont {Zaccanti}}, \bibinfo {author}
  {\bibfnamefont {M.}~\bibnamefont {Modugno}}, \bibinfo {author} {\bibfnamefont
  {M.}~\bibnamefont {Larcher}}, \bibinfo {author} {\bibfnamefont
  {F.}~\bibnamefont {Dalfovo}}, \bibinfo {author} {\bibfnamefont
  {M.}~\bibnamefont {Inguscio}}, \ and\ \bibinfo {author} {\bibfnamefont
  {G.}~\bibnamefont {Modugno}},\ }\href@noop {} {\bibfield  {journal} {\bibinfo
   {journal} {Physical review letters}\ }\textbf {\bibinfo {volume} {106}},\
  \bibinfo {pages} {230403} (\bibinfo {year} {2011})}\BibitemShut {NoStop}%
\bibitem [{\citenamefont {Oganesyan}\ and\ \citenamefont
  {Huse}(2007)}]{huse2007}%
  \BibitemOpen
  \bibfield  {author} {\bibinfo {author} {\bibfnamefont {V.}~\bibnamefont
  {Oganesyan}}\ and\ \bibinfo {author} {\bibfnamefont {D.~A.}\ \bibnamefont
  {Huse}},\ }\href {\doibase 10.1103/PhysRevB.75.155111} {\bibfield  {journal}
  {\bibinfo  {journal} {Phys. Rev. B}\ }\textbf {\bibinfo {volume} {75}},\
  \bibinfo {pages} {155111} (\bibinfo {year} {2007})}\BibitemShut {NoStop}%
\bibitem [{\citenamefont {Pal}\ and\ \citenamefont {Huse}(2010)}]{Pal2010}%
  \BibitemOpen
  \bibfield  {author} {\bibinfo {author} {\bibfnamefont {A.}~\bibnamefont
  {Pal}}\ and\ \bibinfo {author} {\bibfnamefont {D.~A.}\ \bibnamefont {Huse}},\
  }\href {\doibase 10.1103/PhysRevB.82.174411} {\bibfield  {journal} {\bibinfo
  {journal} {Phys. Rev. B}\ }\textbf {\bibinfo {volume} {82}},\ \bibinfo
  {pages} {174411} (\bibinfo {year} {2010})}\BibitemShut {NoStop}%
\bibitem [{\citenamefont {Iyer}\ \emph {et~al.}(2013)\citenamefont {Iyer},
  \citenamefont {Oganesyan}, \citenamefont {Refael},\ and\ \citenamefont
  {Huse}}]{iyer}%
  \BibitemOpen
  \bibfield  {author} {\bibinfo {author} {\bibfnamefont {S.}~\bibnamefont
  {Iyer}}, \bibinfo {author} {\bibfnamefont {V.}~\bibnamefont {Oganesyan}},
  \bibinfo {author} {\bibfnamefont {G.}~\bibnamefont {Refael}}, \ and\ \bibinfo
  {author} {\bibfnamefont {D.~A.}\ \bibnamefont {Huse}},\ }\href {\doibase
  10.1103/PhysRevB.87.134202} {\bibfield  {journal} {\bibinfo  {journal} {Phys.
  Rev. B}\ }\textbf {\bibinfo {volume} {87}},\ \bibinfo {pages} {134202}
  (\bibinfo {year} {2013})}\BibitemShut {NoStop}%
\bibitem [{\citenamefont {Schreiber}\ \emph {et~al.}(2015)\citenamefont
  {Schreiber}, \citenamefont {Hodgman}, \citenamefont {Bordia}, \citenamefont
  {L{\"u}schen}, \citenamefont {Fischer}, \citenamefont {Vosk}, \citenamefont
  {Altman}, \citenamefont {Schneider},\ and\ \citenamefont {Bloch}}]{Bordia}%
  \BibitemOpen
  \bibfield  {author} {\bibinfo {author} {\bibfnamefont {M.}~\bibnamefont
  {Schreiber}}, \bibinfo {author} {\bibfnamefont {S.~S.}\ \bibnamefont
  {Hodgman}}, \bibinfo {author} {\bibfnamefont {P.}~\bibnamefont {Bordia}},
  \bibinfo {author} {\bibfnamefont {H.~P.}\ \bibnamefont {L{\"u}schen}},
  \bibinfo {author} {\bibfnamefont {M.~H.}\ \bibnamefont {Fischer}}, \bibinfo
  {author} {\bibfnamefont {R.}~\bibnamefont {Vosk}}, \bibinfo {author}
  {\bibfnamefont {E.}~\bibnamefont {Altman}}, \bibinfo {author} {\bibfnamefont
  {U.}~\bibnamefont {Schneider}}, \ and\ \bibinfo {author} {\bibfnamefont
  {I.}~\bibnamefont {Bloch}},\ }\href {\doibase 10.1126/science.aaa7432}
  {\bibfield  {journal} {\bibinfo  {journal} {Science}\ }\textbf {\bibinfo
  {volume} {349}},\ \bibinfo {pages} {842} (\bibinfo {year}
  {2015})}\BibitemShut {NoStop}%
\bibitem [{\citenamefont {Deng}\ \emph {et~al.}(2019)\citenamefont {Deng},
  \citenamefont {Ray}, \citenamefont {Sinha}, \citenamefont {Shlyapnikov},\
  and\ \citenamefont {Santos}}]{deng2019one}%
  \BibitemOpen
  \bibfield  {author} {\bibinfo {author} {\bibfnamefont {X.}~\bibnamefont
  {Deng}}, \bibinfo {author} {\bibfnamefont {S.}~\bibnamefont {Ray}}, \bibinfo
  {author} {\bibfnamefont {S.}~\bibnamefont {Sinha}}, \bibinfo {author}
  {\bibfnamefont {G.}~\bibnamefont {Shlyapnikov}}, \ and\ \bibinfo {author}
  {\bibfnamefont {L.}~\bibnamefont {Santos}},\ }\href@noop {} {\bibfield
  {journal} {\bibinfo  {journal} {Physical review letters}\ }\textbf {\bibinfo
  {volume} {123}},\ \bibinfo {pages} {025301} (\bibinfo {year}
  {2019})}\BibitemShut {NoStop}%
\bibitem [{\citenamefont {Roy}\ and\ \citenamefont
  {Sharma}(2021)}]{roy2020prescription}%
  \BibitemOpen
  \bibfield  {author} {\bibinfo {author} {\bibfnamefont {N.}~\bibnamefont
  {Roy}}\ and\ \bibinfo {author} {\bibfnamefont {A.}~\bibnamefont {Sharma}},\
  }\href@noop {} {\bibfield  {journal} {\bibinfo  {journal} {Physical Review
  B}\ }\textbf {\bibinfo {volume} {103}},\ \bibinfo {pages} {075124} (\bibinfo
  {year} {2021})}\BibitemShut {NoStop}%
\bibitem [{\citenamefont {Lieb}\ and\ \citenamefont {Robinson}(1972)}]{lieb}%
  \BibitemOpen
  \bibfield  {author} {\bibinfo {author} {\bibfnamefont {E.~H.}\ \bibnamefont
  {Lieb}}\ and\ \bibinfo {author} {\bibfnamefont {D.~W.}\ \bibnamefont
  {Robinson}},\ }\href {https://projecteuclid.org:443/euclid.cmp/1103858407}
  {\bibfield  {journal} {\bibinfo  {journal} {Comm. Math. Phys.}\ }\textbf
  {\bibinfo {volume} {28}},\ \bibinfo {pages} {251} (\bibinfo {year}
  {1972})}\BibitemShut {NoStop}%
\bibitem [{\citenamefont {Sekino}\ and\ \citenamefont
  {Susskind}(2008)}]{sekino2008fast}%
  \BibitemOpen
  \bibfield  {author} {\bibinfo {author} {\bibfnamefont {Y.}~\bibnamefont
  {Sekino}}\ and\ \bibinfo {author} {\bibfnamefont {L.}~\bibnamefont
  {Susskind}},\ }\href@noop {} {\bibfield  {journal} {\bibinfo  {journal}
  {Journal of High Energy Physics}\ }\textbf {\bibinfo {volume} {2008}},\
  \bibinfo {pages} {065} (\bibinfo {year} {2008})}\BibitemShut {NoStop}%
\bibitem [{\citenamefont {Sachdev}(2015)}]{sachdev2015bekenstein}%
  \BibitemOpen
  \bibfield  {author} {\bibinfo {author} {\bibfnamefont {S.}~\bibnamefont
  {Sachdev}},\ }\href@noop {} {\bibfield  {journal} {\bibinfo  {journal}
  {Physical Review X}\ }\textbf {\bibinfo {volume} {5}},\ \bibinfo {pages}
  {041025} (\bibinfo {year} {2015})}\BibitemShut {NoStop}%
\bibitem [{\citenamefont {Huang}\ \emph {et~al.}(2017)\citenamefont {Huang},
  \citenamefont {Zhang},\ and\ \citenamefont {Chen}}]{huang2017out}%
  \BibitemOpen
  \bibfield  {author} {\bibinfo {author} {\bibfnamefont {Y.}~\bibnamefont
  {Huang}}, \bibinfo {author} {\bibfnamefont {Y.-L.}\ \bibnamefont {Zhang}}, \
  and\ \bibinfo {author} {\bibfnamefont {X.}~\bibnamefont {Chen}},\ }\href@noop
  {} {\bibfield  {journal} {\bibinfo  {journal} {Annalen der Physik}\ }\textbf
  {\bibinfo {volume} {529}},\ \bibinfo {pages} {1600318} (\bibinfo {year}
  {2017})}\BibitemShut {NoStop}%
\bibitem [{\citenamefont {Ch{\'a}vez-Carlos}\ \emph {et~al.}(2019)\citenamefont
  {Ch{\'a}vez-Carlos}, \citenamefont {L{\'o}pez-del Carpio}, \citenamefont
  {Bastarrachea-Magnani}, \citenamefont {Str{\'a}nsk{\`y}}, \citenamefont
  {Lerma-Hern{\'a}ndez}, \citenamefont {Santos},\ and\ \citenamefont
  {Hirsch}}]{chavez2019quantum}%
  \BibitemOpen
  \bibfield  {author} {\bibinfo {author} {\bibfnamefont {J.}~\bibnamefont
  {Ch{\'a}vez-Carlos}}, \bibinfo {author} {\bibfnamefont {B.}~\bibnamefont
  {L{\'o}pez-del Carpio}}, \bibinfo {author} {\bibfnamefont {M.~A.}\
  \bibnamefont {Bastarrachea-Magnani}}, \bibinfo {author} {\bibfnamefont
  {P.}~\bibnamefont {Str{\'a}nsk{\`y}}}, \bibinfo {author} {\bibfnamefont
  {S.}~\bibnamefont {Lerma-Hern{\'a}ndez}}, \bibinfo {author} {\bibfnamefont
  {L.~F.}\ \bibnamefont {Santos}}, \ and\ \bibinfo {author} {\bibfnamefont
  {J.~G.}\ \bibnamefont {Hirsch}},\ }\href@noop {} {\bibfield  {journal}
  {\bibinfo  {journal} {Physical Review Letters}\ }\textbf {\bibinfo {volume}
  {122}},\ \bibinfo {pages} {024101} (\bibinfo {year} {2019})}\BibitemShut
  {NoStop}%
\bibitem [{\citenamefont {Jalabert}\ \emph {et~al.}(2018)\citenamefont
  {Jalabert}, \citenamefont {Garc{\'\i}a-Mata},\ and\ \citenamefont
  {Wisniacki}}]{jalabert2018semiclassical}%
  \BibitemOpen
  \bibfield  {author} {\bibinfo {author} {\bibfnamefont {R.~A.}\ \bibnamefont
  {Jalabert}}, \bibinfo {author} {\bibfnamefont {I.}~\bibnamefont
  {Garc{\'\i}a-Mata}}, \ and\ \bibinfo {author} {\bibfnamefont {D.~A.}\
  \bibnamefont {Wisniacki}},\ }\href@noop {} {\bibfield  {journal} {\bibinfo
  {journal} {Physical Review E}\ }\textbf {\bibinfo {volume} {98}},\ \bibinfo
  {pages} {062218} (\bibinfo {year} {2018})}\BibitemShut {NoStop}%
\bibitem [{\citenamefont {Lin}\ and\ \citenamefont
  {Motrunich}(2018{\natexlab{a}})}]{lin2018out}%
  \BibitemOpen
  \bibfield  {author} {\bibinfo {author} {\bibfnamefont {C.-J.}\ \bibnamefont
  {Lin}}\ and\ \bibinfo {author} {\bibfnamefont {O.~I.}\ \bibnamefont
  {Motrunich}},\ }\href@noop {} {\bibfield  {journal} {\bibinfo  {journal}
  {Physical Review B}\ }\textbf {\bibinfo {volume} {97}},\ \bibinfo {pages}
  {144304} (\bibinfo {year} {2018}{\natexlab{a}})}\BibitemShut {NoStop}%
\bibitem [{\citenamefont {Bao}\ and\ \citenamefont {Zhang}(2019)}]{bao2019out}%
  \BibitemOpen
  \bibfield  {author} {\bibinfo {author} {\bibfnamefont {J.}~\bibnamefont
  {Bao}}\ and\ \bibinfo {author} {\bibfnamefont {C.-Y.}\ \bibnamefont
  {Zhang}},\ }\href@noop {} {\bibfield  {journal} {\bibinfo  {journal} {arXiv
  preprint arXiv:1901.09327}\ } (\bibinfo {year} {2019})}\BibitemShut {NoStop}%
\bibitem [{\citenamefont {Fortes}\ \emph {et~al.}(2020)\citenamefont {Fortes},
  \citenamefont {Garc{\'\i}a-Mata}, \citenamefont {Jalabert},\ and\
  \citenamefont {Wisniacki}}]{fortes2020signatures}%
  \BibitemOpen
  \bibfield  {author} {\bibinfo {author} {\bibfnamefont {E.~M.}\ \bibnamefont
  {Fortes}}, \bibinfo {author} {\bibfnamefont {I.}~\bibnamefont
  {Garc{\'\i}a-Mata}}, \bibinfo {author} {\bibfnamefont {R.~A.}\ \bibnamefont
  {Jalabert}}, \ and\ \bibinfo {author} {\bibfnamefont {D.~A.}\ \bibnamefont
  {Wisniacki}},\ }\href@noop {} {\bibfield  {journal} {\bibinfo  {journal}
  {arXiv preprint arXiv:2004.14440}\ } (\bibinfo {year} {2020})}\BibitemShut
  {NoStop}%
\bibitem [{\citenamefont {Yan}\ \emph {et~al.}(2019)\citenamefont {Yan},
  \citenamefont {Wang},\ and\ \citenamefont {Wang}}]{yan2019similar}%
  \BibitemOpen
  \bibfield  {author} {\bibinfo {author} {\bibfnamefont {H.}~\bibnamefont
  {Yan}}, \bibinfo {author} {\bibfnamefont {J.-Z.}\ \bibnamefont {Wang}}, \
  and\ \bibinfo {author} {\bibfnamefont {W.-G.}\ \bibnamefont {Wang}},\
  }\href@noop {} {\bibfield  {journal} {\bibinfo  {journal} {Communications in
  Theoretical Physics}\ }\textbf {\bibinfo {volume} {71}},\ \bibinfo {pages}
  {1359} (\bibinfo {year} {2019})}\BibitemShut {NoStop}%
\bibitem [{\citenamefont {Riddell}\ and\ \citenamefont
  {S{\o}rensen}(2019)}]{riddell2019out}%
  \BibitemOpen
  \bibfield  {author} {\bibinfo {author} {\bibfnamefont {J.}~\bibnamefont
  {Riddell}}\ and\ \bibinfo {author} {\bibfnamefont {E.~S.}\ \bibnamefont
  {S{\o}rensen}},\ }\href@noop {} {\bibfield  {journal} {\bibinfo  {journal}
  {Physical Review B}\ }\textbf {\bibinfo {volume} {99}},\ \bibinfo {pages}
  {054205} (\bibinfo {year} {2019})}\BibitemShut {NoStop}%
\bibitem [{\citenamefont {Riddell}\ and\ \citenamefont
  {S{\o}rensen}(2020)}]{riddell2020out}%
  \BibitemOpen
  \bibfield  {author} {\bibinfo {author} {\bibfnamefont {J.}~\bibnamefont
  {Riddell}}\ and\ \bibinfo {author} {\bibfnamefont {E.~S.}\ \bibnamefont
  {S{\o}rensen}},\ }\href@noop {} {\bibfield  {journal} {\bibinfo  {journal}
  {Physical Review B}\ }\textbf {\bibinfo {volume} {101}},\ \bibinfo {pages}
  {024202} (\bibinfo {year} {2020})}\BibitemShut {NoStop}%
\bibitem [{\citenamefont {Lee}\ \emph {et~al.}(2019)\citenamefont {Lee},
  \citenamefont {Kim},\ and\ \citenamefont {Kim}}]{lee2019typical}%
  \BibitemOpen
  \bibfield  {author} {\bibinfo {author} {\bibfnamefont {J.}~\bibnamefont
  {Lee}}, \bibinfo {author} {\bibfnamefont {D.}~\bibnamefont {Kim}}, \ and\
  \bibinfo {author} {\bibfnamefont {D.-H.}\ \bibnamefont {Kim}},\ }\href@noop
  {} {\bibfield  {journal} {\bibinfo  {journal} {Physical Review B}\ }\textbf
  {\bibinfo {volume} {99}},\ \bibinfo {pages} {184202} (\bibinfo {year}
  {2019})}\BibitemShut {NoStop}%
\bibitem [{\citenamefont {Bordia}\ \emph {et~al.}(2018)\citenamefont {Bordia},
  \citenamefont {Alet},\ and\ \citenamefont {Hosur}}]{bordia2018out}%
  \BibitemOpen
  \bibfield  {author} {\bibinfo {author} {\bibfnamefont {P.}~\bibnamefont
  {Bordia}}, \bibinfo {author} {\bibfnamefont {F.}~\bibnamefont {Alet}}, \ and\
  \bibinfo {author} {\bibfnamefont {P.}~\bibnamefont {Hosur}},\ }\href@noop {}
  {\bibfield  {journal} {\bibinfo  {journal} {Physical Review A}\ }\textbf
  {\bibinfo {volume} {97}},\ \bibinfo {pages} {030103} (\bibinfo {year}
  {2018})}\BibitemShut {NoStop}%
\bibitem [{\citenamefont {Modugno}(2009)}]{modugno2009exponential}%
  \BibitemOpen
  \bibfield  {author} {\bibinfo {author} {\bibfnamefont {M.}~\bibnamefont
  {Modugno}},\ }\href {http://stacks.iop.org/1367-2630/11/i=3/a=033023}
  {\bibfield  {journal} {\bibinfo  {journal} {New Journal of Physics}\ }\textbf
  {\bibinfo {volume} {11}},\ \bibinfo {pages} {033023} (\bibinfo {year}
  {2009})}\BibitemShut {NoStop}%
\bibitem [{\citenamefont {Bugeaud}(2008)}]{diophantine}%
  \BibitemOpen
  \bibfield  {author} {\bibinfo {author} {\bibfnamefont {Y.}~\bibnamefont
  {Bugeaud}},\ }\href {\doibase 10.1007/s00208-008-0209-4} {\bibfield
  {journal} {\bibinfo  {journal} {Mathematische Annalen}\ }\textbf {\bibinfo
  {volume} {341}},\ \bibinfo {pages} {677} (\bibinfo {year}
  {2008})}\BibitemShut {NoStop}%
\bibitem [{\citenamefont {Cohn}(2006)}]{continuedfraction}%
  \BibitemOpen
  \bibfield  {author} {\bibinfo {author} {\bibfnamefont {H.}~\bibnamefont
  {Cohn}},\ }\href {\doibase 10.1080/00029890.2006.11920278} {\bibfield
  {journal} {\bibinfo  {journal} {The American Mathematical Monthly}\ }\textbf
  {\bibinfo {volume} {113}},\ \bibinfo {pages} {57} (\bibinfo {year} {2006})},\
  \Eprint {http://arxiv.org/abs/https://doi.org/10.1080/00029890.2006.11920278}
  {https://doi.org/10.1080/00029890.2006.11920278} \BibitemShut {NoStop}%
\bibitem [{\citenamefont {Roy}\ and\ \citenamefont
  {Sharma}(2019)}]{roy2019study}%
  \BibitemOpen
  \bibfield  {author} {\bibinfo {author} {\bibfnamefont {N.}~\bibnamefont
  {Roy}}\ and\ \bibinfo {author} {\bibfnamefont {A.}~\bibnamefont {Sharma}},\
  }\href@noop {} {\bibfield  {journal} {\bibinfo  {journal} {Physical Review
  B}\ }\textbf {\bibinfo {volume} {100}},\ \bibinfo {pages} {195143} (\bibinfo
  {year} {2019})}\BibitemShut {NoStop}%
\bibitem [{\citenamefont {McGinley}\ \emph {et~al.}(2019)\citenamefont
  {McGinley}, \citenamefont {Nunnenkamp},\ and\ \citenamefont
  {Knolle}}]{mcginley2019slow}%
  \BibitemOpen
  \bibfield  {author} {\bibinfo {author} {\bibfnamefont {M.}~\bibnamefont
  {McGinley}}, \bibinfo {author} {\bibfnamefont {A.}~\bibnamefont
  {Nunnenkamp}}, \ and\ \bibinfo {author} {\bibfnamefont {J.}~\bibnamefont
  {Knolle}},\ }\href@noop {} {\bibfield  {journal} {\bibinfo  {journal}
  {Physical review letters}\ }\textbf {\bibinfo {volume} {122}},\ \bibinfo
  {pages} {020603} (\bibinfo {year} {2019})}\BibitemShut {NoStop}%
\bibitem [{\citenamefont {Peschel}(2003)}]{peschel2003calculation}%
  \BibitemOpen
  \bibfield  {author} {\bibinfo {author} {\bibfnamefont {I.}~\bibnamefont
  {Peschel}},\ }\href@noop {} {\bibfield  {journal} {\bibinfo  {journal}
  {Journal of Physics A: Mathematical and General}\ }\textbf {\bibinfo {volume}
  {36}},\ \bibinfo {pages} {L205} (\bibinfo {year} {2003})}\BibitemShut
  {NoStop}%
\bibitem [{\citenamefont {Roy}\ and\ \citenamefont {Sharma}(2018)}]{prb}%
  \BibitemOpen
  \bibfield  {author} {\bibinfo {author} {\bibfnamefont {N.}~\bibnamefont
  {Roy}}\ and\ \bibinfo {author} {\bibfnamefont {A.}~\bibnamefont {Sharma}},\
  }\href@noop {} {\bibfield  {journal} {\bibinfo  {journal} {Physical Review
  B}\ }\textbf {\bibinfo {volume} {97}},\ \bibinfo {pages} {125116} (\bibinfo
  {year} {2018})}\BibitemShut {NoStop}%
\bibitem [{\citenamefont {Ro\'osz}\ \emph {et~al.}(2014)\citenamefont
  {Ro\'osz}, \citenamefont {Divakaran}, \citenamefont {Rieger},\ and\
  \citenamefont {Igl\'oi}}]{divakaran}%
  \BibitemOpen
  \bibfield  {author} {\bibinfo {author} {\bibfnamefont {G.~m.~H.}\
  \bibnamefont {Ro\'osz}}, \bibinfo {author} {\bibfnamefont {U.}~\bibnamefont
  {Divakaran}}, \bibinfo {author} {\bibfnamefont {H.}~\bibnamefont {Rieger}}, \
  and\ \bibinfo {author} {\bibfnamefont {F.}~\bibnamefont {Igl\'oi}},\ }\href
  {\doibase 10.1103/PhysRevB.90.184202} {\bibfield  {journal} {\bibinfo
  {journal} {Phys. Rev. B}\ }\textbf {\bibinfo {volume} {90}},\ \bibinfo
  {pages} {184202} (\bibinfo {year} {2014})}\BibitemShut {NoStop}%
\bibitem [{\citenamefont {Hetterich}\ \emph {et~al.}(2017)\citenamefont
  {Hetterich}, \citenamefont {Serbyn}, \citenamefont {Dom{\'\i}nguez},
  \citenamefont {Pollmann},\ and\ \citenamefont
  {Trauzettel}}]{hetterich2017noninteracting}%
  \BibitemOpen
  \bibfield  {author} {\bibinfo {author} {\bibfnamefont {D.}~\bibnamefont
  {Hetterich}}, \bibinfo {author} {\bibfnamefont {M.}~\bibnamefont {Serbyn}},
  \bibinfo {author} {\bibfnamefont {F.}~\bibnamefont {Dom{\'\i}nguez}},
  \bibinfo {author} {\bibfnamefont {F.}~\bibnamefont {Pollmann}}, \ and\
  \bibinfo {author} {\bibfnamefont {B.}~\bibnamefont {Trauzettel}},\
  }\href@noop {} {\bibfield  {journal} {\bibinfo  {journal} {Physical Review
  B}\ }\textbf {\bibinfo {volume} {96}},\ \bibinfo {pages} {104203} (\bibinfo
  {year} {2017})}\BibitemShut {NoStop}%
\bibitem [{\citenamefont {Lerose}\ and\ \citenamefont
  {Pappalardi}(2020)}]{lerose2020origin}%
  \BibitemOpen
  \bibfield  {author} {\bibinfo {author} {\bibfnamefont {A.}~\bibnamefont
  {Lerose}}\ and\ \bibinfo {author} {\bibfnamefont {S.}~\bibnamefont
  {Pappalardi}},\ }\href@noop {} {\bibfield  {journal} {\bibinfo  {journal}
  {Physical Review Research}\ }\textbf {\bibinfo {volume} {2}},\ \bibinfo
  {pages} {012041} (\bibinfo {year} {2020})}\BibitemShut {NoStop}%
\bibitem [{\citenamefont {Buyskikh}\ \emph {et~al.}(2016)\citenamefont
  {Buyskikh}, \citenamefont {Fagotti}, \citenamefont {Schachenmayer},
  \citenamefont {Essler},\ and\ \citenamefont {Daley}}]{daley}%
  \BibitemOpen
  \bibfield  {author} {\bibinfo {author} {\bibfnamefont {A.~S.}\ \bibnamefont
  {Buyskikh}}, \bibinfo {author} {\bibfnamefont {M.}~\bibnamefont {Fagotti}},
  \bibinfo {author} {\bibfnamefont {J.}~\bibnamefont {Schachenmayer}}, \bibinfo
  {author} {\bibfnamefont {F.}~\bibnamefont {Essler}}, \ and\ \bibinfo {author}
  {\bibfnamefont {A.~J.}\ \bibnamefont {Daley}},\ }\href {\doibase
  10.1103/PhysRevA.93.053620} {\bibfield  {journal} {\bibinfo  {journal} {Phys.
  Rev. A}\ }\textbf {\bibinfo {volume} {93}},\ \bibinfo {pages} {053620}
  (\bibinfo {year} {2016})}\BibitemShut {NoStop}%
\bibitem [{\citenamefont {Modak}\ and\ \citenamefont
  {Nag}(2020)}]{modak2020many}%
  \BibitemOpen
  \bibfield  {author} {\bibinfo {author} {\bibfnamefont {R.}~\bibnamefont
  {Modak}}\ and\ \bibinfo {author} {\bibfnamefont {T.}~\bibnamefont {Nag}},\
  }\href@noop {} {\bibfield  {journal} {\bibinfo  {journal} {Physical Review
  Research}\ }\textbf {\bibinfo {volume} {2}},\ \bibinfo {pages} {012074}
  (\bibinfo {year} {2020})}\BibitemShut {NoStop}%
\bibitem [{\citenamefont {Falcon}(2014)}]{fibonacci}%
  \BibitemOpen
  \bibfield  {author} {\bibinfo {author} {\bibfnamefont {S.}~\bibnamefont
  {Falcon}},\ }\href {\doibase https://doi: 10.4236/am.2014.515216} {\bibfield
  {journal} {\bibinfo  {journal} {Applied Mathematics}\ }\textbf {\bibinfo
  {volume} {5}},\ \bibinfo {pages} {2226} (\bibinfo {year} {2014})}\BibitemShut
  {NoStop}%
\bibitem [{\citenamefont {Lin}\ and\ \citenamefont
  {Motrunich}(2018{\natexlab{b}})}]{lin2018out_2}%
  \BibitemOpen
  \bibfield  {author} {\bibinfo {author} {\bibfnamefont {C.-J.}\ \bibnamefont
  {Lin}}\ and\ \bibinfo {author} {\bibfnamefont {O.~I.}\ \bibnamefont
  {Motrunich}},\ }\href@noop {} {\bibfield  {journal} {\bibinfo  {journal}
  {Physical Review B}\ }\textbf {\bibinfo {volume} {98}},\ \bibinfo {pages}
  {134305} (\bibinfo {year} {2018}{\natexlab{b}})}\BibitemShut {NoStop}%
\bibitem [{\citenamefont {Chen}\ \emph {et~al.}(2020)\citenamefont {Chen},
  \citenamefont {Hou}, \citenamefont {Zhou}, \citenamefont {Qian},
  \citenamefont {Shen},\ and\ \citenamefont {Xu}}]{chen2020detecting}%
  \BibitemOpen
  \bibfield  {author} {\bibinfo {author} {\bibfnamefont {B.}~\bibnamefont
  {Chen}}, \bibinfo {author} {\bibfnamefont {X.}~\bibnamefont {Hou}}, \bibinfo
  {author} {\bibfnamefont {F.}~\bibnamefont {Zhou}}, \bibinfo {author}
  {\bibfnamefont {P.}~\bibnamefont {Qian}}, \bibinfo {author} {\bibfnamefont
  {H.}~\bibnamefont {Shen}}, \ and\ \bibinfo {author} {\bibfnamefont
  {N.}~\bibnamefont {Xu}},\ }\href@noop {} {\bibfield  {journal} {\bibinfo
  {journal} {arXiv preprint arXiv:2001.06333}\ } (\bibinfo {year}
  {2020})}\BibitemShut {NoStop}%
\bibitem [{\citenamefont {Da{\u{g}}}\ \emph {et~al.}(2019)\citenamefont
  {Da{\u{g}}}, \citenamefont {Sun},\ and\ \citenamefont
  {Duan}}]{daug2019detection}%
  \BibitemOpen
  \bibfield  {author} {\bibinfo {author} {\bibfnamefont {C.~B.}\ \bibnamefont
  {Da{\u{g}}}}, \bibinfo {author} {\bibfnamefont {K.}~\bibnamefont {Sun}}, \
  and\ \bibinfo {author} {\bibfnamefont {L.-M.}\ \bibnamefont {Duan}},\
  }\href@noop {} {\bibfield  {journal} {\bibinfo  {journal} {Physical review
  letters}\ }\textbf {\bibinfo {volume} {123}},\ \bibinfo {pages} {140602}
  (\bibinfo {year} {2019})}\BibitemShut {NoStop}%
\bibitem [{\citenamefont {Lewis-Swan}\ \emph {et~al.}(2020)\citenamefont
  {Lewis-Swan}, \citenamefont {Muleady},\ and\ \citenamefont
  {Rey}}]{lewis2020detecting}%
  \BibitemOpen
  \bibfield  {author} {\bibinfo {author} {\bibfnamefont {R.}~\bibnamefont
  {Lewis-Swan}}, \bibinfo {author} {\bibfnamefont {S.}~\bibnamefont {Muleady}},
  \ and\ \bibinfo {author} {\bibfnamefont {A.}~\bibnamefont {Rey}},\
  }\href@noop {} {\bibfield  {journal} {\bibinfo  {journal} {Physical Review
  Letters}\ }\textbf {\bibinfo {volume} {125}},\ \bibinfo {pages} {240605}
  (\bibinfo {year} {2020})}\BibitemShut {NoStop}%
\bibitem [{\citenamefont {Nie}\ \emph {et~al.}(2020)\citenamefont {Nie},
  \citenamefont {Wei}, \citenamefont {Chen}, \citenamefont {Zhang},
  \citenamefont {Zhao}, \citenamefont {Qiu}, \citenamefont {Tian},
  \citenamefont {Ji}, \citenamefont {Xin}, \citenamefont {Lu} \emph
  {et~al.}}]{nie2020experimental}%
  \BibitemOpen
  \bibfield  {author} {\bibinfo {author} {\bibfnamefont {X.}~\bibnamefont
  {Nie}}, \bibinfo {author} {\bibfnamefont {B.-B.}\ \bibnamefont {Wei}},
  \bibinfo {author} {\bibfnamefont {X.}~\bibnamefont {Chen}}, \bibinfo {author}
  {\bibfnamefont {Z.}~\bibnamefont {Zhang}}, \bibinfo {author} {\bibfnamefont
  {X.}~\bibnamefont {Zhao}}, \bibinfo {author} {\bibfnamefont {C.}~\bibnamefont
  {Qiu}}, \bibinfo {author} {\bibfnamefont {Y.}~\bibnamefont {Tian}}, \bibinfo
  {author} {\bibfnamefont {Y.}~\bibnamefont {Ji}}, \bibinfo {author}
  {\bibfnamefont {T.}~\bibnamefont {Xin}}, \bibinfo {author} {\bibfnamefont
  {D.}~\bibnamefont {Lu}},  \emph {et~al.},\ }\href@noop {} {\bibfield
  {journal} {\bibinfo  {journal} {Physical Review Letters}\ }\textbf {\bibinfo
  {volume} {124}},\ \bibinfo {pages} {250601} (\bibinfo {year}
  {2020})}\BibitemShut {NoStop}%
\end{thebibliography}%

\bigskip

\appendix

\end{document}